\documentclass[aps,onecolumn,letterpaper,superscriptaddress,notitlepage]{revtex4-2}
\usepackage{CJK}
\usepackage{amsmath}
\usepackage{comment}
\usepackage{eufrak}
\usepackage{natbib}
\usepackage{hyperref}
\usepackage[left=0.75in, right=0.7in, top=0.7in, bottom=1in]{geometry}
\usepackage{xcolor}
\pdfoutput=1

\begin{document}
\begin{CJK*}{GB}{}
\title{Proof of Gauge-Invariance of Magnetic Susceptibility of Topologically Trivial Insulators in the Independent Particle Approximation}
\author{Alistair H. Duff$^{1}$,  and J.E. Sipe}
\affiliation{Department of Physics, University of Toronto, Ontario M5S 1A7, Canada}
\begin{abstract}
    We have derived an expression for the magnetic susceptibility of topologically trivial insulators, however an important consideration for any response tensor is whether it is gauge-invariant. By this we refer to the gauge-freedom in choosing the Bloch functions one uses in a computation since multiplication by an arbitrary $\textbf{k}$-dependent complex phase produces equally valid Bloch functions. Additionally Wannier functions, which can be constructed by a unitary mixing of the Bloch functions, are a useful basis for computation due to their localized properties, but are strongly non-unique. Therefore, we show that the theoretical expression we derived is independent of the choice of Bloch functions or Wannier functions used, and thus is gauge-invariant in this sense. Lastly, we compare our results with a few existing expressions in the literature and settle a dispute on what the correct formula should be.  
\end{abstract}
\maketitle
\end{CJK*}

\begin{widetext}
\section{Introduction}

In our approach to determining the magnetic susceptibility of an insulator in the independent particle approximation, there is a multiband gauge-freedom that is naturally tracked by the Wannier function basis used as an intermediate in calculations. The gauge-freedom is described by a unitary matrix $U(\textbf{k})$ that relates the Bloch and Wannier function bases used. The Wannier functions are strongly non-unique due to the freedom in choosing the matrix $U(\textbf{k})$ that then alters the localization properties of the  resultant Wannier functions. Additionally, the choice of Bloch functions is not unique since one can apply an arbitrary $\textbf{k}$-dependent complex phase and recover equally valid eigenstates. The latter gauge-freedom is commonly considered in the derivation of response tensors, such as the magnetic susceptibility. In the following we will divide up expressions into those parts that depend purely on the dispersion relation and Bloch functions and those that also depend on $U(\textbf{k})$ and its derivatives. If one chooses $U_{n\alpha}(\textbf{k}) = \delta_{n\alpha} e^{-i\phi_{n}(\textbf{k})}$ the additional pieces are exactly what one would get upon changing the phase of the Bloch functions used in the calculation. So, if we can show that the total expression for the magnetic susceptibility does not depend on the $U(\textbf{k})$s and its derivatives, then we have shown that it is gauge-invariant (in that the response does not depend on the choice of Bloch functions or Wannier functions used in the computation). 

Additional details, such as comparisons to the existing literature are also included, proving equivalence between different strategies.

\section{Single Particle Density Matrix}

We use a Green function framework expanded in a Wannier function basis. An important quantity is the single particle density matrix (SPDM), its components are defined as:  

\begin{equation}
\label{SPDM}
\eta_{\alpha\textbf{R};\beta\textbf{R}'}(t) = \langle \hat{a}^\dag_{\alpha\textbf{R}}(t) \hat{a}_{\beta\textbf{R}'}(t) \rangle,
\end{equation}
where $\hat{a}_{\alpha\textbf{R}}$ (and its Hermitian adjoint) are the fermionic annihilation (creation) operators for an electron in a Wannier function state $\alpha$ associated with site $\textbf{R}$. The details of the Hamiltonian and these states are found in the accompanying work \cite{PreprintMagSus}.  

The linear response of the SPDM to the magnetic field is given by \cite{PreprintMagSus}

\begin{equation}
\label{SPDMtoB}
\begin{split}
    \eta^{(B)}_{\alpha\textbf{R};\beta\textbf{R}'}(\omega) =  B^l(\textbf{R}_a,\omega) \frac{\mathcal{V}_{uc}}{(2\pi)^3} \sum_{mn} f_{nm} \int_{BZ} d\textbf{k}  e^{i\textbf{k}\cdot(\textbf{R}-\textbf{R}')}  U^\dag_{\alpha m} 
    \frac{M^l_{mn}(\textbf{k})}{\Delta_{mn}(\textbf{k})-\hbar(\omega+i0^+)}
    U_{n\beta} 
    \\
    + \frac{e\omega}{4c}  B^l(\textbf{R}_a,\omega) \frac{\mathcal{V}_{uc}}{(2\pi)^3} \epsilon^{lab} \sum_{mn} f_{nm} \int_{BZ} \frac{d\textbf{k}}{(2\pi)^3} U^\dag_{\alpha m} \frac{\partial_b (E_{n\textbf{k}}+E_{m\textbf{k}}) \xi^a_{mn}}{(\Delta_{mn}(\textbf{k})-\hbar(\omega+i0^+))^2} U_{n\beta}
    \\
    +\frac{e}{4\hbar c} \epsilon^{lab} B^l(\textbf{R}_a,\omega) \frac{\mathcal{V}_{uc}}{(2\pi)^3} \sum_{mn} f_{nm} \int_{BZ} d\textbf{k} e^{i\textbf{k}\cdot(\textbf{R}-\textbf{R}')} \frac{ \Delta_{mn}(\textbf{k}) \xi^b_{mn} }{\Delta_{mn}(\textbf{k})-\hbar(\omega+i0^+)}  \Big[ \partial_a U^\dag_{\alpha m} U_{n\beta} - U^\dag_{\alpha m} \partial_a U_{n\beta} \Big]
    \\
    +\frac{ie\omega}{4 c}\epsilon^{lab} B^l(\textbf{R}_a,\omega)\frac{\mathcal{V}_{uc}}{(2\pi)^3} \sum_{mn} f_{nm} \Big[
    (R^a-R_a^a)+(R'^a-R^a_a)
    \Big]\int_{BZ} d\textbf{k} e^{i\textbf{k}\cdot(\textbf{R}-\textbf{R}')} \frac{U^\dag_{\alpha m} \xi^b_{mn}U_{n\beta} }{\Delta_{mn}(\textbf{k})-\hbar(\omega+i0^+)},   
\end{split}
\end{equation}
where 
\begin{equation}
\label{SpontMagEq}
\begin{split}
    M^l_{mn} = \epsilon^{lab} \frac{e}{4c} \Bigg( v^b_{ms}\xi^a_{sn} + \xi^a_{ms} v^b_{sn} + \frac{1}{\hbar} \partial_b (E_{m\textbf{k}}+E_{n\textbf{k}}) \xi^a_{mn}\Bigg) + \frac{e}{mc} S^l_{mn} = \Big( M^l_{nm} \Big)^*
\end{split}
\end{equation}
is the spontaneous magnetization matrix element \cite{PreprintMagSus}, and

\begin{equation}
    \Delta_{mn}(\textbf{k}) = E_{m\textbf{k}} - E_{n\textbf{k}} 
\end{equation} 
is the difference of two eigenenergies, with associated cell periodic Bloch functions $u_{n\textbf{k}}$. In equation (\ref{SPDMtoB}) the magnetic field is evaluated at a lattice site and at arbitrary frequency $\omega$. This allows for a generalization to spatially and temporally varying fields. Although our focus here is on the response of an insulator to uniform DC fields, we present some of the equations below in a form that allows for an applied time dependent field, since that will be of interest for future work. The non-Abelian Berry connection in the cell periodic Bloch function basis is
\begin{equation}
    \xi^a_{nm} = i( u_{n\textbf{k}}| \partial_a u_{m\textbf{k}} ),
\end{equation}
where in general for any two functions we take
\begin{equation}
    (g|f) = \frac{1}{\mathcal{V}_{uc}} \int_{\mathcal{V}_{uc}} d\textbf{x} g^*(\textbf{x}) f(\textbf{x}).  
\end{equation}

\section{Magnetic Dipole Moments} 
In our framework the magnetization is broken up into three constituents: the atomic, itinerant, and spin magnetic moments associated with a lattice site \textbf{R}. We will take each to first order in the magnetic field to determine the magnetic susceptibility, 

\begin{equation}
\label{magneticsusceptibility}
    \chi^{il} = \frac{\partial M_\textbf{R}^i}{\partial B^l}, 
\end{equation}
where the magnetization is the sum of the atomic, itinerant and spin contributions,
\begin{equation}
    M^i_{\textbf{R}} = \bar{M}_\textbf{R} + \tilde{M}_\textbf{R} + \breve{M}_\textbf{R}.
\end{equation}
Equation (\ref{magneticsusceptibility}) can be used since we assume an infinite periodic lattice and the magnetic field is taken to be spatially uniform, so the magnetization is independent of the particular site \textbf{R}.  
These expressions are all of a similar form:
\begin{equation}
    \Lambda_\textbf{R}(t) = \sum_{\alpha\beta\textbf{R}'\textbf{R}''} \Lambda_{\beta\textbf{R}'';\alpha\textbf{R}'}(\textbf{R},t) \eta_{\alpha\textbf{R}';\beta\textbf{R}''}(t).
\end{equation}
There will be two ``types" of contributions that will arise at linear response, one from expanding the SPDM to first order in the magnetic field and taking the unperturbed matrix element, and the other from expanding the site matrix elements $\Lambda_{\beta\textbf{R}'';\alpha\textbf{R}'}(\textbf{R},t)$ to first order and using the unperturbed SPDM.
We call these the \emph{dynamical} and \emph{compositional} contributions respectively \cite{PreprintMagSus}. This will become more clear in the following sections.

\subsection{Atomic Magnetization}

\subsubsection{Dynamical contribution} 

The atomic magnetization dynamical contribution can be written as
\begin{equation}
\begin{split}
    \bar{M}^{i}_\text{dyn}(\omega) = 
    \epsilon^{iab} \frac{e}{8c} \sum_{\alpha\beta\gamma\textbf{R}'} \int_{BZ} \frac{d\textbf{k}}{(2\pi)^3} e^{i\textbf{k}\cdot(\textbf{R}-\textbf{R}')} \Big[ \tilde{\xi}^a_{\alpha\gamma} \tilde{v}^b_{\gamma \beta} + \tilde{v}^b_{\alpha\gamma} \tilde{\xi}^a_{\gamma\beta} \Big] \eta^{(B)}_{\beta\textbf{R}';\alpha\textbf{R}}(\omega)
    \\
    +\epsilon^{iab} \frac{e}{8c} \sum_{\alpha\beta\gamma\textbf{R}'} \int_{BZ} \frac{d\textbf{k}}{(2\pi)^3} e^{i\textbf{k}\cdot(\textbf{R}'-\textbf{R})} \Big[ \tilde{\xi}^a_{\alpha\gamma} \tilde{v}^b_{\gamma \beta} + \tilde{v}^b_{\alpha\gamma} \tilde{\xi}^a_{\gamma\beta} \Big] \eta^{(B)}_{\beta\textbf{R};\alpha\textbf{R}'}(\omega)
    \\
    +\epsilon^{iab} \frac{ie}{8c} \sum_{\alpha\beta\gamma\textbf{R}'} \int_{BZ} \frac{d\textbf{k}}{(2\pi)^3} \partial_a v^b_{\alpha\beta} \Big( e^{i\textbf{k}\cdot(\textbf{R}'-\textbf{R})} \eta^{(B)}_{\beta\textbf{R};\alpha\textbf{R}'}(\omega) - e^{i\textbf{k}\cdot(\textbf{R}-\textbf{R}')} \eta^{(B)}_{\beta\textbf{R}';\alpha\textbf{R}}(\omega) \Big), 
\end{split}
\end{equation}
where
\begin{equation}
\begin{split}
    \tilde{\xi}^m_{\alpha\gamma} = i( u_{\alpha\textbf{k}} | \partial_m u_{\gamma\textbf{k}} ) 
\end{split}
\end{equation}
is the non-Abelian Berry connection in the Wannier cell-periodic Basis. For reference, these are given in terms of the unitary matrix U(\textbf{k}) and the eigenstates of the Hamiltonian,
\begin{equation}
\begin{split}
    u_{\alpha\textbf{k}} = \sum_{n} U_{n\alpha}(\textbf{k}) u_{n\textbf{k}}, 
    \\
    \mathcal{H}(\textbf{k}) u_{n\textbf{k}} = E_{n\textbf{k}} u_{n\textbf{k}},
\end{split}
\end{equation}
and so 

\begin{equation}
\label{TransformingMatrixElements}
\begin{split}
    &\tilde{\xi}^a_{\alpha\beta} = \sum_{nm} U^\dag_{\alpha n} \Big(\xi^a_{nm}+\mathcal{W}^a_{nm} \Big) U_{m\beta},
    \\
    &\tilde{v}^a_{\alpha\beta} = \sum_{nm} U^\dag_{\alpha n} v^a_{nm} U_{m\beta} ,
\end{split}
\end{equation}
where $\tilde{v}_{\alpha\gamma}$ are the velocity matrix elements in that same basis and we have defined the quantity $\mathcal{W}^a_{nm}$,

\begin{equation}
    \mathcal{W}^a_{nm} = i \sum_{\alpha} \partial_a U_{n\alpha} U^\dag_{\alpha m}.
\end{equation}
The velocity matrix elements in the Bloch function basis can be related to the non-Abelian Berry connection and band energies by
\begin{equation}
\label{velocity}
\begin{split}
    v^i_{mn} \equiv \delta_{mn} \frac{1}{\hbar} \partial_i E_{n\textbf{k}} + \frac{i}{\hbar} (E_{m\textbf{k}}-E_{n\textbf{k}})\xi^i_{mn}.
\end{split}
\end{equation}

Setting $\omega = 0$ and inserting the 
SPDM response we find 

\begin{equation}
\label{atomicDyn}
\begin{split}
    \bar{M}^{i}_\text{dyn} = \epsilon^{iab} B^l \frac{e}{4c} \sum_{mn} f_{nm} \int_{BZ} \frac{d\textbf{k}}{(2\pi)^3} \frac{ \Big( (\xi^a_{ns}+\mathcal{W}^a_{ns}) v^b_{sm} + v^b_{ns}(\xi^a_{sm}+\mathcal{W}^a_{sm})\Big) M^l_{mn} }{\Delta_{mn}(\textbf{k})} 
    \\
    +\epsilon^{icd} \epsilon^{lab}\frac{ie^2}{16\hbar c^2} B^l \sum_{mn} f_{nm} \int_{BZ} \frac{d\textbf{k}}{(2\pi)^3} \Bigg[\xi^b_{mn} \Big( \xi^c_{nl}v^d_{ls} + v^d_{nl}\xi^c_{ls} \Big)\mathcal{W}^a_{sm} + \mathcal{W}^a_{ns}\Big( \xi^c_{sl}v^d_{lm} + v^d_{sl} \xi^c_{lm} \Big) \Bigg] 
    \\
    + \epsilon^{icd} \epsilon^{lab}\frac{ie^2}{16\hbar c^2} B^l \sum_{mn} f_{nm} \int_{BZ} \frac{d\textbf{k}}{(2\pi)^3} \Bigg[\xi^b_{mn} \Big( \mathcal{W}^c_{nl}v^d_{ls} + v^d_{nl}\mathcal{W}^c_{ls} \Big)\mathcal{W}^a_{sm} + \mathcal{W}^a_{ns}\Big( \mathcal{W}^c_{sl}v^d_{lm} + v^d_{sl} \mathcal{W}^c_{lm} \Big) \Bigg]. 
\end{split}
\end{equation}

\subsubsection{Compositional contribution}
The compositional contribution takes the single particle density matrix to zero order, which is $\eta^{(0)}_{\alpha\textbf{R};\beta\textbf{R}'} = f_{\alpha} \delta_{\alpha\beta} \delta_{\textbf{R}\textbf{R}'}$. The charge current operators depend on the magnetic field, and have a first order perturbative response due both to the Wannier functions and to the modified canonical momentum. The first order response can be expressed as

\begin{equation}
\begin{split}
    j^{k (B)}_{\alpha\textbf{R}';\alpha\textbf{R}'}(\textbf{x},\textbf{R};t) = e\delta_{\textbf{R}\textbf{R}'} \text{Re}\Bigg[
    \chi^{\dag(1)}_{\alpha\textbf{R}'}(\textbf{x},t) \hat{v}^k(\textbf{x}) W_{\alpha\textbf{R}'}(\textbf{x}) + W^\dag_{\alpha\textbf{R}'} (\textbf{x}) \hat{v}^k(\textbf{x}) \chi^{(1)}_{\alpha\textbf{R}'}(\textbf{x},t)
    \Bigg] 
    \\
    -\frac{e^2}{mc}\delta_{\textbf{R}\textbf{R}'} W^\dag_{\alpha\textbf{R}'} \Omega_{\textbf{R}}(\textbf{x},t) W_{\alpha\textbf{R}'}(\textbf{x}).
\end{split}
\end{equation}
The $\dag$ notation is used since the functions are spinors, so it is a complex conjugate and transpose. The modified Wannier spinors are obtained using Lowdin's method of symmetric orthogonalization \cite{Lowdin}. Expanded to first order in the magnetic field they are

\begin{equation}
\label{chi1}
\begin{split}
    \chi^{(1)}_{\alpha\textbf{R}'}(\textbf{x},t) = -\frac{i}{2} \sum_{\beta\textbf{R}''} W_{\beta\textbf{R}'}(\textbf{x}) \int d\textbf{y} W^\dag_{\beta\textbf{R}''}(\textbf{y}) \Delta(\textbf{R}'',\textbf{y},\textbf{R}';t) W_{\alpha\textbf{R}'}(\textbf{y}),
\end{split}
\end{equation}
where
\begin{equation}
\label{DeltaB}
\begin{split}
    \Delta(\textbf{x},\textbf{z},\textbf{y};t) = -\frac{e}{2\hbar c} \epsilon^{lab} B^l(\textbf{y},t) (x^a-y^a)(z^b-y^b) + ...,  
\end{split}
\end{equation}
and
\begin{equation}
\label{OmegaB}
\begin{split}
    \Omega^b_\textbf{R}(\textbf{x},t) = \frac{\epsilon^{lab}}{2} B^l(\textbf{R},t) (x^a-R^a) + ...
\end{split}
\end{equation}

This is then substituted into the expression for the magnetization,
\begin{equation}
\begin{split}
    \bar{M}^{i}_\text{comp}(t) = \frac{1}{2\mathcal{V}_{uc} c} \epsilon^{iab} \sum_{\alpha\textbf{R}'} f_{\alpha} \int d\textbf{x} (x^a-R^a) j^{b(B)}_{\alpha\textbf{R}';\alpha\textbf{R}'}(\textbf{x},\textbf{R};t)
    \\
    =\frac{e}{2\mathcal{V}_{uc} c} \epsilon^{iab} \sum_{\alpha\textbf{R}'} f_{\alpha} \delta_{\textbf{R}\textbf{R}'} \text{Re}\Bigg[ \int d\textbf{x} \chi^{\dag(1)}_{\alpha\textbf{R}'}(\textbf{x},t) (x^a-R^a)\hat{v}^b(\textbf{x}) W_{\alpha\textbf{R}'}(\textbf{x})
    + W^\dag_{\alpha\textbf{R}'}(\textbf{x}) (x^a-R^a) \hat{v}^b(\textbf{x}) \chi^{(1)}_{\alpha\textbf{R}'}(\textbf{x},t)
    \Bigg] 
    \\
    - \frac{e^2}{2mc^2 \mathcal{V}_{uc}} \epsilon^{iab} \sum_{\alpha\textbf{R}'} f_{\alpha} \delta_{\textbf{R}\textbf{R}'} \int d\textbf{x} W^\dag_{\alpha\textbf{R}'}(\textbf{x}) (x^a-R^a) \Omega^b_{\textbf{R}}(\textbf{x},t) W_{\alpha\textbf{R}'}(\textbf{x}).
\end{split}
\end{equation}
Processing this expression we find

\begin{equation}
\begin{split}
    \bar{M}^{i}_\text{comp}(t) = 
    \frac{e}{4c \mathcal{V}_{uc}} \epsilon^{iab} \sum_{\alpha \beta\textbf{R}'} f_{\alpha} \text{Re}\Bigg[ i
    \int d\textbf{x} W^\dag_{\beta\textbf{R}'}(\textbf{x}) (x^a-R^a) \hat{v}^b(\textbf{x}) W_{\alpha\textbf{R}}(\textbf{x}) \int d\textbf{y} W^\dag_{\alpha\textbf{R}}(\textbf{y}) \Delta(\textbf{R}',\textbf{y},\textbf{R};t) W_{\beta\textbf{R}'}(\textbf{y})
    \Bigg]
    \\
    -\frac{e}{4c \mathcal{V}_{uc}} \epsilon^{iab} \sum_{\alpha\beta\textbf{R}'} f_{\alpha} \text{Re}\Bigg[ i
    \int d\textbf{x} W^\dag_{\alpha\textbf{R}}(\textbf{x}) (x^a-R^a) \hat{v}^b(\textbf{x}) W_{\beta\textbf{R}'}(\textbf{x}) \int d\textbf{y} W^\dag_{\beta\textbf{R}'}(\textbf{y}) \Delta(\textbf{R}',\textbf{y},\textbf{R};t) W_{\alpha\textbf{R}}(\textbf{y})
    \Bigg]
    \\
    -\frac{e^2}{4mc^2 \mathcal{V}_{uc}} B^l(\textbf{R},t) \epsilon^{iab} \epsilon^{lcb} \sum_{\alpha} f_{\alpha} \int d\textbf{x} W^\dag_{\alpha\textbf{R}}(\textbf{x}) (x^a-R^a) (x^c-R^c) W_{\alpha\textbf{R}}(\textbf{x}).
\end{split}
\end{equation}
The real space integrals can be evaluated and written as integrals over the BZ 
\begin{equation}
\begin{split}
    \int d\textbf{y} W^\dag_{\alpha\textbf{R}}(\textbf{y}) \Delta(\textbf{R}',\textbf{y},\textbf{R};t) W_{\beta\textbf{R}'}(\textbf{y}) = -\frac{e}{2\hbar c} \epsilon^{lab} B^l(\textbf{R},t) (R'^a-R^a) \int d\textbf{y} W^\dag_{\alpha\textbf{R}}(\textbf{y}) (y^b-R^b) W_{\beta\textbf{R}'}(\textbf{y}) 
    \\
    = -\frac{e}{2\hbar c} \epsilon^{lab} B^l(\textbf{R},t) (R'^a-R^a) \mathcal{V}_{uc} \int_{BZ} \frac{d\textbf{k}}{(2\pi)^3} e^{i\textbf{k}\cdot(\textbf{R}-\textbf{R}')} \tilde{\xi}^b_{\alpha\beta} 
    \\
    = \frac{ie}{2\hbar c} \epsilon^{lab} B^l(\textbf{R},t) \mathcal{V}_{uc} \int_{BZ} \frac{d\textbf{k}}{(2\pi)^3} e^{i\textbf{k}\cdot(\textbf{R}-\textbf{R}')} \partial_a \tilde{\xi}^b_{\alpha\beta}.
\end{split}
\end{equation}
Processing the compositional contribution to the atomic magnetization further we find

\begin{equation}
\begin{split}
    \bar{M}^{i}_\text{comp} = -\frac{e^2}{8\hbar c^2} \epsilon^{icd} \epsilon^{lab} B^l  \sum_{\alpha\beta\textbf{R}'} f_{\alpha} 
    \int_{BZ} \frac{d\textbf{k}}{(2\pi)^3}  
    \text{Re}\Bigg[ 
    \tilde{v}^d_{\beta\gamma} \tilde{\xi}^c_{\gamma\alpha} \partial_a \tilde{\xi}^b_{\alpha\beta} 
    + \tilde{\xi}^c_{\alpha\gamma} \tilde{v}^d_{\gamma\beta} \partial_a \xi^b_{\beta\alpha} 
    \Bigg]
    \\
    -\frac{e^2}{8mc^2} B^l \epsilon^{iab} \epsilon^{lcb} \sum_{\alpha\beta} f_{\alpha} \int_{BZ} \frac{d\textbf{k}}{(2\pi)^3} \Big( \tilde{\xi}^a_{\alpha\beta} \tilde{\xi}^c_{\beta\alpha} +\tilde{\xi}^c_{\alpha\beta} \tilde{\xi}^a_{\beta\alpha}\Big),
\end{split}
\end{equation}
where we assume uniform time-independent fields, $B^l(\textbf{R},t) = B^l$. 
We can identify the ``gauge-dependence" more clearly by making the connection between the matrix elements in the Wannier function basis and the Bloch function basis (\ref{TransformingMatrixElements}), and thus
\begin{equation}
\label{atomicComp}
\begin{split}
    \bar{M}^{i}_\text{comp} = -\frac{e^2}{4\hbar c^2} \epsilon^{icd}\epsilon^{lab} B^l \sum_{mn} f_{n} \text{Re}\Bigg[ \int_{BZ} \frac{d\textbf{k}}{(2\pi)^3} \Big(\partial_a \xi^b_{nm} + i\mathcal{W}^a_{ns}\xi^b_{sm} - i\xi^b_{ns}\mathcal{W}^a_{sm} + \partial_b \mathcal{W}^a_{nm} \Big)v^d_{ml}(\xi^c_{ln}+\mathcal{W}^c_{ln}) 
    \Bigg]
    \\
    -\frac{e^2}{4mc^2} B^l \epsilon^{iab} \epsilon^{lcb} \sum_{mn} f_{n} \int_{BZ} \frac{d\textbf{k}}{(2\pi)^3} \text{Re}\Bigg[ 
    (\xi^a_{nm}+\mathcal{W}^a_{nm})(\xi^c_{mn}+\mathcal{W}^c_{mn}) 
    \Bigg] .
\end{split}
\end{equation}


\subsection{Itinerant Magnetization}
The expression for the itinerant magnetization is found in Section II and Appendix A of Duff et al. \cite{PreprintMagSus}.

\subsubsection{Dynamical contribution}
The dynamical contribution of the itinerant magnetization is

\begin{equation}
\begin{split}
    \tilde{M}^{i}_\text{dyn}(\omega) = -\frac{e}{8\hbar c} \epsilon^{icd} \int_{BZ} \frac{d\textbf{k}}{(2\pi)^3} \Bigg( e^{i\textbf{k}\cdot(\textbf{R}-\textbf{R}')} \eta^{(B)}_{\beta\textbf{R}';\alpha\textbf{R}}(\omega) + e^{i\textbf{k}\cdot(\textbf{R}'-\textbf{R})} \eta^{(B)}_{\beta\textbf{R};\alpha\textbf{R}'}(\omega) \Bigg) \tilde{\xi}^d_{\alpha\gamma} \partial_c \Big( U^\dag_{\gamma n} E_{n\textbf{k}} U_{n\beta} \Big)
    \\
    -\frac{e}{8\hbar c} \epsilon^{icd} \int_{BZ} \frac{d\textbf{k}}{(2\pi)^3} \Bigg( e^{i\textbf{k}\cdot(\textbf{R}-\textbf{R}')} \eta^{(B)}_{\beta\textbf{R}';\alpha\textbf{R}}(\omega) + e^{i\textbf{k}\cdot(\textbf{R}'-\textbf{R})} \eta^{(B)}_{\beta\textbf{R};\alpha\textbf{R}'}(\omega) \Bigg) \partial_c (U^\dag_{\alpha n} E_{n\textbf{k}} U_{n\gamma} \Big) \tilde{\xi}^d_{\gamma\beta}. 
\end{split}
\end{equation}
Setting $\omega = 0$ and substituting in the result for $\eta^{(B)}_{\beta\textbf{R};\alpha\textbf{R}'}(0)$,
\begin{equation}
\begin{split}
    \tilde{M}^{i}_\text{dyn} = -\frac{e}{4\hbar c} \epsilon^{icd} B^l(\textbf{R}) \sum_{mn} f_{nm} \int_{BZ} \frac{d\textbf{k}}{(2\pi)^3} \Bigg[ 
    \tilde{\xi}^d_{\alpha\gamma} \partial_c \Big( U^\dag_{\gamma l} E_{l\textbf{k}} U_{l\beta} \Big) + \partial_c \Big( U^\dag_{\alpha l} E_{l\textbf{k}} U_{l\gamma} \Big) \tilde{\xi}^d_{\gamma\beta} 
    \Bigg]
    \\
    \times\Bigg[ U^\dag_{\beta m} 
    \frac{ M^l_{mn} }{\Delta_{mn}(\textbf{k})} U_{n\alpha} + \frac{e}{4\hbar c} \epsilon^{lab} \xi^b_{mn} \Big(i U^\dag_{\beta s} \mathcal{W}^a_{sm} U_{n\alpha} + i U^\dag_{\beta m} \mathcal{W}^a_{ns} U_{s\alpha} \Big) 
    \Bigg].
\end{split}
\end{equation}
Making the gauge-dependence in terms of the $\mathcal{W}$'s explicit,
\begin{equation}
\label{itinerantDyn}
\begin{split}
    \tilde{M}^{i}_\text{dyn} = -\frac{e}{4\hbar c} \epsilon^{icd} B^l \sum_{mn} f_{nm} \int_{BZ} \frac{d\textbf{k}}{(2\pi)^3} \Bigg[ 
    \Big( \xi^d_{nl} + \mathcal{W}^d_{nl} \Big) \Big( i\mathcal{W}^c_{lm}(E_{m\textbf{k}}- E_{l\textbf{k}}) + \delta_{lm} \partial_c E_{m\textbf{k}} \Big) 
    \\
    + \Big( i\mathcal{W}^c_{nl} ( E_{l\textbf{k}}-E_{n\textbf{k}}) + \delta_{nl} \partial_c E_{n\textbf{k}} \Big) \Big(
    \xi^d_{lm} + \mathcal{W}^d_{lm}
    \Big)
    \Bigg] \frac{ M^l_{mn} }{\Delta_{mn}(\textbf{k})} 
    \\
    -\frac{ie^2}{16\hbar^2 c^2} \epsilon^{icd}\epsilon^{lab} B^l \sum_{mn} f_{nm} \int_{BZ} \frac{d\textbf{k}}{(2\pi)^3}
    \Big( 
    \xi^b_{mn}(\xi^d_{nl}+\mathcal{W}^d_{nl}) \Big( i\mathcal{W}^c_{ls} \mathcal{W}^a_{sm} (E_{s\textbf{k}}-E_{l\textbf{k}})  + \partial_c E_{l\textbf{k}} \mathcal{W}^a_{lm} 
    \Big)
    \\
    -\frac{ie^2}{16\hbar^2 c^2} \epsilon^{icd} \epsilon^{lab} B^l \sum_{mn} f_{nm} \int_{BZ} \frac{d\textbf{k}}{(2\pi)^3} \xi^b_{mn} \mathcal{W}^a_{ns} \Big( \xi^d_{sl} + \mathcal{W}^d_{sl} \Big)\Big( i\mathcal{W}^c_{lm} (E_{m\textbf{k}}-E_{l\textbf{k}}) + \partial_c E_{m\textbf{k}} \delta_{lm} \Big)
    \\
    -\frac{ie^2}{16\hbar^2 c^2} \epsilon^{icd}\epsilon^{lab} B^l\sum_{mn} f_{nm} \int_{BZ} \frac{d\textbf{k}}{(2\pi)^3} \xi^b_{mn} \Big( i\mathcal{W}^c_{nl} (E_{l\textbf{k}}-E_{n\textbf{k}}) + \delta_{nl} \partial_c E_{n\textbf{k}} \Big) \Big( \xi^d_{ls}+\mathcal{W}^d_{ls}\Big) \mathcal{W}^a_{sm} 
    \\
    -\frac{ie^2}{16\hbar^2 c^2} \epsilon^{icd}\epsilon^{lab} B^l \sum_{mn} f_{nm} \int_{BZ} \frac{d\textbf{k}}{(2\pi)^3} \xi^b_{mn} \mathcal{W}^a_{ns} \Big(
    i\mathcal{W}^c_{sl} (E_{l\textbf{k}}-E_{s\textbf{k}}) + \partial_c E_{s\textbf{k}} \delta_{sl}\Big)\Big( \xi^d_{lm} + \mathcal{W}^d_{lm}
    \Big).
\end{split}
\end{equation}

\subsubsection{Compositional contribution}
We can start with a general form that generates all orders of the ``compositional" response of the itinerant magnetization,

\begin{equation}
\begin{split}
    \tilde{M}^{i}_\text{comp} = \frac{e}{2\hbar c} \frac{1}{\mathcal{V}_{uc}} \epsilon^{iab} \sum_{\alpha\beta\textbf{R}'} f_{\alpha} \text{Re}\Bigg[ 
    i \int d\textbf{x} (R'^a-R^a)(x^b-R'^b) e^{-i\Delta(\textbf{R}',\textbf{x},\textbf{R};t)} \chi^\dag_{\alpha\textbf{R}}(\textbf{x},t) \chi_{\beta\textbf{R}'}(\textbf{x},t) \bar{H}_{\beta\textbf{R}';\alpha\textbf{R}}(t) 
    \Bigg].
\end{split}
\end{equation}
There are four contributions to linear order, coming from expanding the left or right spinor, the exponential, or the Hamiltonian matrix to first order in the magnetic field and taking the other terms to zeroth order. To keep track these will be written as $\mathcal{A},\mathcal{B},\mathcal{C},\mathcal{D}$ respectively. For the first three we have

\begin{equation}
\begin{split}
    \mathcal{A} = -\frac{e}{4\hbar c} \frac{1}{\mathcal{V}_{uc}} \epsilon^{icd} \sum_{\alpha\beta\textbf{R}'} f_{\alpha} \text{Re}\Bigg[ 
    (R'^c-R^c) \int d\textbf{y} W^\dag_{\alpha\textbf{R}}(\textbf{y}) \Delta(\textbf{R}'',\textbf{y},\textbf{R}) W_{\mu\textbf{R}''}(\textbf{y}) 
    \int d\textbf{x} W^\dag_{\mu\textbf{R}''}(\textbf{x}) (x^d-R'^d) W_{\beta\textbf{R}'}(\textbf{x}) H^{(0)}_{\beta\textbf{R}';\alpha\textbf{R}}
    \Bigg]
    \\
    = \frac{e^2}{8\hbar^2 c^2} \epsilon^{icd}B^l(\textbf{R},t) \epsilon^{lab} \sum_{\alpha\beta\mu} f_{\alpha} \text{Re}\Bigg[ 
    \int_{BZ} \frac{d\textbf{k}}{(2\pi)^3} \partial_a \tilde{\xi}^b_{\alpha\mu} \tilde{\xi}^d_{\mu\beta} \partial_c \Big( U^\dag_{\beta n} E_{n\textbf{k}} U_{n\alpha} \Big) 
    \Bigg], 
\end{split}
\end{equation}

\begin{equation}
\begin{split}
    \mathcal{B} = \frac{e}{4\hbar c} \frac{1}{\mathcal{V}_{uc}} \epsilon^{icd} \sum_{\alpha\beta\textbf{R}'} f_{\alpha} \text{Re}\Bigg[ 
    (R'^c-R^c) \int d\textbf{x} W^\dag_{\alpha\textbf{R}}(\textbf{x}) (x^d-R'^d) W_{\mu\textbf{R}''}(\textbf{x}) \int d\textbf{y} W^\dag_{\mu\textbf{R}''}(\textbf{y}) \Delta(\textbf{R}'',\textbf{y},\textbf{R}') W_{\beta\textbf{R}'}(\textbf{x}) H^{(0)}_{\beta\textbf{R}';\alpha\textbf{R}}
    \Bigg]
    \\
    = \frac{e^2}{8\hbar^2 c^2} \epsilon^{icd} B^l(\textbf{R}',t) \epsilon^{lab} \sum_{\alpha\beta\mu} f_{\alpha} \text{Re}\Bigg[ 
    \int_{BZ} \frac{d\textbf{k}}{(2\pi)^3} \tilde{\xi}^d_{\alpha \mu} \partial_a \tilde{\xi}^b_{\mu\beta} \partial_c \Big(U^\dag_{\beta n} E_{n\textbf{k}} U_{n\alpha} \Big)
    \Bigg],
\end{split}
\end{equation}
and
\begin{equation}
\begin{split}
    \mathcal{C} = \frac{e}{2\hbar c} \frac{1}{\mathcal{V}_{uc}} \epsilon^{icd} \sum_{\alpha\beta\textbf{R}'} f_{\alpha} \text{Re}\Bigg[ 
    (R'^c-R^c) \int d\textbf{x} W^\dag_{\alpha\textbf{R}}(\textbf{x}) (x^d-R'^d) \Delta(\textbf{R}',\textbf{x},\textbf{R};t) W_{\beta\textbf{R}'}(\textbf{x},t) H^{(0)}_{\beta\textbf{R}';\alpha\textbf{R}}
    \Bigg]
    \\
    = \frac{e^2}{4\hbar^2 c^2} \epsilon^{icd} B^l(\textbf{R},t) \epsilon^{lab} \sum_{\alpha\beta\gamma} f_{\alpha} \text{Re}\Bigg[ 
    \int_{BZ} \frac{d\textbf{k}}{(2\pi)^3} \tilde{\xi}^d_{\alpha\gamma} \tilde{\xi}^b_{\gamma\beta} \partial_a \partial_c \Big(U^\dag_{\beta n} E_{n\textbf{k}} U_{n\alpha} \Big)
    \Bigg]. 
\end{split}
\end{equation}

Expanding the Hamiltonian matrix element, to
first order in a static magnetic field we have
\begin{equation}
\begin{split}
    \bar{H}^{(B)}_{\beta\textbf{R}';\alpha\textbf{R}} = -\frac{e}{2c} \int d\textbf{x} W^\dag_{\beta\textbf{R}'}(\textbf{x}) \Big(
    \Omega_{\textbf{R}}(\textbf{x}) \cdot \hat{\textbf{v}}(\textbf{x}) + \frac{1}{mc} \frac{\hbar}{2} \boldsymbol\sigma \cdot\textbf{B} \Big)
    W_{\alpha\textbf{R}}(\textbf{x}) 
    \\
    -\frac{e}{2c} \int d\textbf{x} \Big( \Big(\Omega_{\textbf{R}'}(\textbf{x})\cdot\hat{\textbf{v}}(\textbf{x}) + \frac{1}{mc} \frac{\hbar}{2} \boldsymbol\sigma \cdot \textbf{B} \Big) W_{\beta\textbf{R}'}(\textbf{x}) \Big)^\dag W_{\alpha\textbf{R}}(\textbf{x})
    \\
    + \int d\textbf{x} \Bigg( \chi^{\dag(1)}_{\beta\textbf{R}'}(\textbf{x}) H^{(0)}(\textbf{x}) W_{\alpha\textbf{R}}(\textbf{x}) + W^\dag_{\beta\textbf{R}'}(\textbf{x}) H^{(0)}(\textbf{x}) \chi^{(1)}_{\alpha\textbf{R}}(\textbf{x}) \Bigg) 
    \\
    +\frac{i}{2} \int d\textbf{x} W^\dag_{\beta\textbf{R}'}(\textbf{x}) \Big( H^{(0)}(\textbf{x}) \Delta(\textbf{R}',\textbf{x},\textbf{R}) + \Delta(\textbf{R}',\textbf{x},\textbf{R}) H^{(0)}(\textbf{x}) \Big) W_{\alpha\textbf{R}}(\textbf{x}).
\end{split}
\end{equation}
We can then substitute in equations (\ref{chi1}), (\ref{DeltaB}), and (\ref{OmegaB}) to find
\begin{equation}
\begin{split}
    \bar{H}^{(B)}_{\beta\textbf{R}';\alpha\textbf{R}} = -\frac{e}{4c} \epsilon^{lab} B^l \int d\textbf{x} W^\dag_{\beta\textbf{R}'}(\textbf{x}) (x^a-R^a) v^b(\textbf{x}) W_{\alpha\textbf{R}}(\textbf{x}) 
    \\
    - \frac{e}{4c} \epsilon^{lab} B^l \int d\textbf{x} W^\dag_{\beta\textbf{R}'}(\textbf{x}) v^b(\textbf{x}) (x^a-R'^a) W_{\alpha\textbf{R}}(\textbf{x})
    \\
    -\frac{e}{mc^2} B^l \int d\textbf{x} W^\dag_{\beta\textbf{R}'}(\textbf{x}) \frac{\hbar}{2} \sigma^l W_{\alpha\textbf{R}}(\textbf{x})
    \\
    +\frac{i}{2} \sum_{\gamma\textbf{R}''} \int d\textbf{y} W^\dag_{\beta\textbf{R'}}(\textbf{y}) \Delta(\textbf{R}'',\textbf{y},\textbf{R}') W_{\gamma\textbf{R}''}(\textbf{y})  \int d\textbf{x} W^\dag_{\gamma\textbf{R}''}(\textbf{x}) H^{(0)}(\textbf{x}) W_{\alpha\textbf{R}}(\textbf{x}) 
    \\
    -\frac{i}{2} \sum_{\gamma\textbf{R}''} \int d\textbf{x} W^\dag_{\beta\textbf{R}'}(\textbf{x}) H^{(0)}(\textbf{x}) W_{\gamma\textbf{R}''}(\textbf{x}) \int d\textbf{y} W^\dag_{\gamma\textbf{R}''}(\textbf{y}) \Delta(\textbf{R}'',\textbf{y},\textbf{R}) W_{\alpha\textbf{R}}(\textbf{y}) 
    \\
    -\frac{ie}{4\hbar c} \epsilon^{lab} B^l (R'^a-R^a) \int d\textbf{x} W^\dag_{\beta\textbf{R}'}(\textbf{x}) \Big( H^{(0)}(\textbf{x}) (x^b-R^b) + (x^b-R^b) H^{(0)}(\textbf{x}) \Big) W_{\alpha\textbf{R}}(\textbf{x}).
\end{split}
\end{equation}
Evaluating the real space integrals we can write the above as
\begin{equation}    
\label{HB}
\begin{split}
    \bar{H}^{(B)}_{\beta\textbf{R}';\alpha\textbf{R}} = -\frac{e}{4c} \epsilon^{lab} B^l \mathcal{V}_{uc} \int_{BZ} \frac{d\textbf{k}}{(2\pi)^3} e^{i\textbf{k}\cdot(\textbf{R}'-\textbf{R})} \Big( \tilde{v}^b_{\beta\gamma} \tilde{\xi}^a_{\gamma\alpha} + \tilde{\xi}^a_{\beta\gamma} \tilde{v}^b_{\gamma\alpha} \Big) 
    \\
    - \frac{e}{mc^2} B^l \mathcal{V}_{uc} \int_{BZ} \frac{d\textbf{k}}{(2\pi)^3} e^{i\textbf{k}\cdot(\textbf{R}'-\textbf{R})} S^l_{\beta\alpha}
    \\
    -\frac{e}{4\hbar c} \epsilon^{lab} B^l \mathcal{V}_{uc} \int_{BZ} \frac{d\textbf{k}}{(2\pi)^3} e^{i\textbf{k}\cdot(\textbf{R}'-\textbf{R})} \Big( \partial_a \tilde{\xi}^b_{\beta\gamma} U^\dag_{\gamma n} E_{n\textbf{k}} U_{n\alpha}  
    + U^\dag_{\beta n} E_{n\textbf{k}} U_{n\gamma} \partial_a \tilde{\xi}^b_{\gamma\alpha} \Big)
    \\
    +\frac{e}{4\hbar c} \epsilon^{lab} B^l \mathcal{V}_{uc} \int_{BZ} \frac{d\textbf{k}}{(2\pi)^3} e^{i\textbf{k}\cdot(\textbf{R}'-\textbf{R})} \partial_a \Bigg( U^\dag_{\beta n} E_{n\textbf{k}} U_{n\gamma} \tilde{\xi}^b_{\gamma\alpha} + \tilde{\xi}^b_{\beta\gamma} U^\dag_{\gamma n} E_{n\textbf{k}} U_{n\alpha} \Bigg),
\end{split}
\end{equation}
and so the last piece $\mathcal{D}$, 
\begin{equation}
\begin{split}
    \mathcal{D} =
    \frac{e}{2\hbar c} \frac{1}{\mathcal{V}_{uc}} \epsilon^{icd} \sum_{\alpha\beta\textbf{R}'} f_{\alpha} \text{Re}\Bigg[ 
    i(R'^c-R^c) \int d\textbf{x} W^\dag_{\alpha\textbf{R}}(\textbf{x}) (x^d-R'^d) W_{\beta\textbf{R}'}(\textbf{x}) \bar{H}^{(B)}_{\beta\textbf{R}';\alpha\textbf{R}}
    \Bigg]
    \\
    = \frac{e}{2\hbar c} \epsilon^{icd} \sum_{\alpha\beta\textbf{R}'} f_{\alpha} \text{Re}\Bigg[ 
    \int_{BZ} \frac{d\textbf{k}}{(2\pi)^3} \partial_c \tilde{\xi}^d_{\alpha\beta} \bar{H}^{(B)}_{\beta\textbf{R}';\alpha\textbf{R}}
    \Bigg], 
\end{split}
\end{equation}
is obtained by inserting what we found for $\bar{H}^{(B)}_{\beta\textbf{R}';\alpha\textbf{R}}$ (equation (\ref{HB}))
\begin{equation}
\begin{split}
    \mathcal{D} = -\frac{e^2}{8\hbar c^2} \epsilon^{icd} B^l \epsilon^{lab} \sum_{\alpha\beta\gamma} f_{\alpha} \text{Re}\Bigg[  \int_{BZ} \frac{d\textbf{k}}{(2\pi)^3}
    \partial_c \tilde{\xi}^d_{\alpha\beta} \Bigg( 
    \tilde{v}^b_{\beta\gamma} \tilde{\xi}^a_{\gamma\alpha} + \tilde{\xi}^a_{\beta\gamma} \tilde{v}^b_{\gamma\alpha} - \frac{1}{\hbar} \partial_a \Big(U^\dag_{\beta n} E_{n\textbf{k}} U_{n\gamma}\Big)  \tilde{\xi}^b_{\gamma\alpha} - \frac{1}{\hbar} \tilde{\xi}^b_{\beta\gamma} \partial_a \Big( U^\dag_{\gamma n} E_{n\textbf{k}} U_{n\alpha} \Big)
    \Bigg)
    \Bigg]
    \\
    -\frac{e^2}{2m\hbar c^2} B^l \epsilon^{icd} \sum_{\alpha\beta} f_{\alpha} \text{Re}\Bigg[ \int_{BZ} \frac{d\textbf{k}}{(2\pi)^3} 
    \partial_c \tilde{\xi}^d_{\alpha\beta} \tilde{S}^l_{\beta\alpha}
    \Bigg].
\end{split}
\end{equation}
Combining all these pieces, assuming uniform fields
\begin{equation}
\begin{split}
    \tilde{M}^{i}_\text{comp} = \mathcal{A} + \mathcal{B} + \mathcal{C} + \mathcal{D} 
    = \frac{e^2}{4\hbar^2 c^2} \epsilon^{icd} B^l \epsilon^{lab} \sum_{\alpha\gamma\beta} f_{\alpha} \text{Re}\Bigg[ 
    \int_{BZ} \frac{d\textbf{k}}{(2\pi)^3} \tilde{\xi}^d_{\alpha\gamma} \tilde{\xi}^b_{\gamma\beta} \partial_a \partial_c \Big( U^\dag_{\beta n} E_{n\textbf{k}} U_{n\alpha} \Big) 
    \Bigg]
    \\
    +\frac{e^2}{8\hbar^2 c^2} \epsilon^{icd} B^l \epsilon^{lab} \sum_{\alpha\beta\gamma} f_{\alpha} \text{Re}\Bigg[ 
    \int_{BZ} \frac{d\textbf{k}}{(2\pi)^3} \Big( \tilde{\xi}^d_{\alpha\gamma} \partial_a \tilde{\xi}^b_{\gamma\beta} + \partial_a \tilde{\xi}^b_{\alpha\gamma} \tilde{\xi}^d_{\gamma\beta} \Big) \partial_c \Big( U^\dag_{\beta n} E_{n\textbf{k}} U_{n\alpha} \Big) 
    \Bigg]
    \\
    -\frac{e^2}{8\hbar c^2} \epsilon^{icd} B^l \epsilon^{lab} \sum_{\alpha\beta\gamma} f_{\alpha} \text{Re}\Bigg[  \int_{BZ} \frac{d\textbf{k}}{(2\pi)^3}
    \partial_c \tilde{\xi}^d_{\alpha\beta} \Bigg( 
    \tilde{v}^b_{\beta\gamma} \tilde{\xi}^a_{\gamma\alpha} + \tilde{\xi}^a_{\beta\gamma} \tilde{v}^b_{\gamma\alpha} - \frac{1}{\hbar} \partial_a \Big(U^\dag_{\beta n} E_{n\textbf{k}} U_{n\gamma}\Big)  \tilde{\xi}^b_{\gamma\alpha} - \frac{1}{\hbar} \tilde{\xi}^b_{\beta\gamma} \partial_a \Big( U^\dag_{\gamma n} E_{n\textbf{k}} U_{n\alpha} \Big)
    \Bigg)
    \Bigg]
    \\
    -\frac{e^2}{2m\hbar c^2} B^l \epsilon^{icd} \sum_{\alpha\beta} f_{\alpha} \text{Re}\Bigg[ \int_{BZ} \frac{d\textbf{k}}{(2\pi)^3} 
    \partial_c \tilde{\xi}^d_{\alpha\beta} \tilde{S}^l_{\beta\alpha}
    \Bigg].
\end{split}
\end{equation}
Making the gauge-dependence splitting explicit
\begin{equation}
\label{itinComp}
\begin{split}
    \tilde{M}^{i}_\text{comp} = \frac{e^2}{4\hbar^2 c^2} \epsilon^{icd}  \epsilon^{lab} B^l \sum_{mns} f_{n} \text{Re}\Bigg[ 
    \int_{BZ} \frac{d\textbf{k}}{(2\pi)^3} (\xi^d_{nm}+\mathcal{W}^d_{nm})(\xi^b_{ms} + \mathcal{W}^b_{ms})\Big( \partial_a \partial_c E_{n\textbf{k}} \delta_{sn} + i\partial_a (E_{n\textbf{k}}-E_{s\textbf{k}}) \mathcal{W}^c_{sn}
    \\
    + i \partial_c (E_{n\textbf{k}}-E_{s\textbf{k}}) \mathcal{W}^a_{sn} 
    + \frac{i}{2}(E_{n\textbf{k}}-E_{s\textbf{k}})( \partial_a \mathcal{W}^c_{sn} + \partial_c \mathcal{W}^a_{sn} \Big) + \frac{1}{2}(2E_{l\textbf{k}}-E_{s\textbf{k}}-E_{n\textbf{k}})\Big( \mathcal{W}^a_{sl}\mathcal{W}^c_{ln} + \mathcal{W}^c_{sl}\mathcal{W}^a_{ln} \Big)
    \Bigg]
    \\
    +\frac{e^2}{4\hbar^2 c^2} \epsilon^{icd}\epsilon^{lab} B^l \sum_{mns} f_{n} \text{Re}\Bigg[ 
    \int_{BZ} \frac{d\textbf{k}}{(2\pi)^3} (\xi^d_{ns}+\mathcal{W}^d_{ns}) \Big( \partial_a \xi^b_{sm} + i\mathcal{W}^a_{sl}\xi^b_{lm} - i\xi^b_{sl}\mathcal{W}^a_{lm} + \partial_b \mathcal{W}^a_{sm}\Big) 
    \\
    \times \Big(
    i(E_{n\textbf{k}}-E_{m\textbf{k}}) \mathcal{W}^c_{mn} + \delta_{nm} \partial_c E_{n\textbf{k}}
    \Big)
    \Bigg]
    \\
    -\frac{e^2}{8\hbar c^2} \epsilon^{icd} \epsilon^{lab} B^l \sum_{mns} f_{n} \text{Re}\Bigg[ 
    \int_{BZ} \frac{d\textbf{k}}{(2\pi)^3} \Big( \partial_c \xi^d_{nm} + i\mathcal{W}^c_{nl}\xi^d_{lm} - i \xi^d_{nl}\mathcal{W}^c_{lm} + \partial_d \mathcal{W}^c_{nm} \Big)
    \\
    \times \Big( v^b_{ms}(\xi^a_{sn}+\mathcal{W}^a_{sn}) + (\xi^a_{ms}+\mathcal{W}^a_{ms})v^b_{sn} 
    -\frac{1}{\hbar} \partial_a (E_{m\textbf{k}}+E_{n\textbf{k}}) (\xi^b_{mn}+\mathcal{W}^b_{mn})
    \\
    -\frac{i}{\hbar}(E_{s\textbf{k}}-E_{m\textbf{k}})\mathcal{W}^a_{ms}(\xi^b_{sn}+\mathcal{W}^b_{sn})
    -\frac{i}{\hbar}(E_{n\textbf{k}}-E_{s\textbf{k}}) (\xi^b_{ms}+\mathcal{W}^b_{ms})\mathcal{W}^a_{sn}
    \Big)
    \Bigg]
    \\
    -\frac{e^2}{2m\hbar c^2} B^l \epsilon^{icd} \sum_{mn} f_{n} \text{Re}\Bigg[ 
    \int_{BZ} \frac{d\textbf{k}}{(2\pi)^3} \Big( \partial_c \xi^d_{nm} + i\mathcal{W}^c_{ns}\xi^d_{sm} -i\xi^d_{ns}\mathcal{W}^c_{sm} + \partial_d \mathcal{W}^c_{nm}\Big) S^l_{mn} 
    \Bigg].
\end{split}
\end{equation}

\subsection{Spin Magnetization}

The spin magnetization is defined as

\begin{equation}
\begin{split}
    \breve{M}_\textbf{R}^{i}(t) = \frac{e}{2mc} \sum_{\alpha\beta\textbf{R}'\textbf{R}''} \Big( \delta_{\textbf{R}\textbf{R}'}+\delta_{\textbf{R}\textbf{R}''}\Big) \int d\textbf{x} e^{i\Delta(\textbf{R}',\textbf{x},\textbf{R}'';t)} \chi^\dag_{\beta\textbf{R}'}(\textbf{x},t) \frac{\hbar\sigma^i}{2} \chi_{\alpha\textbf{R}''}(\textbf{x},t) \eta_{\alpha\textbf{R}'';\beta\textbf{R}'}(t).
\end{split}
\end{equation}

\subsubsection{Dynamical contribution}
The dynamical contribution to the linear response is
\begin{equation}
\begin{split}
    \breve{M}^{i}_\text{dyn}(\omega) = \frac{e}{2mc} \sum_{\alpha\beta\textbf{R}'} \int_{BZ} \frac{d\textbf{k}}{(2\pi)^3} \tilde{S}^i_{\beta\alpha} \Big( e^{i\textbf{k}\cdot(\textbf{R}-\textbf{R}')} \eta^{(B)}_{\alpha\textbf{R}';\beta\textbf{R}}(\omega) + e^{i\textbf{k}\cdot(\textbf{R}'-\textbf{R})} \eta^{(B)}_{\alpha\textbf{R};\beta\textbf{R}'}(\omega) \Big). 
\end{split}
\end{equation}
Inserting the SPDM response, setting $\omega = 0$, and assuming a uniform magnetic field,

\begin{equation}
\label{spinDyn}
\begin{split}
    \breve{M}^{i}_\text{dyn} = & \frac{e}{mc} B^l \sum_{mn} f_{nm} \int_{BZ} \frac{d\textbf{k}}{(2\pi)^3} \frac{ S^i_{nm} M^l_{mn} }{\Delta_{mn}(\textbf{k})}  
    \\
    + &\frac{ie^2}{4\hbar mc^2} B^l \sum_{mn} f_{nm} \int_{BZ} \frac{d\textbf{k}}{(2\pi)^3} \xi^b_{mn} \Big(S^i_{ns} \mathcal{W}^a_{sm} + \mathcal{W}^a_{ns} S^i_{sm} \Big).
\end{split}
\end{equation}

\subsubsection{Compositional contribution}
This comes from expanding the spinors and using the zeroth order single particle density matrix,

\begin{equation}
\label{spincomp}
\begin{split}
    \breve{M}^{i}_\text{comp}(t) = \frac{e}{mc} \sum_{\alpha} f_{\alpha} \int d\textbf{x} \Big( \chi^{(1)\dag}_{\alpha\textbf{R}}(\textbf{x},t) \frac{\hbar\sigma^i}{2} W_{\alpha\textbf{R}}(\textbf{x})+ W^\dag_{\alpha \textbf{R}}(\textbf{x}) \frac{\hbar \sigma^i}{2} \chi^{(1)}_{\alpha\textbf{R}}(\textbf{x},t) \Big)
    \\
    = 2\frac{e}{mc} \sum_{\alpha} f_{\alpha} \text{Re}\Bigg[ 
    \int d\textbf{x} W^\dag_{\alpha\textbf{R}} \frac{\hbar \sigma^i}{2} \chi^{(1)}_{\alpha\textbf{R}}(\textbf{x},t)
    \Bigg]
    \\
    = -\frac{e}{mc} \sum_{\alpha\beta\textbf{R}'} f_{\alpha} \text{Re}\Bigg[ 
    i\int d\textbf{x} W^\dag_{\alpha\textbf{R}}(\textbf{x}) \frac{\hbar\sigma^i}{2} W_{\beta\textbf{R}'}(\textbf{x}) \int d\textbf{y} W^\dag_{\beta\textbf{R}'}(\textbf{y}) \Delta(\textbf{R}',\textbf{y},\textbf{R};t) W_{\alpha\textbf{R}}(\textbf{y})
    \Bigg]
    \\
    = -\frac{e^2}{2m\hbar c^2} \epsilon^{lab} B^l(t) \sum_{\alpha\beta} f_{\alpha} \text{Re}\Bigg[ \int_{BZ} \frac{d\textbf{k}}{(2\pi)^3} 
    \tilde{S}^i_{\alpha\beta} \partial_a \tilde{\xi}^b_{\beta\alpha} \Bigg].
\end{split}
\end{equation}
Making the gauge dependence explicit, for a magnetic field independent of time, we have

\begin{equation}
\begin{split}
    \breve{M}^{i}_\text{comp} = -\frac{e^2}{2m\hbar c^2} \epsilon^{lab} B^l \sum_{mn} f_{n} \text{Re}\Bigg[ 
    \int_{BZ} \frac{d\textbf{k}}{(2\pi)^3} S^i_{nm} \Big( \partial_a \xi^b_{mn} + i\mathcal{W}^a_{ms}\xi^b_{sn} - i\xi^b_{ms}\mathcal{W}^a_{sn} + \partial_b \mathcal{W}^a_{mn} \Big)
    \Bigg]
\end{split}
\end{equation}

\section{The total magnetization - Split based on ``gauge dependence"}

The ``gauge dependence" we refer to is tracked by the $\mathcal{W}^a_{nm}$'s. A way to show the gauge-invariance in this sense methodically requires going through the orders of gauge-dependence. Henceforth the response is evaluated at zero-frequency and we consider a uniform applied magnetic field. Additionally, we restrict to topologically trivial insulators such that integration by parts can be employed without picking up boundary terms dependent upon the topological index. 

We start with the lowest order gauge-dependence and ``ratchet up", in a way that will become clear. In the following, integration by parts is crucial to show the gauge-dependence vanishes. Additionally, at times real part operations are made explicit and band indices are relabelled. This is done to show that the integrands are symmetric upon exchange of the Cartesian indices $a$ and $b$; when multiplied by the antisymmetric $\epsilon^{lab}$ the result is thus zero.

\subsection{The Susceptibility}
In the associated paper \cite{PreprintMagSus} we show that we can write 
\begin{equation}
    \chi^{il}=\chi^{il}_{\text{VV}}+\chi^{il}_{\text{occ}}+\chi^{il}_{\text{geo}},
\end{equation}
if 
\begin{equation}
    X^{il}_{\text{dyn}}(\mathcal{W})+X^{il}_{\text{comp}}(\mathcal{W})=0.
\end{equation}
That is, the sum of the ``extra" terms that arise in the dynamical and compositional contributions that depend on $\mathcal{W}$, and are explicitly gauge-dependent, should vanish.  We show here that they do. To do this it is useful to break up the contributions according to the powers of $\mathcal{W}$ that are involved; since terms involving derivatives of $\mathcal{W}$ involve products of $\mathcal{W}s$, in counting powers we take the derivative to raise the power by one. 

\subsection{Initial Linear Gauge-Dependence}
We begin by assembling the terms in $X^{il}_{\text{dyn}}$ that are explicitly linear in $\mathcal{W}$.  Combining the terms from (equations (\ref{atomicDyn}), (\ref{itinerantDyn}), and (\ref{spinDyn})), we find that the sum of these ``linearly gauge-dependent'' terms is
\begin{equation}
\begin{split}
    \text{Linear:Dyn} = \epsilon^{iab} B^l \frac{ie}{4\hbar c} \sum_{mn} f_{nm} \int_{BZ} \frac{d\textbf{k}}{(2\pi)^3} \Big((E_{l\textbf{k}}-E_{m\textbf{k}}) \mathcal{W}^a_{nl}\xi^b_{lm} + (E_{n\textbf{k}}-E_{l\textbf{k}})\xi^b_{nl}\mathcal{W}^a_{lm} \Big)
    \frac{ M^l_{mn} }{\Delta_{mn}(\textbf{k})} 
    \\
    + \epsilon^{icd} \epsilon^{lab} B^l \frac{ie^2}{16\hbar c^2} B^l \sum_{mn} f_{nm} \int_{BZ} \frac{d\textbf{k}}{(2\pi)^3} \Bigg[\xi^b_{mn} \Big( \xi^c_{nl}v^d_{ls} + v^d_{nl}\xi^c_{ls} \Big)\mathcal{W}^a_{sm} + \xi^b_{mn} \mathcal{W}^a_{ns}\Big( \xi^c_{sl}v^d_{lm} + v^d_{sl} \xi^c_{lm} \Big) \Bigg]
    \\
    -\frac{ie}{4\hbar c} \epsilon^{iab} B^l \sum_{mn} f_{nm} \int_{BZ} \frac{d\textbf{k}}{(2\pi)^3} \Big( (E_{m\textbf{k}}-E_{l\textbf{k}})\xi^b_{nl}\mathcal{W}^a_{lm} + (E_{l\textbf{k}}-E_{n\textbf{k}})\mathcal{W}^a_{nl}\xi^b_{lm} \Big) \frac{M^l_{mn}}{\Delta_{mn}(\textbf{k})}
    \\
    -\frac{ie^2}{16\hbar^2 c^2} \epsilon^{icd}\epsilon^{lab} B^l \sum_{mn} f_{nm} \int_{BZ} \frac{d\textbf{k}}{(2\pi)^3} \Bigg( \xi^b_{mn}\xi^d_{nl} \mathcal{W}^a_{lm}\partial_c (E_{l\textbf{k}}+E_{n\textbf{k}}) + \xi^b_{mn} \mathcal{W}^a_{nl} \xi^d_{lm} \partial_c (E_{m\textbf{k}}+E_{l\textbf{k}})
     \Big)
     \\
     +\frac{ie^2}{4\hbar mc^2} \epsilon^{lab}B^l \sum_{mn} f_{nm} \int_{BZ} \frac{d\textbf{k}}{(2\pi)^3} \xi^b_{mn} \Big( S^i_{ns}\mathcal{W}^a_{sm} + \mathcal{W}^a_{ns}S^i_{sm} \Big).
\end{split}
\end{equation}
Using $f_{nm}\mathcal{W}^a_{nm} = 0$, we can rewrite the above as

\begin{equation}
\begin{split}
    \text{Linear:Dyn} = -\epsilon^{iab} B^l \frac{ie}{4\hbar c} \sum_{mn} f_{nm} \int_{BZ} \frac{d\textbf{k}}{(2\pi)^3} \Big(\mathcal{W}^a_{nl}\xi^b_{lm} + \xi^b_{nl}\mathcal{W}^a_{lm} \Big) M^l_{mn}
    \\
    + \epsilon^{icd} \epsilon^{lab} B^l \frac{ie^2}{16\hbar c^2} \sum_{mn} f_{nm} \int_{BZ} \frac{d\textbf{k}}{(2\pi)^3} \Big( \xi^c_{nl}v^d_{lm} + v^d_{nl}\xi^c_{lm} - \frac{1}{\hbar} \xi^d_{nm} \partial_c (E_{m\textbf{k}}+E_{n\textbf{k}}) \Big)\Big( \mathcal{W}^a_{ms}\xi^b_{sn} + \xi^b_{ms}\mathcal{W}^a_{sn} \Big)
    \\
    +\frac{ie^2}{4\hbar mc^2} B^l \epsilon^{lab} \sum_{mn} f_{nm} \int_{BZ} \frac{d\textbf{k}}{(2\pi)^3} S^i_{nm} \Big( \mathcal{W}^a_{ms}\xi^b_{sn}+\xi^b_{ms}\mathcal{W}^a_{sn}\Big).
\end{split}
\end{equation}
We can next identify the Hermitian magnetization matrix element, and write the above as an explicitly real quantity by expanding $f_{nm} = f_{n} - f_{m}$,
\begin{equation}
\begin{split}
    \text{Linear:Dyn} = - \frac{e}{2\hbar c} \epsilon^{iab} B^l \sum_{n} f_{n} \int_{BZ} \frac{d\textbf{k}}{(2\pi)^3} \text{Re}\Bigg[ 
    i\Big(\mathcal{W}^a_{ns}\xi^b_{sm} + \xi^b_{ns} \mathcal{W}^a_{sm} \Big) M^l_{mn} 
    \Bigg]
    \\
    + \frac{e}{2\hbar c} \epsilon^{lab} B^l \sum_{n} f_{n} \int_{BZ} \frac{d\textbf{k}}{(2\pi)^3} \text{Re}\Bigg[ 
    iM^i_{nm} \Big(\mathcal{W}^a_{ms}\xi^b_{sn} + \xi^b_{ms}\mathcal{W}^a_{sn} \Big)
    \Bigg].
\end{split}
\end{equation}
We then expand the magnetization and velocity matrix elements to obtain an expression written in terms of the band energies,

\begin{equation}
\label{LinDyn}
\begin{split}
    \text{Linear:Dyn} =
    \\
    \epsilon^{icd} \epsilon^{lab} B^l \frac{e^2}{8\hbar^2 c^2} \sum_{n} f_{n} \int_{BZ} \frac{d\textbf{k}}{(2\pi)^3} \text{Re}\Bigg[ 
    \Big( \mathcal{W}^c_{nl}\xi^d_{lm} + \xi^d_{nl}\mathcal{W}^c_{lm} \Big)\Big( \xi^a_{ms}\xi^b_{sn}(2E_{s\textbf{k}}-E_{n\textbf{k}}-E_{m\textbf{k}}) - 2i \xi^a_{mn} \partial_b (E_{n\textbf{k}}+E_{m\textbf{k}}) \Big)
    \Bigg]
    \\
    +\epsilon^{icd}\epsilon^{lab} B^l\frac{e^2}{8\hbar^2 c^2}  \sum_{n} f_{n} \int_{BZ} \frac{d\textbf{k}}{(2\pi)^3} \text{Re}\Bigg[\Big(\xi^c_{ns}\xi^d_{sm}(E_{n\textbf{k}}+E_{m\textbf{k}}-2E_{s\textbf{k}}) + 2i\xi^c_{nm}\partial_d (E_{n\textbf{k}}+E_{m\textbf{k}})\Big)\Big(\mathcal{W}^a_{ml}\xi^b_{ln} + \xi^b_{ml}\mathcal{W}^a_{ln}\Big) \Bigg]
    \\
    + \frac{e^2}{2\hbar m c^2} B^l \sum_{n} f_{n} \int_{BZ} \frac{d\textbf{k}}{(2\pi)^3} \text{Re}\Bigg[ 
    i\epsilon^{lab} S^i_{nm}\Big( \mathcal{W}^a_{ms}\xi^b_{sn}+\xi^b_{ms}\mathcal{W}^a_{sn}\Big)
    -i\epsilon^{iab} \Big(\mathcal{W}^a_{ns}\xi^b_{sm} + \xi^b_{ns}\mathcal{W}^a_{sm}\Big) S^l_{mn}
    \Bigg].
\end{split}
\end{equation}
Next, we collect the different contributions to $X^{il}_{\text{comp}}$ that are linear in $\mathcal{W}$. The linearly gauge-dependent part of equation (\ref{atomicComp}) is

\begin{equation}
\label{Lincomp1}
\begin{split}
    \text{Linear:atomic comp} = - \epsilon^{icd} \epsilon^{lbd} B^l \frac{e^2}{4mc^2} \sum_{mn} f_{n} \int_{BZ} \frac{d\textbf{k}}{(2\pi)^3} \text{Re}\Bigg[ 
    \mathcal{W}^c_{nm} \xi^b_{mn} + \xi^c_{nm} \mathcal{W}^b_{mn} 
    \Bigg]
    \\
    -\frac{e^2}{4\hbar c^2} \epsilon^{icd} \epsilon^{lab} B^l \sum_{n} f_{n} \int_{BZ} \frac{d\textbf{k}}{(2\pi)^3} \text{Re}\Bigg[ \partial_a \xi^b_{nm}v^d_{ms}\mathcal{W}^c_{sn}  
    + iv^d_{ml}\xi^c_{ln} \Big(\mathcal{W}^a_{ns}\xi^b_{sm} - \xi^b_{ns}\mathcal{W}^a_{sm} \Big)
    \Bigg],
\end{split}
\end{equation}
while the linearly gauge-dependent part of equation (\ref{itinComp}) is
\begin{equation}
\label{Lincomp2}
\begin{split}
    \text{Linear:itinerant comp} = \frac{e^2}{4\hbar^2 c^2} \epsilon^{icd} \epsilon^{lab} B^l \sum_{mns} f_{n} \text{Re}\Bigg[ \int_{BZ} \frac{d\textbf{k}}{(2\pi)^3}
    \Big( \mathcal{W}^d_{nm}\xi^b_{mn} + \xi^d_{nm} \mathcal{W}^b_{mn} \Big) \partial_a \partial_c E_{n\textbf{k}}
    \Bigg]
    \\
    + \frac{e^2}{4\hbar^2 c^2} \epsilon^{icd} \epsilon^{lab} B^l \sum_{mns} f_{n} \text{Re}\Bigg[ \int_{BZ} \frac{d\textbf{k}}{(2\pi)^3} \xi^d_{nm}\xi^b_{ms}\Big( i\partial_a (E_{n\textbf{k}}-E_{s\textbf{k}}) \mathcal{W}^c_{sn} + i\partial_c (E_{n\textbf{k}}-E_{s\textbf{k}}) \mathcal{W}^a_{sn} \Big) \Bigg]
    \\
    +\frac{e^2}{4\hbar^2 c^2} \epsilon^{icd} \epsilon^{lab} B^l \sum_{mns} f_{n} \text{Re}\Bigg[ \int_{BZ} \frac{d\textbf{k}}{(2\pi)^3} 
    i\partial_c E_{n\textbf{k}} \Big( \xi^d_{nm}\Big(\mathcal{W}^a_{ml}\xi^b_{ln} - \xi^b_{ml}\mathcal{W}^a_{ln}\Big) + \mathcal{W}^d_{nm}\partial_a\xi^b_{mn} \Big)
    \Bigg]
    \\
    +\frac{e^2}{8\hbar^2 c^2} \epsilon^{icd}\epsilon^{lab} B^l \sum_{mns} f_{n} \text{Re}\Bigg[ 
    \int_{BZ} \frac{d\textbf{k}}{(2\pi)^3} i\partial_a (E_{m\textbf{k}}+E_{n\textbf{k}}) \Big( \mathcal{W}^c_{nl} \xi^d_{lm} - \xi^d_{nl}\mathcal{W}^c_{lm} \Big) \xi^b_{mn}
    \Bigg]
    \\
    +\frac{e^2}{8\hbar^2 c^2} \epsilon^{icd}\epsilon^{lab} B^l \sum_{mns} f_{n} \text{Re}\Bigg[ 
    \int_{BZ} \frac{d\textbf{k}}{(2\pi)^3} \partial_a (E_{n\textbf{k}}+E_{m\textbf{k}})\partial_c \xi^d_{nm} \mathcal{W}^b_{mn} 
    \Bigg]
    \\
    +\frac{e^2}{4\hbar^2 c^2} \epsilon^{icd} \epsilon^{lab} B^l \sum_{mns} f_{n} \text{Re}\Bigg[ \int_{BZ} \frac{d\textbf{k}}{(2\pi)^3} i(E_{n\textbf{k}}-E_{m\textbf{k}}) \xi^d_{ns}\partial_a \xi^b_{sm} \mathcal{W}^c_{mn} \Bigg]
    \\
    +\frac{e^2}{8\hbar^2 c^2} \epsilon^{icd} \epsilon^{lab} B^l \sum_{mns} f_{n} \text{Re}\Bigg[ 
    \int_{BZ} \frac{d\textbf{k}}{(2\pi)^3} i\partial_c \xi^d_{nm} \Big( 
    (E_{s\textbf{k}}-E_{m\textbf{k}}) \mathcal{W}^a_{ms}\xi^b_{sn} 
    + (E_{n\textbf{k}}-E_{s\textbf{k}}) \xi^b_{ms}\mathcal{W}^a_{sn} \Big)
    \Bigg] 
    \\
    -\frac{e^2}{8\hbar c^2} \epsilon^{icd}\epsilon^{lab}
    B^l \sum_{mns} f_{n} \text{Re}\Bigg[ 
    \int_{BZ} \frac{d\textbf{k}}{(2\pi)^3}  
    \partial_c \xi^d_{nm} \Big( v^b_{ms}\mathcal{W}^a_{sn} + \mathcal{W}^a_{ms}v^b_{sn} \Big) + i\Big(\mathcal{W}^c_{nl} \xi^d_{lm} - \xi^d_{nl}\mathcal{W}^c_{lm} \Big) \Big(v^b_{ms}\xi^a_{sn} + \xi^a_{ms}v^b_{sn} \Big)
    \Bigg]
    \\
    -\frac{e^2}{2m\hbar c^2} B^l \epsilon^{icd} \sum_{mn} f_{n} \text{Re}\Bigg[ 
    \int_{BZ} \frac{d\textbf{k}}{(2\pi)^3} \Big(i\mathcal{W}^c_{ns}\xi^d_{sm} - i\xi^d_{ns}\mathcal{W}^c_{sm} \Big) S^l_{mn}
    \Bigg],
\end{split}
\end{equation}
and lastly the linearly gauge-dependent term in (\ref{spincomp}) is
\begin{equation}
\label{Lincomp3}
\begin{split}
    \text{Linear:spin comp} = -\frac{e^2}{2m\hbar c^2} \epsilon^{lab }B^l \sum_{mn} f_{n} \text{Re}\Bigg[ 
    \int_{BZ} \frac{d\textbf{k}}{(2\pi)^3} S^i_{nm} \Big(i\mathcal{W}^a_{ms}\xi^b_{sn} - i\xi^b_{ms}
    \mathcal{W}^a_{sn} \Big)
    \Bigg].
\end{split}
\end{equation}
Combining all the contributions (equations (\ref{LinDyn}), (\ref{Lincomp1}), (\ref{Lincomp2}), and (\ref{Lincomp3}))  we collect those that are of a similar form, involving features such as partial derivatives of band energies, band energy dependence, and explicit spin dependence. This partitioning allows us to systematically identify cancellations in what is a very large expression. The full term that depends linearly on $\mathcal{W}$ can then be written as
\begin{equation}
    \text{Linear}_\text{GD} = \mathcal{A}_\text{linear} 
    +\mathcal{B}_\text{linear} 
    +\mathcal{C}_\text{linear} 
    +\mathcal{D}_\text{linear},
\end{equation}
where the terms involving the partial derivative of band energies combine to produce

\begin{equation}
\begin{split}
    \mathcal{A}_\text{linear}=-\frac{e^2}{2\hbar^2 c^2} \epsilon^{icd}\epsilon^{lab} B^l \sum_{n} f_{n}
    \int_{BZ} \frac{d\textbf{k}}{(2\pi)^3} \text{Re}\Bigg[ \partial_d E_{n\textbf{k}} \mathcal{W}^c_{nm} \partial_a \xi^b_{mn} + \mathcal{W}^a_{nm} \partial_c \xi^d_{mn} \partial_b E_{n\textbf{k}} \Bigg] 
    \\
    +\frac{e^2}{4\hbar^2 c^2} \epsilon^{icd}\epsilon^{lab} B^l \sum_{n} f_{n} \int_{BZ} \frac{d\textbf{k}}{(2\pi)^3} \text{Re}\Bigg[ 
    i\partial_b (2E_{m\textbf{k}}+E_{n\textbf{k}}+E_{s\textbf{k}}) \xi^a_{nm}\xi^d_{ms}\mathcal{W}^c_{sn} 
    \Bigg] 
    \\
    +\frac{e^2}{4\hbar^2 c^2} \epsilon^{icd}\epsilon^{lab} B^l \sum_{n} f_{n} \int_{BZ} \frac{d\textbf{k}}{(2\pi)^3} \text{Re}\Bigg[ 
    i\partial_d (2E_{m\textbf{k}}+E_{n\textbf{k}}+E_{s\textbf{k}}) \xi^c_{nm}\xi^b_{ms}\mathcal{W}^a_{sn}
    \Bigg],
\end{split}
\end{equation}
the energy dependent terms combine to produce
\begin{equation}
\begin{split}
    \mathcal{B}_\text{linear}= \frac{e^2}{4\hbar^2 c^2} \epsilon^{icd}\epsilon^{lab} B^l \sum_{n} f_{n} \text{Re}\Bigg[ 
    2E_{s\textbf{k}}\mathcal{W}^c_{nl}\xi^d_{lm}\xi^a_{ms}\xi^b_{sn}
    +2iE_{m\textbf{k}}\mathcal{W}^c_{nl}\xi^d_{lm}\partial_a\xi^b_{mn} 
    \\
    -2iE_{m\textbf{k}}\partial_c\xi^d_{nm} \xi^b_{ml}\mathcal{W}^a_{ln}
    -2E_{s\textbf{k}}\xi^c_{ns}\xi^d_{sm} \xi^b_{ml}\mathcal{W}^a_{ln}
    \Bigg],
\end{split}    
\end{equation}
where we have used 
\begin{equation}
\label{CurlConnection}
    \epsilon^{lab}\partial_a \xi^b_{nm} = i\epsilon^{lab}\xi^a_{ns}\xi^b_{sm},
\end{equation}
and the explicitly spin dependent terms are
\begin{equation}
\begin{split}
    \mathcal{C}_\text{linear}=\frac{e^2}{m\hbar c^2} \epsilon^{lab }B^l \sum_{mn} f_{n} \text{Re}\Bigg[ 
    \int_{BZ} \frac{d\textbf{k}}{(2\pi)^3} iS^i_{nm} \xi^b_{ms}
    \mathcal{W}^a_{sn}
    \Bigg]
    \\
    -\frac{e^2}{m\hbar c^2} B^l \epsilon^{icd} \sum_{mn} f_{n} \text{Re}\Bigg[ 
    \int_{BZ} \frac{d\textbf{k}}{(2\pi)^3}i\mathcal{W}^c_{ns}\xi^d_{sm}S^l_{mn}
    \Bigg].
\end{split}
\end{equation}
Lastly, a final set of terms that will be useful in implementing the inverse effective mass tensor sum rule are
\begin{equation}
\label{Dlinear}
\begin{split}
    \mathcal{D}_\text{linear}= \frac{e^2}{4\hbar^2 c^2} \epsilon^{icd} \epsilon^{lab} B^l \sum_{mns} f_{n} \text{Re}\Bigg[ \int_{BZ} \frac{d\textbf{k}}{(2\pi)^3}
    \Big( \mathcal{W}^d_{nm}\xi^b_{mn} + \xi^d_{nm} \mathcal{W}^b_{mn} \Big) \Big(\partial_a \partial_c E_{n\textbf{k}} - \frac{\hbar^2}{m} \delta_{ca} \Big)
    \Bigg].
\end{split}
\end{equation}

To put this in a form that will be useful, we take a derivative with respect to $\bf{k}$ of the expression (Eq. 32 of \cite{PreprintMagSus}) for the velocity matrix element to get an expression for $\partial_a\partial_c E_{n\textbf{k}}$. With this expression in hand we can convert equation (\ref{Dlinear}) to
\begin{equation}
\begin{split}
    \mathcal{D}_\text{linear} = \frac{e^2}{8\hbar c^2} \epsilon^{icd} \epsilon^{lab} B^l \sum_{mns} f_{n} \text{Re}\Bigg[ 
    \int_{BZ} \frac{d\textbf{k}}{(2\pi)^3} \Big(\xi^d_{sm} \mathcal{W}^b_{mn} + \xi^b_{sm} \mathcal{W}^d_{mn} \Big)\Big( \partial_a v^c_{ns} + \partial_c v^a_{ns} - \frac{2\hbar}{m} \delta_{ca}\delta_{ns}
    \\
    -\frac{i}{\hbar}\partial_a (E_{n\textbf{k}}-E_{s\textbf{k}})\xi^c_{ns} 
    -\frac{i}{\hbar}\partial_c(E_{n\textbf{k}}-E_{s\textbf{k}}) \xi^a_{ns}  
    -\frac{i}{\hbar}(E_{n\textbf{k}}-E_{s\textbf{k}}) ( \partial_c \xi^a_{ns} + \partial_a \xi^c_{ns})
    \Big)
    \Bigg].
\end{split}
\end{equation}
Next we use the sum rule
\begin{equation}
\begin{split}
    \partial_a v^c_{ns} = i\xi^a_{nl}v^c_{ls} -iv^c_{nl}\xi^a_{ls} + \frac{\hbar}{m} \delta_{ca} \delta_{ns},
\end{split}
\end{equation}
and process this equation further to write
\begin{equation}
\begin{split}
    \mathcal{D}_\text{linear} = \frac{e^2}{8\hbar^2 c^2} \epsilon^{icd} \epsilon^{lab} B^l \sum_{mns} f_{n} \text{Re}\Bigg[ 
    \int_{BZ} \frac{d\textbf{k}}{(2\pi)^3} \Big( \xi^d_{sm}\mathcal{W}^b_{mn} + \xi^b_{sm}\mathcal{W}^d_{mn} \Big)
    \Bigg( (E_{n\textbf{k}}+E_{s\textbf{k}}-2E_{l\textbf{k}})\xi^a_{nl}\xi^c_{ls} 
    \\
    + (E_{n\textbf{k}}+E_{s\textbf{k}}-2E_{l\textbf{k}})\xi^c_{nl}\xi^a_{ls}
    + 2i \partial_c (E_{s\textbf{k}}-E_{n\textbf{k}}) \xi^a_{ns} + 2i\partial_a (E_{s\textbf{k}}-E_{n\textbf{k}}) \xi^c_{ns} + i(E_{s\textbf{k}}-E_{n\textbf{k}}) \Big(\partial_c \xi^a_{ns} + \partial_a \xi^c_{ns}\Big)
    \Bigg)
    \Bigg].
\end{split}
\end{equation}

The inverse effective mass tensor sum rule could have been applied in many different ways. We have done it this way here because we can now combine $\mathcal{D}_{\text{linear}}$ with $\mathcal{A}_{\text{linear}}$ and $\mathcal{B}_{\text{linear}}$ to get terms that involve at most first derivatives of the band energies,   

\begin{equation}
\label{primeAB}
    \mathcal{A}_\text{linear} + \mathcal{B}_\text{linear} + \mathcal{D}_\text{linear} = \mathcal{A}'_\text{linear} + \mathcal{B}'_\text{linear},
\end{equation}
with
\begin{equation}
\begin{split}
    \mathcal{A}'_\text{linear} = \frac{e^2}{4\hbar^2 c^2} \epsilon^{icd} \epsilon^{lab} B^l \sum_{mns} f_{n} \text{Re}\Bigg[ 
    i\partial_d (E_{m\textbf{k}}+E_{n\textbf{k}}+2E_{s\textbf{k}}) \xi^c_{nm}\xi^b_{ms}\mathcal{W}^a_{sn}
    -i\partial_b (E_{n\textbf{k}}+E_{s\textbf{k}}) \mathcal{W}^a_{nm} \xi^c_{ms} \xi^d_{sn} 
    \\
    +i\partial_b (E_{m\textbf{k}}+E_{n\textbf{k}}+2E_{s\textbf{k}}) \xi^a_{nm}\xi^d_{ms}\mathcal{W}^c_{sn} 
    -i\partial_d (E_{n\textbf{k}}+E_{s\textbf{k}}) \mathcal{W}^c_{nm} \xi^a_{ms} \xi^b_{sn}
    \Bigg],
\end{split}
\end{equation}
and 
\begin{equation}
\begin{split}
    \mathcal{B}'_\text{linear} = \frac{e^2}{8\hbar^2 c^2} \epsilon^{icd}\epsilon^{lab} B^l \sum_{n} f_{n} \text{Re}\Bigg[
    4iE_{m\textbf{k}}\mathcal{W}^c_{nl}\xi^d_{lm}\partial_a\xi^b_{mn} 
    -4iE_{m\textbf{k}}\partial_c\xi^d_{nm} \xi^b_{ml}\mathcal{W}^a_{ln}
    -3E_{s\textbf{k}}\xi^c_{ns}\xi^d_{sm} \xi^b_{ml}\mathcal{W}^a_{ln}
    +3E_{s\textbf{k}}\mathcal{W}^c_{nl}\xi^d_{lm}\xi^a_{ms}\xi^b_{sn}
    \\
    +i(E_{s\textbf{k}}-E_{n\textbf{k}}) \Big( \partial_c \xi^a_{ns} + \partial_a \xi^c_{ns} \Big)\Big(\xi^d_{sm}\mathcal{W}^b_{mn} + \xi^b_{sm} \mathcal{W}^d_{mn} \Big)
    +(E_{n\textbf{k}}+E_{s\textbf{k}}-2E_{l\textbf{k}}) \xi^a_{nl}\xi^c_{ls}\xi^b_{sm}\mathcal{W}^d_{mn}\Big)
    \\
    -i(E_{n\textbf{k}}-2E_{l\textbf{k}}) \xi^a_{nl}\partial_c \xi^d_{lm}\mathcal{W}^b_{mn}
    + (E_{n\textbf{k}}+E_{s\textbf{k}}-2E_{l\textbf{k}}) \xi^c_{nl}\xi^a_{ls}\xi^d_{sm}\mathcal{W}^b_{mn} 
    -i (E_{n\textbf{k}}-2E_{l\textbf{k}}) \xi^c_{nl}\partial_a \xi^b_{lm}\mathcal{W}^d_{mn}
    \Bigg].
\end{split}
\end{equation}

We then integrate the terms involving partial derivatives of band energies by parts,
\begin{equation}
\begin{split}
    \mathcal{A}'_\text{linear} =-\frac{e^2}{8\hbar^2 c^2} \epsilon^{icd}\epsilon^{lab} B^l \sum_{mns} f_{n} \text{Re}\Bigg[ 
    i(2E_{m\textbf{k}}+2E_{n\textbf{k}}+4E_{s\textbf{k}}) \Big( \partial_d \xi^c_{nm} \xi^b_{ms}\mathcal{W}^a_{sn} + \xi^c_{nm} \partial_d \xi^b_{ms} \mathcal{W}^a_{sn} + \xi^c_{nm}\xi^b_{ms} \partial_d \mathcal{W}^a_{sn} \Big)
    \\
    -i(2E_{n\textbf{k}}+2E_{s\textbf{k}})\Big( \partial_b \mathcal{W}^a_{nm} \xi^c_{ms} \xi^d_{sn} + \mathcal{W}^a_{nm} \partial_b \xi^c_{ms} \xi^d_{sn} + \mathcal{W}^a_{nm} \xi^c_{ms} \partial_b \xi^d_{sn} \Big)
    \\
    +i(2E_{m\textbf{k}} + 2E_{n\textbf{k}} + 4E_{s\textbf{k}} ) \Big( 
    \partial_b \xi^a_{nm} \xi^d_{ms}\mathcal{W}^c_{sn} +  \xi^a_{nm} \partial_b \xi^d_{ms}\mathcal{W}^c_{sn}
    +  \xi^a_{nm} \xi^d_{ms} \partial_b \mathcal{W}^c_{sn}
    \Big)
    \\
    -i(2E_{n\textbf{k}}+2E_{s\textbf{k}}) \Big( \partial_d \mathcal{W}^c_{nm} \xi^a_{ms}\xi^b_{sn} 
    + \mathcal{W}^c_{nm} \partial_d\xi^a_{ms}\xi^b_{sn}  
    +\mathcal{W}^c_{nm} \xi^a_{ms} \partial_d\xi^b_{sn} \Big)
    \Bigg]
\end{split}
\end{equation}
where now the linear gauge-dependence has been manipulated to be dependent only on the energies, although at the cost of picking up a new quadratic gauge-dependence in the process, 

\begin{equation}
\begin{split}
    \mathcal{A}'_\text{linear} + \mathcal{B}'_\text{linear} = \frac{e^2}{8\hbar^2 c^2} \epsilon^{icd} \epsilon^{lab} B^l \sum_{mns} f_{n} \int_{BZ} \frac{d\textbf{k}}{(2\pi)^3}  \text{Re}\Bigg[ 
    i(2E_{m\textbf{k}}+3E_{s\textbf{k}}+3E_{n\textbf{k}})  \Big(\partial_c \xi^a_{ns}-\partial_a \xi^c_{ns}\Big)\xi^d_{sm} \mathcal{W}^b_{mn}  
    \\
    +i(2E_{m\textbf{k}} + 3E_{s\textbf{k}} + 3E_{n\textbf{k}} ) \Big(   \partial_a \xi^c_{ns}-\partial_c \xi^a_{ns}\Big)\xi^b_{sm}\mathcal{W}^d_{mn}
    \\
    +(2E_{n\textbf{k}}+3E_{l\textbf{k}}+3E_{s\textbf{k}}) \xi^c_{ns} \xi^d_{sm} \xi^a_{ml}\mathcal{W}^b_{ln} 
    +(2E_{n\textbf{k}} + 3E_{s\textbf{k}}+3E_{l\textbf{k}} ) 
    \xi^a_{nl} \xi^b_{lm} \xi^c_{ms}\mathcal{W}^d_{sn}
    \\
    +(E_{n\textbf{k}}+E_{s\textbf{k}}-2E_{l\textbf{k}}) \xi^a_{nl}\xi^c_{ls}\xi^b_{sm}\mathcal{W}^d_{mn}
    + (E_{n\textbf{k}}+E_{s\textbf{k}}-2E_{l\textbf{k}}) \xi^c_{nl}\xi^a_{ls}\xi^d_{sm}\mathcal{W}^b_{mn} 
    \Bigg]
    \\
    + \mathcal{E}_\text{quadratic},
\end{split}
\end{equation}
where
\begin{equation}
\label{Equadratic}
\begin{split}
    \mathcal{E}_\text{quadratic} = -\frac{e^2}{8\hbar^2 c^2} \epsilon^{icd}\epsilon^{lab} B^l \sum_{mns} f_{n} \text{Re}\Bigg[ 
    i(2E_{m\textbf{k}}+2E_{n\textbf{k}}+4E_{s\textbf{k}}) \xi^c_{nm}\xi^b_{ms} \partial_d \mathcal{W}^a_{sn} 
    -i(2E_{n\textbf{k}}+2E_{s\textbf{k}})\partial_b \mathcal{W}^a_{nm} \xi^c_{ms} \xi^d_{sn}  
    \\
    +i(2E_{m\textbf{k}} + 2E_{n\textbf{k}} + 4E_{s\textbf{k}} ) 
    \xi^a_{nm} \xi^d_{ms} \partial_b \mathcal{W}^c_{sn}
    -i(2E_{n\textbf{k}}+2E_{s\textbf{k}}) 
    \partial_d \mathcal{W}^c_{nm} \xi^a_{ms}\xi^b_{sn} 
    \Bigg].
\end{split}
\end{equation}
We then use
\begin{equation}
\label{Dconnection}
\begin{split}
    i\partial_c \xi^a_{ns} - i\partial_a \xi^c_{ns} = \xi^a_{nl}\xi^c_{ls} - \xi^c_{nl}\xi^a_{ls} ,
\end{split}
\end{equation}
so
\begin{equation}
\begin{split}
    \mathcal{A}'_\text{linear} + \mathcal{B}'_\text{linear} = \frac{e^2}{8\hbar^2 c^2} \epsilon^{icd} \epsilon^{lab} B^l \sum_{mns} f_{n} \int_{BZ} \frac{d\textbf{k}}{(2\pi)^3}  \text{Re}\Bigg[ 
    -(2E_{l\textbf{k}}+2E_{m\textbf{k}}+2E_{s\textbf{k}}+2E_{n\textbf{k}})  \xi^c_{nl}\xi^a_{ls}\xi^d_{sm}\mathcal{W}^b_{mn} 
    \\
    -(2E_{l\textbf{k}}+2E_{m\textbf{k}}+2E_{s\textbf{k}}+2E_{n\textbf{k}})
    \xi^a_{nl}\xi^c_{ls}\xi^b_{sm}\mathcal{W}^d_{mn}
    \Bigg] + \mathcal{E}_\text{quadratic}
    \\
    = -\frac{e^2}{8\hbar^2 c^2} \epsilon^{icd} \epsilon^{lab} B^l \sum_{mns} f_{n} \int_{BZ} \frac{d\textbf{k}}{(2\pi)^3}  \text{Re}\Bigg[(E_{l\textbf{k}}+E_{m\textbf{k}}+E_{s\textbf{k}}+E_{n\textbf{k}})  \Big(\xi^c_{nl}\xi^a_{ls}\xi^d_{sm}\mathcal{W}^b_{mn}
    +\xi^d_{nl}\xi^a_{ls}\xi^c_{sm}\mathcal{W}^b_{mn}\Big) 
    \\
    +(E_{l\textbf{k}}+E_{m\textbf{k}}+E_{s\textbf{k}}+E_{n\textbf{k}})
    \Big(\xi^a_{nl}\xi^c_{ls}\xi^b_{sm}\mathcal{W}^d_{mn}
    +\xi^b_{nl}\xi^c_{ls}\xi^a_{sm}\mathcal{W}^d_{mn} \Big)
    \Bigg] + \mathcal{E}_\text{quadratic}
    \\
    = \mathcal{E}_\text{quadratic},
\end{split}
\end{equation}
where to obtain the result the real part operation is made explicit and the band indices interchanged as needed. 
The integrand is either symmetric upon exchange of $c$ and $d$, or of $a$ and $b$. 

The spin term $\mathcal{C}_{\text{linear}}$ can also be manipulated and converted by partial integration to a form that exhibits quadratic gauge dependence as well. We have

\begin{equation}
\begin{split}
    \mathcal{C}_\text{linear} = \frac{e^2}{2m\hbar c^2} \epsilon^{lab }B^l \sum_{mn} f_{n} \text{Re}\Bigg[ 
    \int_{BZ} \frac{d\textbf{k}}{(2\pi)^3} i\mathcal{W}^a_{ns} \Big(S^i_{sm} \xi^b_{mn}
    - \xi^b_{sm}S^i_{mn}
    \Big)
    \Bigg]
    \\
    -\frac{e^2}{2m\hbar c^2} B^l \epsilon^{icd} \sum_{mn} f_{n} \text{Re}\Bigg[
    \int_{BZ} \frac{d\textbf{k}}{(2\pi)^3} i\mathcal{W}^c_{ns}\Big(\xi^d_{sm}S^l_{mn} - S^l_{sm}\xi^d_{mn} \Big)
    \Bigg],
\end{split}
\end{equation}
and then using
\begin{equation}
    \partial_b S^i_{sn} = i\xi^b_{sm}S^i_{mn} - S^i_{sm}\xi^b_{mn}, 
\end{equation}
we have
\begin{equation}
\label{Clinear}
\begin{split}
    \mathcal{C}_\text{linear} = -\frac{e^2}{2 m \hbar c^2} \sum_{ns} f_{n} \text{Re}\Bigg[ 
    \int_{BZ} \frac{d\textbf{k}}{(2\pi)^3} \Bigg( \epsilon^{lab} \mathcal{W}^a_{ns} \partial_b S^i_{sn} + \epsilon^{icd} \mathcal{W}^c_{ns} \partial_d S^l_{sn} \Bigg)
    \Bigg]
    \\
    = \frac{e^2}{2 m \hbar c^2} \sum_{ns} f_{n} \text{Re}\Bigg[ 
    \int_{BZ} \frac{d\textbf{k}}{(2\pi)^3} \Bigg( \epsilon^{lab} \partial_b\mathcal{W}^a_{ns}  S^i_{sn} + \epsilon^{icd} \partial_d \mathcal{W}^c_{ns} S^l_{sn} \Bigg)
    \Bigg].
\end{split}
\end{equation}
We have thus shown that
\begin{equation}
\label{LtoNL}
    \text{Linear}_\text{GD} \rightarrow \mathcal{E}_{\text{quadratic}}+\mathcal{C}_{\text{linear}},
\end{equation}
where we keep that ``linear" notation on both $\text{Linear}_\text{GD}$ and $\mathcal{C}_{\text{linear}}$, although those terms are now written in a way that involves only quadratic dependence on $\mathcal{W}$.
\subsection{Quadratic Gauge-Dependence}
We now add to the above terms (\ref{LtoNL}) the terms in $X^{il}_{\text{dyn}}$ and $X^{il}_{\text{comp}}$ that are initially quadratic in their dependence on $\mathcal{W}$. The combined quadratically gauge-dependent dynamical contributions are
\begin{equation}
\label{QuadDyn}
\begin{split}
    \text{Quadratic:Dyn} =\epsilon^{icd} \epsilon^{lab}\frac{ie^2}{16\hbar c^2} B^l \sum_{mn} f_{nm} \int_{BZ} \frac{d\textbf{k}}{(2\pi)^3} \Bigg[\xi^b_{mn} \Big( \mathcal{W}^c_{nl}v^d_{ls} + v^d_{nl}\mathcal{W}^c_{ls} \Big)\mathcal{W}^a_{sm} + \mathcal{W}^a_{ns}\Big( \mathcal{W}^c_{sl}v^d_{lm} + v^d_{sl} \mathcal{W}^c_{lm} \Big)\xi^b_{mn} \Bigg]
    \\
    -\frac{ie^2}{16\hbar^2 c^2} \epsilon^{icd}\epsilon^{lab} B^l \sum_{mn} f_{nm} \int_{BZ} \frac{d\textbf{k}}{(2\pi)^3}
    i\xi^b_{mn}\xi^d_{nl} \mathcal{W}^c_{ls} \mathcal{W}^a_{sm} (E_{s\textbf{k}}-E_{l\textbf{k}}) 
    \\
    -\frac{ie^2}{16\hbar^2 c^2} \epsilon^{icd} \epsilon^{lab} B^l \sum_{mn} f_{nm} \int_{BZ} \frac{d\textbf{k}}{(2\pi)^3} i\xi^b_{mn} \mathcal{W}^a_{ns}\xi^d_{sl}\mathcal{W}^c_{lm} (E_{m\textbf{k}}-E_{l\textbf{k}})
    \\
    -\frac{ie^2}{16\hbar^2 c^2} \epsilon^{icd}\epsilon^{lab} B^l \sum_{mn} f_{nm} \int_{BZ} \frac{d\textbf{k}}{(2\pi)^3} i\xi^b_{mn} \mathcal{W}^c_{nl} (E_{l\textbf{k}}-E_{n\textbf{k}}) \xi^d_{ls} \mathcal{W}^a_{sm} 
    \\
    -\frac{ie^2}{16\hbar^2 c^2} \epsilon^{icd}\epsilon^{lab} B^l \sum_{mn} f_{nm} \int_{BZ} \frac{d\textbf{k}}{(2\pi)^3} \xi^b_{mn} \mathcal{W}^a_{ns}
    i\mathcal{W}^c_{sl} (E_{l\textbf{k}}-E_{s\textbf{k}}) \xi^d_{lm} 
    \\
    = \epsilon^{icd}\epsilon^{lab} \frac{e^2}{8\hbar c^2} B^l \sum_{mn} f_{nm} \int_{BZ} \frac{d\textbf{k}}{(2\pi)^3}  \text{Re}\Bigg[ 
    (E_{s\textbf{k}}-E_{n\textbf{k}})\xi^b_{mn} \xi^d_{nl} \mathcal{W}^c_{ls}\mathcal{W}^a_{sm} + (E_{m\textbf{k}}-E_{s\textbf{k}})\xi^b_{mn}\mathcal{W}^a_{ns}\xi^d_{sl} \mathcal{W}^c_{lm}  
    \Bigg], 
\end{split}
\end{equation}
the atomic compositional contributions are
\begin{equation}
\label{Quadcomp1}
\begin{split}
    \text{Quadratic:atomic comp}-\frac{e^2}{4\hbar c^2} \epsilon^{icd}\epsilon^{lab} B^l \sum_{mn} f_{n} \text{Re}\Bigg[ 
    \int_{BZ} \frac{d\textbf{k}}{(2\pi)^3} \Big(i\mathcal{W}^a_{ns}\xi^b_{sm} - i\xi^b_{ns}\mathcal{W}^a_{sm} \Big) v^d_{ml}\mathcal{W}^c_{ln} + \partial_b \mathcal{W}^a_{nm} v^d_{ml} \xi^c_{ln}
    \Bigg]
    \\
    -\frac{e^2}{4mc^2} B^l \epsilon^{iab} \epsilon^{lcb} \sum_{mn} f_{n} \int_{BZ} \frac{d\textbf{k}}{(2\pi)^3} \text{Re}\Bigg[ 
    \mathcal{W}^a_{nm}\mathcal{W}^c_{mn}
    \Bigg],
\end{split}
\end{equation}
the spin compositional contributions are
\begin{equation}
\label{QuadComp2}
\begin{split}
    \text{Quadratic:spin comp}-\frac{e^2}{2m\hbar c^2} \epsilon^{lab} B^l \sum_{mn} f_{n} \text{Re}\Bigg[ 
    \int_{BZ} \frac{d\textbf{k}}{(2\pi)^3} S^i_{nm} \partial_b \mathcal{W}^a_{mn} 
    \Bigg]
\end{split}
\end{equation}
and the itinerant compositional contributions are

\begin{equation}
\label{QuadComp3}
\begin{split}
    \text{Quadratic:itinerant comp} = \frac{e^2}{4\hbar^2 c^2} \epsilon^{icd} \epsilon^{lab} B^l \sum_{mns} f_{n} \text{Re}\Bigg[ 
    \int_{BZ} \frac{d\textbf{k}}{(2\pi)^3} \mathcal{W}^d_{nm} \mathcal{W}^b_{ms} \partial_a\partial_c E_{n\textbf{k}} \delta_{sn} 
    \Bigg]
    \\
    +\frac{e^2}{4\hbar^2 c^2} \epsilon^{icd} \epsilon^{lab} B^l \sum_{mns} f_{n} \text{Re}\Bigg[ 
    \int_{BZ} \frac{d\textbf{k}}{(2\pi)^3} 
    i\Big(\xi^d_{nm}\mathcal{W}^b_{ms} + \mathcal{W}^d_{nm} \xi^b_{ms}\Big)\Big(\partial_a (E_{n\textbf{k}}-E_{s\textbf{k}}) \mathcal{W}^c_{sn} + \partial_c (E_{n\textbf{k}}-E_{s\textbf{k}}) \mathcal{W}^a_{sn} \Big) 
    \\
    + \frac{i}{2} (E_{n\textbf{k}}-E_{s\textbf{k}}) \xi^d_{nm} \xi^b_{ms} \Big(\partial_a \mathcal{W}^c_{sn} + \partial_c \mathcal{W}^a_{sn} \Big) \Bigg]
    + \frac{1}{2} \Big(2E_{l\textbf{k}}-E_{s\textbf{k}}-E_{n\textbf{k}}\Big) \xi^d_{nm}\xi^b_{ms} \Big( \mathcal{W}^a_{sl}\mathcal{W}^c_{ln} + \mathcal{W}^c_{sl}\mathcal{W}^a_{ln} \Big)
    \Bigg] 
    \\
    +\frac{e^2}{4\hbar^2 c^2}\epsilon^{icd} \epsilon^{lab} B^l \sum_{mns} f_{n} 
    \int_{BZ} \frac{d\textbf{k}}{(2\pi)^3} \text{Re}\Bigg[  i(E_{n\textbf{k}}-E_{m\textbf{k}}) \mathcal{W}^d_{ns} \partial_a \xi^b_{sm} \mathcal{W}^c_{mn} 
    +(E_{m\textbf{k}}-E_{n\textbf{k}}) \xi^d_{ns} \Big(\mathcal{W}^a_{sl}\xi^b_{lm} - \xi^b_{sl}\mathcal{W}^a_{lm}\Big) \mathcal{W}^c_{mn}
    \\
    + i\partial_c E_{n\textbf{k}} \mathcal{W}^d_{ns} \Big( \mathcal{W}^a_{sl}\xi^b_{ln} - \xi^b_{sl}\mathcal{W}^a_{ln} \Big) 
    +\partial_c E_{n\textbf{k}} \xi^d_{ns} \partial_b \mathcal{W}^a_{sn}
    \Bigg]
    \\
    -\frac{e^2}{8\hbar c^2} \epsilon^{icd} \epsilon^{lab} B^l \sum_{mns} f_{n} \text{Re}\Bigg[ 
    \int_{BZ} \frac{d\textbf{k}}{(2\pi)^3} i\Big(\mathcal{W}^c_{nl}\xi^d_{lm} - \xi^d_{nl}\mathcal{W}^c_{lm} \Big)\Big( v^b_{ms} \mathcal{W}^a_{sn} + \mathcal{W}^a_{ms}v^b_{sn} -\frac{1}{\hbar} \partial_a (E_{m\textbf{k}}+E_{n\textbf{k}}) \mathcal{W}^b_{mn} 
    \\
    - \frac{i}{\hbar} (E_{s\textbf{k}}-E_{m\textbf{k}}) \mathcal{W}^a_{ms}\xi^b_{sn} 
    - \frac{i}{\hbar}(E_{n\textbf{k}}-E_{s\textbf{k}}) \xi^b_{ms}\mathcal{W}^a_{sn} \Big)
    +\partial_d \mathcal{W}^c_{nm} \Big( v^b_{ms} \xi^a_{sn} + \xi^a_{ms} v^b_{sn} - \frac{1}{\hbar} \partial_a(E_{m\textbf{k}}+E_{n\textbf{k}}) \xi^b_{mn} \Big)
    \\
    -\frac{i}{\hbar} (E_{s\textbf{k}}-E_{m\textbf{k}}) \partial_c \xi^d_{nm} \mathcal{W}^a_{ms} \mathcal{W}^b_{sn} - \frac{i}{\hbar} (E_{n\textbf{k}}-E_{s\textbf{k}}) \partial_c \xi^d_{nm} \mathcal{W}^b_{ms}\mathcal{W}^a_{sn} 
    \Bigg]
    \\
    -\frac{e^2}{2m\hbar c^2} B^l \epsilon^{icd} \sum_{mn} f_{n} \text{Re}\Bigg[ 
    \int_{BZ} \frac{d\textbf{k}}{(2\pi)^3} \partial_d \mathcal{W}^c_{nm} S^l_{mn}
    \Bigg].
\end{split}
\end{equation}

We group all these contributions (equations (\ref{QuadDyn}), (\ref{Quadcomp1}), (\ref{QuadComp2}), and (\ref{QuadComp3})), and also add in the quadratically gauge-dependent terms equations (\ref{Equadratic}) and (\ref{Clinear}) that arose from processing the initially linearly gauge-dependent terms. Thus the total quadratic gauge-dependence is

\begin{equation}
\label{QuadTotal}
    \text{Quadratic}_\text{GD} = \mathcal{A}_\text{quadratic} + \mathcal{B}_\text{quadratic} + \mathcal{C}_\text{quadratic} + \mathcal{D}_\text{quadratic}
    +\mathcal{E}_\text{quadratic}
    + \mathcal{C}_\text{linear},
\end{equation}
where $\mathcal{A}_{\text{quadratic}}$ involves the band energies; using 
\begin{equation}
    \epsilon^{lab} \partial_a \mathcal{W}^b_{nm} = i\epsilon^{lab}\mathcal{W}^b_{ns}\mathcal{W}^a_{sm},
\end{equation}
it can be written as
\begin{equation}
\begin{split}
    \mathcal{A}_\text{quadratic} = 
    \epsilon^{icd} \epsilon^{lab} \frac{e^2}{8\hbar c^2} B^l \sum_{mn} f_{n} \int_{BZ} \frac{d\textbf{k}}{(2\pi)^3} \text{Re}\Bigg[ (E_{s\textbf{k}}-E_{n\textbf{k}})\xi^b_{mn} \xi^d_{nl} \mathcal{W}^c_{ls}\mathcal{W}^a_{sm} + (E_{m\textbf{k}}-E_{s\textbf{k}})\xi^b_{mn}\mathcal{W}^a_{ns}\xi^d_{sl} \mathcal{W}^c_{lm} 
    \\
    -(E_{s\textbf{k}}-E_{m\textbf{k}})\xi^b_{nm} \xi^d_{ml} \mathcal{W}^c_{ls}\mathcal{W}^a_{sn} - (E_{n\textbf{k}}-E_{s\textbf{k}})\xi^b_{nm}\mathcal{W}^a_{ms}\xi^d_{sl} \mathcal{W}^c_{ln} 
    \\
    +(2E_{m\textbf{k}}-2E_{l\textbf{k}})\Big(\mathcal{W}^a_{ns}\xi^b_{sm} - \xi^b_{ns}\mathcal{W}^a_{sm} \Big) \xi^d_{ml}\mathcal{W}^c_{ln} 
    + i(2E_{l\textbf{k}}-2E_{m\textbf{k}})\partial_b \mathcal{W}^a_{nm} \xi^d_{ml}\xi^c_{ln}
    \\
    +i(E_{n\textbf{k}}-E_{s\textbf{k}}) \xi^d_{nm}\xi^b_{ms}\Big(\partial_a \mathcal{W}^c_{sn} + \partial_c \mathcal{W}^a_{sn}\Big) 
    + \Big(2E_{l\textbf{k}}-E_{s\textbf{k}}-E_{n\textbf{k}}\Big) \xi^d_{nm} \xi^b_{ms} \Big( \mathcal{W}^a_{sl} \mathcal{W}^c_{ln} + \mathcal{W}^c_{sl}\mathcal{W}^a_{ln}\Big)
    \\
    +i(2E_{n\textbf{k}}-2E_{m\textbf{k}}) \mathcal{W}^d_{ns}\partial_a \xi^b_{sm} \mathcal{W}^c_{mn} + (2E_{m\textbf{k}}-2E_{n\textbf{k}}) \xi^d_{ns} \Big(\mathcal{W}^a_{sl}\xi^b_{lm} - \xi^b_{sl} \mathcal{W}^a_{lm}\Big) \mathcal{W}^c_{mn} 
    \\
    +\Big(\mathcal{W}^c_{nl}\xi^d_{lm} - \xi^d_{nl}\mathcal{W}^c_{lm}\Big)\Big( \xi^b_{ms}\mathcal{W}^a_{sn} (E_{m\textbf{k}}-E_{n\textbf{k}}) + \mathcal{W}^a_{ms}\xi^b_{sn} (E_{m\textbf{k}}-E_{n\textbf{k}}) \Big)
    \\
    +i(E_{m\textbf{k}}+E_{n\textbf{k}}-2E_{s\textbf{k}}) \partial_d \mathcal{W}^c_{nm} \xi^a_{ms}\xi^b_{sn}
    +i (2E_{s\textbf{k}} - E_{n\textbf{k}}-E_{m\textbf{k}}) \partial_c \xi^d_{nm} \mathcal{W}^a_{ms}\mathcal{W}^b_{sn}
    \Bigg],
\end{split}
\end{equation}
which can be more compactly written as
\begin{equation}
\label{Aquadratic}
\begin{split}
    \mathcal{A}_\text{quadratic} = \epsilon^{icd} \epsilon^{lab} \frac{e^2}{8\hbar c^2} B^l \sum_{mn} f_{n} \int_{BZ} \frac{d\textbf{k}}{(2\pi)^3} \text{Re}\Bigg[ 
    (4E_{m\textbf{k}}-E_{l\textbf{k}}-E_{n\textbf{k}}-2E_{s\textbf{k}})\xi^b_{nm}\xi^d_{ml}\mathcal{W}^c_{ls}\mathcal{W}^a_{sn} 
    \\
    +
    (2E_{l\textbf{k}}-E_{s\textbf{k}}-E_{n\textbf{k}}) \xi^d_{nm} \xi^b_{ms}\mathcal{W}^c_{sl}\mathcal{W}^a_{ln}
    + (2E_{s\textbf{k}}-2E_{l\textbf{k}}+2E_{m\textbf{k}}- 2E_{n\textbf{k}})\mathcal{W}^a_{ns} \mathcal{W}^b_{sm} \xi^d_{ml}\xi^c_{ln}
    \\
    +(E_{m\textbf{k}}-E_{n\textbf{k}}+2E_{l\textbf{k}}-2E_{s\textbf{k}}) \xi^a_{nl} \xi^b_{lm} \mathcal{W}^c_{ms}\mathcal{W}^d_{sn} 
    +i(E_{n\textbf{k}}-E_{s\textbf{k}}) \xi^d_{nm}\xi^b_{ms}\Big(\partial_a \mathcal{W}^c_{sn} + \partial_c \mathcal{W}^a_{sn}\Big)
    \Bigg].
\end{split}
\end{equation}
The terms involving partial derivatives of the band energies are
\begin{equation}
\label{Bquadratic}
\begin{split}
    \mathcal{B}_\text{quadratic} = \frac{e^2}{4\hbar^2 c^2} \epsilon^{icd} \epsilon^{lab} B^l \sum_{mn} f_{n} \int_{BZ} \frac{d\textbf{k}}{(2\pi)^3} \text{Re}\Bigg[  
    - i\partial_d (E_{n\textbf{k}}+E_{m\textbf{k}}) \mathcal{W}^a_{ln} \mathcal{W}^b_{nm} \xi^c_{ml}
    + i\partial_a (E_{n\textbf{k}}+E_{m\textbf{k}})\xi^b_{ms} \mathcal{W}^c_{sn}\mathcal{W}^d_{nm} 
    \\
    +i\partial_d (2E_{m\textbf{k}}+E_{n\textbf{k}}+E_{l\textbf{k}}) \xi^b_{ml}\mathcal{W}^a_{ln}\mathcal{W}^c_{nm} 
    +i \partial_a (2E_{m\textbf{k}}+E_{l\textbf{k}}+E_{n\textbf{k}}) \xi^d_{nm}\mathcal{W}^b_{ml}\mathcal{W}^c_{ln} 
    \Bigg],
\end{split}
\end{equation}
and the terms that will be processed further by using the general effective mass tensor sum rule are
\begin{equation}
\begin{split}
    \mathcal{D}_\text{quadratic} = \frac{e^2}{4\hbar^2 c^2} \epsilon^{icd} \epsilon^{lab} B^l \sum_{mns} f_{n} \text{Re}\Bigg[ 
    \int_{BZ} \frac{d\textbf{k}}{(2\pi)^3} \mathcal{W}^d_{nm} \mathcal{W}^b_{ms} \delta_{sn}\Big(  \partial_a\partial_c E_{n\textbf{k}} - \frac{\hbar^2}{m} \delta_{ac}\Big)
    \Bigg].
\end{split}
\end{equation}
Addressing $\mathcal{D}_{\text{quadratic}}$ in the same way that we addressed $\mathcal{D}_{\text{linear}}$, we find $\mathcal{D}_\text{quadratic}$ is rewritten
\begin{equation}
\label{Dquadratic}
\begin{split}
    \mathcal{D}_\text{quadratic} = \frac{e^2}{8\hbar^2 c^2} \epsilon^{icd} \epsilon^{lab} B^l \sum_{mns} f_{n} \text{Re}\Bigg[ 
    \int_{BZ} \frac{d\textbf{k}}{(2\pi)^3} \mathcal{W}^d_{nm} \mathcal{W}^b_{ms} \Big( \hbar\partial_a v^c_{sn} + \hbar\partial_c v^a_{sn} - \frac{2\hbar^2}{m} \delta_{ac} \delta_{sn} 
    \\
    - i \partial_a (E_{s\textbf{k}}-E_{n\textbf{k}}) \xi^c_{sn} -i\partial_c (E_{s\textbf{k}}-E_{n\textbf{k}})\xi^a_{sn} -i(E_{s\textbf{k}}-E_{n\textbf{k}}) \Big(\partial_a \xi^c_{sn} + \partial_c \xi^a_{sn} \Big) 
    \Big)
    \\
    = \frac{e^2}{8\hbar^2 c^2} \epsilon^{icd} \epsilon^{lab} B^l \sum_{mn} f_{n} \int_{BZ} \frac{d\textbf{k}}{(2\pi)^3} \text{Re}\Bigg[
    \mathcal{W}^d_{nm}\mathcal{W}^b_{ms} \Big( (E_{n\textbf{k}}-2E_{l\textbf{k}}+E_{s\textbf{k}} )\Big(\xi^c_{sl}\xi^a_{ln} 
    +\xi^a_{sl}\xi^c_{ln} \Big)
    \\
    + 2i \partial_a (E_{n\textbf{k}}-E_{s\textbf{k}}) \xi^c_{sn} 
    + 2i\partial_c (E_{n\textbf{k}}-E_{s\textbf{k}})\xi^a_{sn} 
    +i(E_{n\textbf{k}}-E_{s\textbf{k}}) \Big(\partial_a \xi^c_{sn} + \partial_c \xi^a_{sn} \Big)
    \Big)
    \Bigg].
\end{split}
\end{equation}
Finally, we have
\begin{equation}
   \mathcal{C}_\text{quadratic}= - \frac{e^2}{2m\hbar c^2} B^l \sum_{mn} f_{n} \text{Re}\Bigg[ 
    \int_{BZ} \frac{d\textbf{k}}{(2\pi)^3} \epsilon^{lab} S^i_{nm} \partial_b \mathcal{W}^a_{mn} + \epsilon^{icd} S^l_{nm} \partial_d \mathcal{W}^c_{mn}
    \Bigg] 
\end{equation}

Combining the spin contributions we find
\begin{equation}  \mathcal{C}_{\text{linear}}+\mathcal{C}_{\text{quadratic}}=0,
\end{equation}
and so we can write 
\begin{equation}
\label{QuadGD}
\text{Quadratic}_\text{GD} =\mathcal{A}'_{\text{quadratic}} + \mathcal{E}_{\text{quadratic}},    
\end{equation}
where 
\begin{equation}
    \mathcal{A}'_{\text{quadratic}} = \mathcal{A}_{\text{quadratic}}
    + \mathcal{B}_{\text{quadratic}}
    + \mathcal{D}_{\text{quadratic}}.
\end{equation}
We find  
\begin{equation}
\begin{split}
    \mathcal{A}'_\text{quadratic}= \frac{e^2}{4\hbar^2 c^2} \epsilon^{icd} \epsilon^{lab} B^l \sum_{mn} f_{n} \int_{BZ} \frac{d\textbf{k}}{(2\pi)^3} \text{Re}\Bigg[
    i\partial_a (E_{n\textbf{k}}+E_{m\textbf{k}})\xi^b_{ms} \mathcal{W}^c_{sn}\mathcal{W}^d_{nm} 
    - i\partial_d (E_{n\textbf{k}}+E_{m\textbf{k}}) \mathcal{W}^a_{ln} \mathcal{W}^b_{nm} \xi^c_{ml}
    \\
    +i\partial_d (E_{m\textbf{k}}+E_{n\textbf{k}}+2E_{l\textbf{k}}) \xi^b_{ml}\mathcal{W}^a_{ln}\mathcal{W}^c_{nm} 
    +i \partial_a (E_{m\textbf{k}}+E_{l\textbf{k}}+2E_{n\textbf{k}}) \xi^d_{nm}\mathcal{W}^b_{ml}\mathcal{W}^c_{ln}
    \Bigg].
\end{split}
\end{equation}
We then integrate by parts and find terms depending on the band energies and quadratically on $\mathcal{W}$, and terms that depend cubically on $\mathcal{W}$,
\begin{equation}
\label{Aprimequad}
\begin{split}
    \mathcal{A}'_\text{quadratic} = -\frac{e^2}{4\hbar^2 c^2} \epsilon^{icd}\epsilon^{lab} B^l \sum_{mn} f_{n} \int_{BZ} \frac{d\textbf{k}}{(2\pi)^3} \text{Re}\Bigg[ i(E_{n\textbf{k}}+E_{m\textbf{k}}) \partial_a \xi^b_{ms} \mathcal{W}^c_{sn} \mathcal{W}^d_{nm}
    -i(E_{n\textbf{k}}+E_{m\textbf{k}}) \partial_d \xi^c_{ml} \mathcal{W}^a_{ln}\mathcal{W}^b_{nm}
    \\
    +i(E_{m\textbf{k}}+E_{n\textbf{k}}+2E_{l\textbf{k}}) \partial_d \xi^b_{ml} \mathcal{W}^a_{ln} \mathcal{W}^c_{nm} +i(E_{m\textbf{k}}+E_{l\textbf{k}}+2E_{n\textbf{k}}) \partial_a \xi^d_{nm}\mathcal{W}^b_{ml}\mathcal{W}^c_{ln} 
    \\
    +i(E_{n\textbf{k}}+E_{m\textbf{k}}) \xi^b_{ms} \Big( \partial_a \mathcal{W}^c_{sn} \mathcal{W}^d_{nm} +\mathcal{W}^c_{sn} \partial_a \mathcal{W}^d_{nm} \Big)
    -i(E_{n\textbf{k}}+E_{m\textbf{k}}) \xi^c_{ml} \Big(\partial_d \mathcal{W}^a_{ln}\mathcal{W}^b_{nm}
    +\mathcal{W}^a_{ln}\partial_d \mathcal{W}^b_{nm} \Big)
    \\
    +i(E_{m\textbf{k}}+E_{n\textbf{k}}+2E_{l\textbf{k}}) \xi^b_{ml} \Big(\partial_d \mathcal{W}^a_{ln} \mathcal{W}^c_{nm}
    + \mathcal{W}^a_{ln} \partial_d \mathcal{W}^c_{nm}\Big)
    +i(E_{m\textbf{k}}+E_{l\textbf{k}}+2E_{n\textbf{k}}) \xi^d_{nm}\Big(\partial_a \mathcal{W}^b_{ml}\mathcal{W}^c_{ln} 
    + \mathcal{W}^b_{ml}\partial_a\mathcal{W}^c_{ln} \Big)
    \Bigg].
\end{split}
\end{equation}
Using this in (\ref{QuadGD}) we find

\begin{equation}
\begin{split}
    \text{Quadratic}_\text{GD} 
    = \epsilon^{icd} \epsilon^{lab} \frac{e^2}{8\hbar c^2} B^l \sum_{mn} f_{n} \int_{BZ} \frac{d\textbf{k}}{(2\pi)^3} \text{Re}\Bigg[ 
    (E_{n\textbf{k}}-E_{m\textbf{k}})\mathcal{W}^a_{ns} \mathcal{W}^b_{sm} \xi^d_{ml}\xi^c_{ln} 
    +(E_{n\textbf{k}}-E_{m\textbf{k}}) \xi^a_{nl} 
    \xi^b_{lm} \mathcal{W}^c_{ms} \mathcal{W}^d_{sn} 
    \\
     +i(3E_{l\textbf{k}}+2E_{n\textbf{k}}+3E_{m\textbf{k}}) \Big(\partial_a \xi^c_{ml}-\partial_c\xi^a_{ml}\Big)\mathcal{W}^b_{ln}\mathcal{W}^d_{nm}
     +i(2E_{m\textbf{k}}+3E_{n\textbf{k}}+3E_{s\textbf{k}}) \xi^c_{nm}\xi^a_{ms} \Big(\partial_d \mathcal{W}^b_{sn} - \partial_b \mathcal{W}^d_{sn}\Big) 
     \\
     +(2E_{l\textbf{k}}-2E_{m\textbf{k}} )\xi^c_{nm}\xi^a_{ms} \mathcal{W}^d_{sl}\mathcal{W}^b_{ln}
     +(2E_{m\textbf{k}}-2E_{s\textbf{k}}) \xi^a_{nm}\xi^c_{ml}\mathcal{W}^d_{ls}\mathcal{W}^b_{sn}
    \Bigg]
    + \mathcal{E}_\text{cubic},
\end{split}
\end{equation}
where 
\begin{equation}
\label{Ecubic}
\begin{split}
    \mathcal{E}_\text{cubic} = -\frac{e^2}{4\hbar^2 c^2} \epsilon^{icd}\epsilon^{lab} B^l \sum_{mn} f_{n} \int_{BZ} \frac{d\textbf{k}}{(2\pi)^3} \text{Re}\Bigg[ 
    i(E_{n\textbf{k}}+E_{m\textbf{k}}) \xi^b_{ms} \Big( \partial_a \mathcal{W}^c_{sn} \mathcal{W}^d_{nm} +\mathcal{W}^c_{sn} \partial_a \mathcal{W}^d_{nm} \Big)
    \\
    -i(E_{n\textbf{k}}+E_{m\textbf{k}}) \xi^c_{ml} \Big(\partial_d \mathcal{W}^a_{ln}\mathcal{W}^b_{nm}
    +\mathcal{W}^a_{ln}\partial_d \mathcal{W}^b_{nm} \Big)
    +i(E_{m\textbf{k}}+E_{n\textbf{k}}+2E_{l\textbf{k}}) \xi^b_{ml} \Big(\partial_d \mathcal{W}^a_{ln} \mathcal{W}^c_{nm}
    + \mathcal{W}^a_{ln} \partial_d \mathcal{W}^c_{nm}\Big)
    \\+i(E_{m\textbf{k}}+E_{l\textbf{k}}+2E_{n\textbf{k}}) \xi^d_{nm}\Big(\partial_a \mathcal{W}^b_{ml}\mathcal{W}^c_{ln} 
    + \mathcal{W}^b_{ml}\partial_a\mathcal{W}^c_{ln} \Big)
    \Bigg]
\end{split}
\end{equation}
We then use equation (\ref{Dconnection}) and
\begin{equation}
\label{DW}
\begin{split}
    i\partial_d \mathcal{W}^b_{ns} - i\partial_b \mathcal{W}^d_{ns} = \mathcal{W}^d_{nl}\mathcal{W}^b_{ls} - \mathcal{W}^b_{nl}\mathcal{W}^d_{ls}
\end{split}
\end{equation}
to rewrite the remaining quadratic gauge-dependence terms and show that they vanish, leaving the new cubic dependence introduced,
\begin{equation}
\begin{split}
\text{Quadratic}_\text{GD}
     = \epsilon^{icd} \epsilon^{lab} \frac{e^2}{8\hbar c^2} B^l \sum_{mn} f_{n} \int_{BZ} \frac{d\textbf{k}}{(2\pi)^3} \text{Re}\Bigg[ 
     (3E_{l\textbf{k}}+2E_{n\textbf{k}}+3E_{m\textbf{k}}) \Big(\xi^c_{ms}\xi^a_{sl}-\xi^a_{ms}\xi^c_{sl}\Big)\mathcal{W}^b_{ln}\mathcal{W}^d_{nm}
     \\
     +(2E_{m\textbf{k}}+3E_{n\textbf{k}}+3E_{s\textbf{k}}) \xi^c_{nm}\xi^a_{ms} \Big(\mathcal{W}^d_{sl} \mathcal{W}^b_{ln} - \mathcal{W}^b_{sl} \mathcal{W}^d_{ln}\Big) 
     \\
     +(2E_{l\textbf{k}}-2E_{m\textbf{k}} )\xi^c_{nm}\xi^a_{ms} \mathcal{W}^d_{sl}\mathcal{W}^b_{ln}
     +(2E_{m\textbf{k}}-2E_{s\textbf{k}}) \xi^a_{nm}\xi^c_{ml}\mathcal{W}^d_{ls}\mathcal{W}^b_{sn}
    \Bigg] + \mathcal{E}_\text{cubic}
    \\
    = \mathcal{E}_\text{cubic}.
\end{split}
\end{equation}
To get to the last line the real part is made explicit and band indices are relabelled to show the integrand is symmetric in either interchanges of $c$ and $d$, or $a$ and $b$. 
So only the term that has a cubic gauge-dependence remains,
\begin{equation}
  \text{Quadratic}_\text{GD} \rightarrow \mathcal{E}_{\text{cubic}}.  
\end{equation}

\subsection{Cubic Gauge-Dependence}
We will now combine the cubic gauge-dependent terms generated above with the cubic gauge-dependent terms that arise directly from the $X^{il}_{\text{dyn/comp}}(\mathcal{W})$. In fact, we only have cubic gauge-dependent terms in $X^{il}_{\text{comp}}(\mathcal{W})$, and they are

\begin{equation}
\begin{split}
     -\frac{e^2}{4\hbar c^2} \epsilon^{icd} \epsilon^{lab} B^l \sum_{mn} f_{n} \int_{BZ} \frac{d\textbf{k}}{(2\pi)^3} \text{Re}\Bigg[ \partial_d E_{m\textbf{k}}
    \partial_b \mathcal{W}^a_{nm} \mathcal{W}^c_{mn}
    +i(E_{m\textbf{k}}-E_{l\textbf{k}})\partial_b \mathcal{W}^a_{nm} \xi^d_{ml} \mathcal{W}^c_{ln}
    \Bigg]
    \\
    +\frac{e^2}{4\hbar^2 c^2} \epsilon^{icd} \epsilon^{lab} \sum_{mns} f_{n} 
    \int_{BZ} \frac{d\textbf{k}}{(2\pi)^3}
    \text{Re}\Bigg[ 
    \mathcal{W}^d_{nm}\mathcal{W}^b_{ms}\Big( i\partial_a (E_{n\textbf{k}}-E_{s\textbf{k}}) \mathcal{W}^c_{sn} + i\partial_c (E_{n\textbf{k}}-E_{s\textbf{k}}) \mathcal{W}^a_{sn} \Big)
    \\
    + \frac{1}{2}\Big(\xi^d_{nm}\mathcal{W}^b_{ms} + \mathcal{W}^d_{nm}\xi^b_{ms}\Big)\Bigg(i(E_{n\textbf{k}}-E_{s\textbf{k}}) (\partial_a \mathcal{W}^c_{sn} + \partial_c \mathcal{W}^a_{sn}) + (2E_{l\textbf{k}}-E_{s\textbf{k}}-E_{n\textbf{k}})(\mathcal{W}^a_{sl}\mathcal{W}^c_{ln}+\mathcal{W}^c_{sl}\mathcal{W}^a_{ln} ) \Bigg)
    \\
    + \partial_c E_{n\textbf{k}} \mathcal{W}^d_{ns} \partial_b \mathcal{W}^a_{sn} 
    + i(E_{n\textbf{k}}-E_{m\textbf{k}})\mathcal{W}^d_{ns}\Big(i\mathcal{W}^a_{sl}\xi^b_{lm} - i\xi^b_{sl}\mathcal{W}^a_{lm}\Big) \mathcal{W}^c_{mn}
    +i(E_{n\textbf{k}}-E_{m\textbf{k}}) \mathcal{W}^c_{mn} \xi^d_{ns} \partial_b \mathcal{W}^a_{sm}
    \Bigg]
    \\
    -\frac{e^2}{8\hbar^2 c^2} \epsilon^{icd} \epsilon^{lab} B^l \sum_{mns} f_{n}  \int_{BZ} \frac{d\textbf{k}}{(2\pi)^3} \text{Re}\Bigg[ 
    \partial_d \mathcal{W}^c_{nm} \Big( 2\partial_b (E_{m\textbf{k}}+E_{n\textbf{k}}) \mathcal{W}^a_{mn}
    +i (E_{m\textbf{k}}-E_{n\textbf{k}})\mathcal{W}^a_{ms} \xi^b_{sn} 
    +i (E_{m\textbf{k}}-E_{n\textbf{k}}) \xi^b_{ms}\mathcal{W}^a_{sn}
    \\
    + (2E_{s\textbf{k}}-E_{m\textbf{k}}-E_{n\textbf{k}})\Big(\mathcal{W}^c_{nl}\xi^d_{lm} - \xi^d_{nl}\mathcal{W}^c_{lm}\Big) \mathcal{W}^a_{ms}\mathcal{W}^b_{sn} \Bigg].
\end{split}
\end{equation}
These can be divided into terms that involve partial derivatives of band energies, the sum of which we call $\mathcal{A}_\text{cubic}$, and terms that involve only band energy dependence, the sum of which we call $\mathcal{B}_\text{cubic}$. Assembling the terms, we find 
\begin{equation}
\begin{split}
    \mathcal{A}_\text{cubic} = -\frac{e^2}{4\hbar^2 c^2} \epsilon^{icd}\epsilon^{lab} B^l \sum_{mn} f_{n} \int_{BZ} \frac{d\textbf{k}}{(2\pi)^3} \text{Re}\Bigg[ i\partial_b (E_{m\textbf{k}}+E_{l\textbf{k}})\mathcal{W}^c_{nl} \mathcal{W}^d_{lm}\mathcal{W}^a_{mn}
    +i\partial_d(E_{m\textbf{k}}+E_{s\textbf{k}})  \mathcal{W}^a_{ms}\mathcal{W}^b_{sn}\mathcal{W}^c_{nm}  
    \Bigg]. 
\end{split}
\end{equation}
Integrating by parts, we get 
\begin{equation}
\label{Acubic}
\begin{split}
    \mathcal{A}_\text{cubic} = \frac{e^2}{4\hbar^2 c^2} \epsilon^{icd}\epsilon^{lab} B^l \sum_{mn} f_{n} \int_{BZ} \frac{d\textbf{k}}{(2\pi)^3} \text{Re}\Bigg[
    i(E_{m\textbf{k}}+E_{l\textbf{k}})  \Big( 
    \partial_b \mathcal{W}^c_{nl} \mathcal{W}^d_{lm} \mathcal{W}^a_{mn}
    +\mathcal{W}^c_{nl}\partial_b  \mathcal{W}^d_{lm} \mathcal{W}^a_{mn}
    +\mathcal{W}^c_{nl} \mathcal{W}^d_{lm} \partial_b \mathcal{W}^a_{mn}
    \\
    +\partial_d \mathcal{W}^a_{ml}\mathcal{W}^b_{ln} \mathcal{W}^c_{nm}
    +\mathcal{W}^a_{ml}\partial_d \mathcal{W}^b_{ln} \mathcal{W}^c_{nm}
    +\mathcal{W}^a_{ml}\mathcal{W}^b_{ln}\partial_d \mathcal{W}^c_{nm}
    \Big)
    \Bigg].
\end{split}
\end{equation}
This has now been converted to a term with quartic gauge-dependence, so we put it aside until the next section.

The terms that involve dependence on the band energies sum to
\begin{equation}
\begin{split}
    \mathcal{B}_\text{cubic} = \frac{e^2}{8\hbar^2 c^2} \epsilon^{icd}\epsilon^{lab} B^l \sum_{mn} f_{n} \int_{BZ} \frac{d\textbf{k}}{(2\pi)^3} \text{Re}\Bigg[ 
    (3E_{s\textbf{k}}-2E_{n\textbf{k}}-E_{l\textbf{k}})\xi^d_{ml} \mathcal{W}^c_{ln}\mathcal{W}^a_{ns}\mathcal{W}^b_{sm}
    +(2E_{m\textbf{k}}-3E_{s\textbf{k}}+E_{l\textbf{k}})\mathcal{W}^c_{ns}\mathcal{W}^d_{sm}\mathcal{W}^a_{ml}\xi^b_{ln} 
    \\
    + i(E_{n\textbf{k}}-E_{s\textbf{k}}) \xi^d_{ms}\Big(\partial_a \mathcal{W}^c_{sn} + \partial_c \mathcal{W}^a_{sn}\Big)\mathcal{W}^b_{nm}  
    + i(E_{n\textbf{k}}-E_{s\textbf{k}})\xi^b_{ms}\Big(\partial_a \mathcal{W}^c_{sn} + \partial_c \mathcal{W}^a_{sn}\Big)\mathcal{W}^d_{nm}
    \\
    + (2E_{l\textbf{k}}-E_{s\textbf{k}}-E_{n\textbf{k}})\xi^d_{nm} \mathcal{W}^b_{ms}\mathcal{W}^c_{sl}\mathcal{W}^a_{ln} +(2E_{l\textbf{k}}-E_{s\textbf{k}}-E_{n\textbf{k}}) \mathcal{W}^c_{sl}\mathcal{W}^a_{ln} \mathcal{W}^d_{nm}\xi^b_{ms}
    \Bigg]
\end{split}
\end{equation}
and we add these to the earlier term (\ref{Ecubic}) exhibiting cubic dependence that resulted from processing the terms that involved quadratic gauge-dependence.  First, however, we manipulate that earlier term (\ref{Ecubic}) by using equation (\ref{DW}) and the fact that it is under a real part operation, to get 

\begin{equation}
\begin{split}
    \mathcal{E}_\text{cubic} = -\frac{e^2}{4\hbar^2 c^2} \epsilon^{icd}\epsilon^{lab} B^l \sum_{mn} f_{n} \int_{BZ} \frac{d\textbf{k}}{(2\pi)^3} \text{Re}\Bigg[ 
     i(2E_{n\textbf{k}}+E_{m\textbf{k}}+E_{s\textbf{k}}) \xi^b_{ms}\partial_a \mathcal{W}^c_{sn} \mathcal{W}^d_{nm}
    \\
    +i(2E_{n\textbf{k}}+E_{m\textbf{k}}+E_{s\textbf{k}})  \xi^d_{ms}\partial_c \mathcal{W}^a_{sn}\mathcal{W}^b_{nm}
    \\
     -i(E_{m\textbf{k}}+E_{n\textbf{k}}+2E_{s\textbf{k}}) \xi^b_{ms}\partial_c \mathcal{W}^a_{sn} \mathcal{W}^d_{nm}
    -(E_{m\textbf{k}}+E_{n\textbf{k}}+2E_{l\textbf{k}}) \xi^b_{ml}\mathcal{W}^a_{ln}  \mathcal{W}^c_{ns}\mathcal{W}^d_{sm}
    \\
    -(E_{m\textbf{k}}+E_{l\textbf{k}}+2E_{n\textbf{k}}) \xi^d_{nm}\mathcal{W}^b_{ms}\mathcal{W}^a_{sl}\mathcal{W}^c_{ln} 
    -i(E_{m\textbf{k}}+E_{n\textbf{k}}+2E_{s\textbf{k}}) \xi^d_{ms}\partial_a\mathcal{W}^c_{sn}\mathcal{W}^b_{nm}
    \Bigg].
\end{split}
\end{equation}
The combination of all remaining cubic gauge-dependent terms then gives
\begin{equation}
\begin{split}
    \mathcal{B}_\text{cubic} + \mathcal{E}_\text{cubic} = \frac{e^2}{8\hbar^2 c^2} \epsilon^{icd}\epsilon^{lab} B^l \sum_{mn} f_{n} \int_{BZ} \frac{d\textbf{k}}{(2\pi)^3} \text{Re}\Bigg[ i(2E_{m\textbf{k}}+3E_{n\textbf{k}}+3E_{s\textbf{k}}) \xi^d_{ms}\Big(\partial_a\mathcal{W}^c_{sn}-\partial_c \mathcal{W}^a_{sn}\Big)\mathcal{W}^b_{nm}
    \\
    +i(2E_{m\textbf{k}}+3E_{n\textbf{k}}+3E_{s\textbf{k}}) \xi^b_{ms}\Big(\partial_c \mathcal{W}^a_{sn} - \partial_a \mathcal{W}^c_{sn}\Big) \mathcal{W}^d_{nm}
    \\
    +(2E_{m\textbf{k}}+3E_{l\textbf{k}}+3E_{s\textbf{k}}) \xi^d_{ml}\mathcal{W}^c_{ln}\mathcal{W}^a_{ns}\mathcal{W}^b_{sm} 
    +(2E_{m\textbf{k}}+3E_{l\textbf{k}}+3E_{s\textbf{k}}) \xi^b_{ml}\mathcal{W}^a_{ln}  \mathcal{W}^c_{ns}\mathcal{W}^d_{sm}
    \\
    +(2E_{l\textbf{k}}-E_{s\textbf{k}}-E_{n\textbf{k}})\xi^d_{nm} \mathcal{W}^b_{ms}\mathcal{W}^c_{sl}\mathcal{W}^a_{ln}
    +(2E_{l\textbf{k}}-E_{s\textbf{k}}-E_{n\textbf{k}}) \mathcal{W}^c_{sl}\mathcal{W}^a_{ln} \mathcal{W}^d_{nm}\xi^b_{ms} \hspace{3pt}
    \Bigg]
\end{split}
\end{equation}
Then using equation (\ref{DW}) again
\begin{equation}
\begin{split}
     \mathcal{B}_\text{cubic} + \mathcal{E}_\text{cubic} =
    \frac{e^2}{8\hbar^2 c^2} \epsilon^{icd}\epsilon^{lab} B^l \sum_{mn} f_{n} \int_{BZ} \frac{d\textbf{k}}{(2\pi)^3} \text{Re}\Bigg[ (2E_{m\textbf{k}}+2E_{n\textbf{k}}+2E_{s\textbf{k}}+2E_{l\textbf{k}})\Big( \xi^d_{ms}\mathcal{W}^a_{sl}\mathcal{W}^c_{ln}\mathcal{W}^b_{nm}
    +\xi^b_{ms}\mathcal{W}^c_{sl} \mathcal{W}^a_{ln}\mathcal{W}^d_{nm}\Big)
    \Bigg],
\end{split}
\end{equation}
and explicitly evaluating the real part operation, and relabeling the free band indices, we find
\begin{equation}
    \mathcal{B}_\text{cubic} + \mathcal{E}_\text{cubic} = 0.
\end{equation}
Thus only quartic gauge dependence remains.  

\subsection{Quartic Gauge-Dependence}

The quartic gauge-dependence only comes from the itinerant compositional contribution

\begin{equation}
\begin{split}
    \mathcal{B}_\text{quartic} = \frac{e^2}{8\hbar^2 c^2} \epsilon^{icd} \epsilon^{lab} B^l \sum_{mns} f_{n} \text{Re}\Bigg[ 
    \int_{BZ} \frac{d\textbf{k}}{(2\pi)^3} i(E_{n\textbf{k}}-E_{s\textbf{k}}) \mathcal{W}^d_{nm}\mathcal{W}^b_{ms}\Big(\partial_a \mathcal{W}^c_{sn} + \partial_c \mathcal{W}^a_{sn} \Big)
    \\
    + \Big(2E_{l\textbf{k}}-E_{s\textbf{k}}-E_{n\textbf{k}}\Big)\mathcal{W}^d_{nm}\mathcal{W}^b_{ms} \mathcal{W}^c_{sl} \mathcal{W}^a_{ln}
    -(E_{s\textbf{k}}+E_{l\textbf{k}}-2E_{n\textbf{k}})\mathcal{W}^c_{ns}\mathcal{W}^d_{sm}  \mathcal{W}^a_{ml}\mathcal{W}^b_{ln} 
    \Bigg],
\end{split}
\end{equation}

and the integrated by parts terms manipulated
\begin{equation}
\begin{split}
    \mathcal{A}_\text{cubic} = \frac{e^2}{8\hbar^2 c^2} \epsilon^{icd}\epsilon^{lab} B^l \sum_{mn} f_{n} \int_{BZ} \frac{d\textbf{k}}{(2\pi)^3} \text{Re}\Bigg[
    i(4E_{s\textbf{k}}+2E_{n\textbf{k}}+2E_{m\textbf{k}})\mathcal{W}^d_{nm}\mathcal{W}^b_{ms} \partial_c \mathcal{W}^a_{sn}
    \\
    - i(2E_{m\textbf{k}}+2E_{s\textbf{k}}+4E_{n\textbf{k}})\mathcal{W}^d_{nm} \mathcal{W}^b_{ms}\partial_a \mathcal{W}^c_{sn}
    \\
    -(4E_{n\textbf{k}}+2E_{l\textbf{k}}+2E_{s\textbf{k}})\mathcal{W}^c_{ns}\mathcal{W}^d_{sm}\mathcal{W}^a_{ml}\mathcal{W}^b_{ln}
    \Bigg].
\end{split}
\end{equation}
Combining all the quartic dependence we get:

\begin{equation}
\begin{split}
    \mathcal{A}_\text{cubic} + \mathcal{B}_\text{quartic} = \frac{e^2}{8\hbar^2 c^2} \epsilon^{icd}\epsilon^{lab} B^l \sum_{mn} f_{n} \int_{BZ} \frac{d\textbf{k}}{(2\pi)^3} \text{Re}\Bigg[ 
    i(3E_{s\textbf{k}}+3E_{n\textbf{k}}+2E_{m\textbf{k}}) \mathcal{W}^d_{nm}\mathcal{W}^b_{ms} \Big(\partial_c \mathcal{W}^a_{sn} - \partial_a \mathcal{W}^c_{sn}\Big)
    \\
    +(2E_{l\textbf{k}}-E_{s\textbf{k}}-E_{n\textbf{k}}) \mathcal{W}^d_{nm}\mathcal{W}^b_{ms}\mathcal{W}^c_{sl}\mathcal{W}^a_{ln}
    -(2E_{n\textbf{k}}+3E_{l\textbf{k}}+3E_{s\textbf{k}}) \mathcal{W}^c_{ns}\mathcal{W}^d_{sm} \mathcal{W}^a_{ml}\mathcal{W}^b_{ln}
    \Bigg]
\end{split}
\end{equation}
We then use the identity (\ref{DW}) to find
\begin{equation}
\begin{split}
    \mathcal{A}_\text{cubic} + \mathcal{B}_\text{quartic} = \frac{e^2}{8\hbar^2 c^2} \epsilon^{icd}\epsilon^{lab} B^l \sum_{mn} f_{n} \int_{BZ} \frac{d\textbf{k}}{(2\pi)^3} \text{Re}\Bigg[ (2E_{s\textbf{k}}+2E_{n\textbf{k}}+2E_{m\textbf{k}}+2E_{l\textbf{k}}) \mathcal{W}^d_{nm}\mathcal{W}^b_{ms}\mathcal{W}^c_{sl}\mathcal{W}^a_{ln} 
    \Bigg]
    \\
    = 0
\end{split}
\end{equation}
To see that this is zero we make the real part operation explicit and perform a relabelling of the band indices and find the result is symmetric in the Cartesian indices $a$ and $b$. Thus the quartic gauge-dependence is shown to vanish.  

\section{Discussion on Gauge-Transformations}

There are many different ways to partition the total magnetic susceptibility, and this in fact leads to some of the confusion in the literature on the correct theoretical expression \cite{PreprintMagSus}. It is only the total susceptibility, as we have shown in the preceding sections, that is gauge-invariant. If one were to evaluate one of the purported contributions individually the result may depend on the choice of Wannier/Bloch functions used in the computation. In this section we will examine a specific example of a gauge-transformation that can be applied to the cell periodic Bloch functions to show how an example individual contribution to the magnetic susceptibility would change. In addition, the choice will demonstrate how the freedom in choosing an arbitrary complex phase to multiply the Bloch functions is a special case of the larger multiband gauge-freedom in choosing the Wannier functions, which is tracked by the unitary matrix U(\textbf{k}). 

Referring to sections V and VI from \cite{PreprintMagSus}, there gauge-dependence is separated into the dynamical and compositional contributions $X^{il}_\text{dyn}$, and $X^{il}_\text{comp}$. As has been shown in the preceding sections in this work the gauge-dependence cancels, so 

\begin{equation}
\label{X}
    X^{il}_\text{dyn}(\mathcal{W}) = -X^{il}_\text{comp}(\mathcal{W})
\end{equation}
It is easier to work with the dynamical contribution so we will focus on that, with the knowledge that equation (\ref{X}) holds for the set of $\mathcal{W}'s$ that can be integrated by parts and obey $f_{nm} \mathcal{W}^a_{nm} = 0$.  

The dynamical contribution can be split into the Van Vleck paramagnetism and the part that depends on the $\mathcal{W}$'s \cite{PreprintMagSus}, 
\begin{equation}
\begin{split}
    \chi^{il}_\text{dyn} = \chi^{il}_\text{VV} + X^{il}_\text{dyn}(\mathcal{W}),
\end{split}
\end{equation}
where 
\begin{equation}
    \chi^{il}_\text{VV} = \sum_{mn} f_{nm} \int_{BZ} \frac{d\textbf{k}}{(2\pi)^3} \frac{ M^i_{nm} M^l_{mn} }{\Delta_{mn}(\textbf{k})},
\end{equation}
and
\begin{equation}
\label{Xdyn}
\begin{split}
     X^{il}_\text{dyn}(\mathcal{W}) = -\frac{ie}{4\hbar c}  \sum_{mn} f_{nm} \int_{BZ} \frac{d\textbf{k}}{(2\pi)^3} \epsilon^{icd} \Bigg( \mathcal{W}^c_{nl}\xi^d_{lm} + \xi^d_{nl} \mathcal{W}^c_{lm} \Bigg) M^l_{mn} 
    \\
    +\frac{ie}{4\hbar c} \epsilon^{lab} \sum_{mn} f_{nm} \int_{BZ} \frac{d\textbf{k}}{(2\pi)^3} M^i_{nm}\Big( \mathcal{W}^a_{ms}\xi^b_{sn} + \xi^b_{ms}\mathcal{W}^a_{sn}\Big) 
    \\
    -\epsilon^{icd}\epsilon^{lab} \frac{e^2}{8\hbar^2 c^2} \sum_{mn} f_{nm} \int_{BZ} \frac{d\textbf{k}}{(2\pi)^3} \text{Re}\Bigg[ (E_{n\textbf{k}}-E_{m\textbf{k}}) \mathcal{W}^c_{nl} \xi^d_{lm} \mathcal{W}^a_{ms}\xi^b_{sn} + (E_{n\textbf{k}}-E_{m\textbf{k}}) \mathcal{W}^c_{nl} \xi^d_{lm} \xi^b_{ms} \mathcal{W}^a_{sn} \Bigg].
\end{split}
\end{equation}

If one were to perform a computation based on a band structure calculation it is only $\chi^{il}_\text{VV}$ that one would use. However, since the phases of those Bloch functions are arbitrary two computations may produce different results. The cell periodic part of the Bloch functions in the two hypothetical computations are related to each other by

\begin{equation}
\begin{split}
    u'_{n\textbf{k}}(\textbf{x}) = e^{-i\phi_{n}(\textbf{k})} u_{n\textbf{k}}(\textbf{x}) .
\end{split}
\end{equation}
This alters the Berry connection non-trivially, 

\begin{equation}
    \xi'^i_{nm} = e^{i(\phi_{n}(\textbf{k})-\phi_{m}(\textbf{k}))} \Big( \xi^i_{nm}(\textbf{k}) + \partial_i \phi_n(\textbf{k}) \delta_{nm} \Big).
\end{equation}

Thus, the two evaluations of the Van Vleck paramagnetism would be related by 
\begin{equation}
\begin{split}
    \chi'^{il}_\text{VV} = \chi^{il}_\text{VV} - \frac{ie}{4\hbar c} \epsilon^{lcd} \sum_{nm} f_{nm} \int_{BZ} \frac{d\textbf{k}}{(2\pi)^3} M^i_{nm} \xi^c_{mn} \partial_d\Big(\phi_{n}+\phi_{m}\Big) 
    \\
    + \frac{ie}{4\hbar c} \epsilon^{iab} \sum_{nm} f_{nm} \int_{BZ} \frac{d\textbf{k}}{(2\pi)^3} \xi^a_{nm} \partial_b \Big(\phi_{n}+\phi_{m}\Big) M^l_{mn}
    \\
    - \frac{e^2}{16\hbar^2c^2} \epsilon^{iab}\epsilon^{lcd} \sum_{mn} f_{nm} \int_{BZ} \frac{d\textbf{k}}{(2\pi)^3} (E_{n\textbf{k}}-E_{m\textbf{k}}) \xi^a_{nm}\xi^c_{mn} \partial_b (\phi_{n} + \phi_{m} )\partial_d (\phi_{n} + \phi_{m} )
\end{split}
\end{equation}

So far the choice of the matrix elements $U_{n\alpha}(\textbf{k})$ has been restricted to those transformations that connect the cell periodic Bloch functions to cell periodic functions associated with exponentially localized Wannier functions, however we note that a $U(\textbf{k})$ can be chosen that relates the two Bloch functions above, it is $U'_{nn'}(\textbf{k}) = e^{-i\phi_{n}(\textbf{k})} \delta_{nn'}$. The form of the $\mathcal{W}'$'s is then $\mathcal{W}'^a_{nn'} = \delta_{nn'} \partial_a \phi_{n}(\textbf{k})$. Using this $\mathcal{W}'$ in equation (\ref{Xdyn}) produces
\begin{equation}
\begin{split}
     X^{il}_\text{dyn}(\mathcal{W}') = \frac{ie}{4\hbar c} \epsilon^{iab}\sum_{mn} f_{nm} \int_{BZ} \frac{d\textbf{k}}{(2\pi)^3}  \partial_b( \phi_{n}+\phi_{m}) \xi^a_{nm} M^l_{mn} 
    \\
    -\frac{ie}{4\hbar c} \epsilon^{lcd} \sum_{mn} f_{nm} \int_{BZ} \frac{d\textbf{k}}{(2\pi)^3} \partial_d(\phi_{n} +  \phi_{m}) M^i_{nm} \xi^c_{mn} 
    \\
    -\epsilon^{icd}\epsilon^{lab} \frac{e^2}{16\hbar^2 c^2} \sum_{mn} f_{nm} \int_{BZ} \frac{d\textbf{k}}{(2\pi)^3} (E_{n\textbf{k}}-E_{m\textbf{k}}) \partial_d (\phi_{n}+\phi_{m}) \partial_b (\phi_{n} + \phi_{m}) \xi^a_{nm} \xi^c_{mn},
\end{split}
\end{equation}
In the last line the real part was made explicit and a relabelling of $n\leftrightarrow m$ was made, from which it is clear that
\begin{equation}
    \chi'^{il}_\text{VV} - \chi^{il}_\text{VV} = X^{il}_\text{dyn}(\mathcal{W}')
\end{equation}
This illustrates how $X^{il}_\text{dyn}(\mathcal{W})$ captures the ``gauge-dependence" of the dynamical susceptibility. Changing the Bloch functions or the Wannier functions has an associated $(U(\textbf{k}),\mathcal{W}(\textbf{k}))$, and how the evaluation of $\chi^{il}_\text{VV}$ is altered by such a choice is tracked by plugging the associated $\mathcal{W}$ into equation (\ref{Xdyn}). It can also be shown that performing successive transformations leads to a difference characterized by the sum of the $\mathcal{W}$'s that describe the successive transformations.

\section{Generalized effective mass tensor sum rule}\label{Appendix:GeneralSumRule} 

The inverse effective mass tensor sum rule is used at many key points in the preceding and subsequent derivations, so here we give a proof of the sum rule. There are two methods that we can use to determine the generalized effective mass tensor sum rule. 
The first method is to use \textbf{k}$\cdot$\textbf{p} perturbation theory. Here we begin with the unperturbed Hamiltonian
\begin{equation}
\begin{split}
    \mathcal{H}^{(0)}(\textbf{x}) = \frac{ \mathfrak{p}(\textbf{x})^2}{2m} + \text{V}(\textbf{x}) + \frac{\hbar}{4m^2c^2} \boldsymbol\sigma\cdot\nabla\text{V}(\textbf{x})\times\mathfrak{p}(\textbf{x)},
\end{split}
\end{equation}
where $\mathfrak{p}(\textbf{x})$ is defined as
\begin{equation}
\label{pscriptstatic}
 \mathfrak{p}(\textbf{x})=-i\hbar\nabla-\frac{e}{c}\textbf{A}_\text{static}(\textbf{x}),
\end{equation}
where $\textbf{A}_\text{static}(\textbf{x})$ is a static cell periodic vector potential used to break time reversal symmetry but not the periodicity of the Hamiltonian \cite{Haldane_Julen}. The eigenstates are Bloch functions
\begin{equation}
    \mathcal{H}^{(0)}(\textbf{x})\psi_{n\textbf{k}}(\textbf{x}) = E_{n\textbf{k}}\psi_{n\textbf{k}}(\textbf{x}).
\end{equation}
The cell periodic part of the Bloch functions obey the equation:
\begin{equation}
    H(\textbf{k}) u_{n\textbf{k}}(\textbf{x}) = E_{n\textbf{k}}u_{n\textbf{k}}(\textbf{x}),
\end{equation}
with
\begin{equation}
    H(\textbf{k}) = \frac{ (\mathfrak{p}(\textbf{x})+\hbar\textbf{k})^2}{2m} + \text{V}(\textbf{x}) + \frac{\hbar}{4m^2c^2} \boldsymbol\sigma\cdot\nabla\text{V}(\textbf{x}) \times(\mathfrak{p}+\hbar\textbf{k}).
\end{equation}
The velocity operator is
\begin{equation}
    \hat{v}^i(\textbf{k}) = \frac{1}{\hbar} \frac{\partial H(\textbf{k})}{\partial k^i} = \frac{(\mathfrak{p}^i(\textbf{x})+\hbar k^i)}{m} + \frac{\hbar}{4m^2c^2} \epsilon^{iab} \sigma^a \frac{\partial \text{V}(\textbf{x})}{\partial x^b}.
\end{equation}

We evaluate the Hamiltonian at a small shift from \textbf{k}$_0$, where we assume we have known eigenstates and eigenenergies, and shift a small amount $\delta\textbf{k}$, treating this as a perturbation: 
\begin{equation}
    H(\textbf{k}_0+\delta\textbf{k}) = H(\textbf{k}_0) + \hbar\delta\textbf{k} \cdot \hat{\textbf{v}}(\textbf{k}_0) + \frac{\hbar^2}{2m} \delta\textbf{k}^2.
\end{equation}
This perturbation leads to a first order in $\delta\textbf{k}$ correction to the states, as well from the Sternheimer equation there is the additional change of phase of the state $|u_{\textbf{n}\textbf{k}}\rangle$, so the linear response of the state is \cite{ModernTheoryVanderbilt}
\begin{equation}
\begin{split}
    |u^{(1)}_{n\textbf{k}_0+\delta\textbf{k}} \rangle = &-i\delta\textbf{k} \cdot \boldsymbol\xi_{nn}(\textbf{k}_0)|u_{n\textbf{k}_0}\rangle
    + \sum_{m\neq n} \frac{\hbar \delta\textbf{k} \cdot \textbf{v}_{mn}(\textbf{k}_0)}{E_{n\textbf{k}_0}-E_{m\textbf{k}_0}}|u_{m\textbf{k}_0}\rangle 
\end{split}
\end{equation}
The velocity operator itself is modified as $\textbf{k}_0\rightarrow \textbf{k}_0+\delta\textbf{k}$,
\begin{equation}
    \hat{\textbf{v}}(\textbf{k}_0+\delta\textbf{k}) = \hat{\textbf{v}}(\textbf{k}_0) + \hat{\textbf{v}}^{(1)}(\textbf{k}_0+\delta\textbf{k}),
\end{equation}
where the first order in $\delta\textbf{k}$ correction to the velocity operator is
\begin{equation}
    \hat{\textbf{v}}^{(1)}(\textbf{k}_0+\delta\textbf{k}) = \frac{\hbar}{m} \delta\textbf{k}.
\end{equation}
The first order correction to the velocity matrix elements -- indicated by a superscript $(1)$ -- is found by taking the velocity operator and the Bloch functions up to first order in $\delta\textbf{k}$ and discarding any higher order terms,

\begin{equation}
    v^i_{nm}(\textbf{k}_0+\delta\textbf{k}) = v^i_{nm}(\textbf{k}_0) +  v^{i(1)}_{nm}(\textbf{k}_0+\delta\textbf{k}) + O(\delta\textbf{k}^2),
\end{equation}
where
\begin{equation}
\label{vperturbation}
\begin{split}
    &v^{i(1)}_{nm}(\textbf{k}_0+\delta\textbf{k}) = \langle u_{n\textbf{k}_0} | v^{i,(1)}(\textbf{k}_0+\delta\textbf{k}) |u_{m\textbf{k}_0}\rangle 
    + \langle u^{(1)}_{n\textbf{k}_0+\delta\textbf{k}}| v^i(\textbf{k}_0)|u_{m\textbf{k}_0}\rangle + \langle u_{n\textbf{k}_0}| v^i(\textbf{k}_0)|u^{(1)}_{m\textbf{k}_0+\delta\textbf{k}}\rangle  
    \\
    &= \frac{\hbar}{m} \delta k^i \delta_{nm} + \hbar \delta k^j \sum_{l\neq n} \frac{ v^j_{nl}(\textbf{k}_0) v^i_{lm}(\textbf{k}_0)}{E_{n\textbf{k}_0}-E_{l\textbf{k}_0}} + \hbar \delta k^j \sum_{l\neq m} \frac{ v^i_{nl}(\textbf{k}_0)v^j_{lm}(\textbf{k}_0)}{E_{m\textbf{k}_0}-E_{l\textbf{k}_0}} 
    +\delta k^j\Big(i \xi^j_{nn}(\textbf{k}_0) v^i_{nm}(\textbf{k}_0)-iv^i_{nm}(\textbf{k}_0)\xi^j_{mm}(\textbf{k}_0) \Big)
    \\
    &= \frac{\hbar}{m} \delta k^i \delta_{nm} + i\delta k^j \sum_{l}\Big(\xi^j_{nl}v^i_{lm} - v^i_{nl}\xi^j_{lm}\Big).
\end{split}
\end{equation}

We can compare this first 
perturbative correction to the Taylor expansion of the velocity matrix elements,

\begin{equation}
\label{vtaylor}
    v^i_{nm}(\textbf{k}_0+\delta\textbf{k}) = v^i_{nm}(\textbf{k}_0) + \delta k^j \frac{\partial v^i_{nm}(\textbf{k})}{\partial k^j}\Big|_{\textbf{k}_0} + ...
\end{equation}
Comparing equations (\ref{vperturbation}) and (\ref{vtaylor}), since $\textbf{k}_0$ is arbitrary, we find an expression for the derivative of the velocity matrix elements,
\begin{equation}
\label{vsumrule1}
    \frac{\partial v^i_{nm}(\textbf{k})}{\partial k^j} = \frac{\hbar}{m} \delta_{ij}\delta_{nm} + i \sum_{l}\Big(\xi^j_{nl}v^i_{lm}-v^i_{nl}\xi^j_{lm}\Big).
\end{equation}
Only considering the diagonal elements ($n=m$) this is simply the effective mass tensor sum rule. Note that the sum `$l$' is over all states.

Instead we could employ 
a more direct approach, and obtain 
the same result, taking the derivative of the velocity matrix elements 
\begin{equation}
\begin{split}
    v^i_{nm}(\textbf{k}) = \frac{1}{\mathcal{V}_{uc}} \int_{\mathcal{V}_{uc}} d\textbf{x} u^\dag_{n\textbf{k}}(\textbf{x}) \hat{v}^i(\textbf{k}) u_{m\textbf{k}}(\textbf{x}),
\end{split}
\end{equation}
directly,
\begin{equation}
\begin{split}
    \frac{\partial v
    ^i_{nm}(\textbf{k})}{\partial k^j} =& \frac{1}{\mathcal{V}_{uc}} \int_{\mathcal{V}_{uc}} d\textbf{x} \Bigg[ \partial_j u^\dag_{n\textbf{k}}(\textbf{x}) \hat{v}^i(\textbf{k}) u_{m\textbf{k}}(\textbf{x})
    + u^\dag_{n\textbf{k}}(\textbf{x}) \partial_j \hat{v}^i(\textbf{k}) u_{m\textbf{k}}(\textbf{x}) 
    + u^\dag_{n\textbf{k}}(\textbf{x}) \hat{v}^i(\textbf{k}) \partial_j u_{m\textbf{k}}(\textbf{x})\Bigg].
\end{split}
\end{equation}
Inserting resolutions of identity this becomes
\begin{equation}
\begin{split}
    &\frac{\partial v^i_{nm}(\textbf{k})}{\partial k^j} = \frac{\hbar}{m} \frac{\partial k^i}{\partial k^j} \frac{1}{\mathcal{V}_{uc}} \int_{\mathcal{V}_{uc}} d\textbf{x} u^\dag_{n\textbf{k}}(\textbf{x}) u_{m\textbf{k}}(\textbf{x})
    \\
    &+ \frac{1}{\mathcal{V}_{uc}^2} \sum_{l} \int_{\mathcal{V}_{uc}} d\textbf{x} \partial_j u^\dag_{n\textbf{k}}(\textbf{x}) u_{l\textbf{k}}(\textbf{x}) \int_{\mathcal{V}_{uc}} d\textbf{y} u^\dag_{l\textbf{k}}(\textbf{y}) \hat{v}^i(\textbf{k}) u_{m\textbf{k}}(\textbf{y})
    \\
    &+\frac{1}{\mathcal{V}_{uc}^2} \sum_l \int_{\mathcal{V}_{uc}} d\textbf{x} u^\dag_{n\textbf{k}}(\textbf{x}) \hat{v}^i(\textbf{k}) u_{l\textbf{k}}(\textbf{x}) \int_{\mathcal{V}_{uc}} d\textbf{y} u^\dag_{l\textbf{k}}(\textbf{y}) \partial_j u_{m\textbf{k}}(\textbf{y})
    \\
    &= \frac{\hbar}{m}\delta_{ij}\delta_{nm} + i\sum_{l}\Big(\xi^j_{nl}v^i_{lm} - v^i_{nl}\xi^j_{lm}\Big),
\end{split}
\end{equation}
which 
is the same result (\ref{vsumrule1}) 
found via the other method. 

\section{Comparison to Ogata \cite{OgataMagSus2017}}\label{SectionOgataComparison}

In the definition of a quantum mechanical formula for the magnetic susceptibility one can identify a ``spontaneous magnetization matrix element" that will feature heavily. Ogata \cite{OgataMagSus2017} defines theirs, written in the notation of our formalism as
\begin{equation}
\label{MagOgata}
\begin{split}
    \mathcal{M}^{l}_{nm} = \frac{e}{2c} \epsilon^{lab} \Big( \sum_{l}  \xi^a_{nl} v^b_{lm} + \frac{1}{\hbar} \partial_b E_{n\textbf{k}} \xi^a_{nm} \Big) + \frac{e}{mc} S^l_{nm} .
\end{split}
\end{equation}
It is 
not Hermitian, $\mathcal{M}^i_{nm} \neq (\mathcal{M}^i_{mn})^*$. Our definition (equation (\ref{SpontMagEq})) is clearly Hermitian. However, our different definitions can be related according to 
\begin{equation}
\label{SpontMagRelation}
    M^l_{nm} = \mathcal{M}^l_{nm} - \frac{e}{4\hbar c} (E_{n\textbf{k}}-E_{m\textbf{k}}) \Omega^l_{nm}, 
\end{equation}
which can be shown using the sum rule shown in section \ref{Appendix:GeneralSumRule} equation (\ref{vsumrule1}):

\begin{equation}
\begin{split}
    M^l_{nm} = \frac{e}{4c} \epsilon^{lab} \Bigg( \xi^a_{nl}v^b_{ln} + v^b_{nl} \xi^a_{lm} + \frac{1}{\hbar} \partial_b (E_{n\textbf{k}} + E_{m\textbf{k}}) \xi^a_{nm} \Bigg) + \frac{e}{mc} S^l_{nm}
    \\
    = \mathcal{M}^l_{nm} - \frac{e}{4c} \Bigg( \xi^a_{nl}v^b_{lm} - v^b_{nl} \xi^a_{lm} + \frac{1}{\hbar} \partial_b (E_{n\textbf{k}}-E_{m\textbf{k}}) \xi^a_{nm} \Bigg)
    \\
    = \mathcal{M}^l_{nm} - \frac{e}{4c} \epsilon^{lab} \Bigg( \delta_{nm} \delta^{ab} - i\partial_a v^b_{nm} + \frac{1}{\hbar} \partial_b (E_{n\textbf{k}}-E_{m\textbf{k}})\xi^a_{nm}  \Bigg)
    \\
    = \mathcal{M}^l_{nm} - \frac{e}{4c} \epsilon^{lab} \Bigg( -\frac{i}{\hbar} \partial_a \Bigg( i(E_{n\textbf{k}}-E_{m\textbf{k}}) \xi^b_{nm} + \delta_{nm} \partial_b E_{n\textbf{k}} \Bigg) + \frac{1}{\hbar} \partial_b (E_{n\textbf{k}} - E_{m\textbf{k}}) \xi^a_{nm} \Bigg)
    \\
    = \mathcal{M}^l_{nm} -\frac{e}{4\hbar c} \epsilon^{lab} \Bigg( \partial_a (E_{n\textbf{k}}-E_{m\textbf{k}})\xi^b_{nm} + \partial_b (E_{n\textbf{k}}-E_{m\textbf{k}})\xi^a_{nm} + (E_{n\textbf{k}}-E_{m\textbf{k}}) \partial_a \xi^b_{nm} \Bigg) 
    \\
    =\mathcal{M}^l_{nm} - \frac{e}{4\hbar c} \epsilon^{lab} (E_{n\textbf{k}}-E_{m\textbf{k}}) \partial_a \xi^b_{nm} 
    \\
    = \mathcal{M}^l_{nm} - \frac{e}{4\hbar c} (E_{n\textbf{k}}-E_{m\textbf{k}}) \Omega^l_{nm} 
\end{split}
\end{equation}

For an insulator there are only three contributions to the total magnetization response that we need to consider. These are the ``Van Vleck" contribution, the ``geometric" contribution, and the ``occupied" contribution, the last being an extension of atomic diamagnetism to crystals. Their functional forms are found in the associated manuscript \cite{PreprintMagSus}, and copied here

\begin{equation}
\label{ChiVanVleck}
\begin{split}
    \chi^{il}_\text{VV} = \sum_{m\neq n} f_{nm} \int_{BZ} \frac{d\textbf{k}}{(2\pi)^3} \frac{ M^i_{nm} M^l_{mn} }{E_{m\textbf{k}}-E_{n\textbf{k}}},
\end{split}
\end{equation}
\begin{equation}
\begin{split}\label{ChiGeometric}
    \chi^{il}_\text{geo} = 
    -\frac{e}{2\hbar c} \sum_{nm} f_{n} \int_{BZ} \frac{d\textbf{k}}{(2\pi)^3}
    \text{Re}\Bigg[ 
    \Omega^i_{nm} \Big( M^l_{mn} + \frac{e}{8\hbar c} (E_{n\textbf{k}}-E_{m\textbf{k}})\Omega^l_{mn}) 
    \\
    + \Big(M^i_{nm} + \frac{e}{8\hbar c}(E_{n\textbf{k}}-E_{m\textbf{k}})\Omega^i_{nm} \Big) \Omega^l_{mn} 
    \Bigg],
\end{split}
\end{equation}
and
\begin{equation}
\label{ChiOcc}
\begin{split}
    \chi^{il}_\text{occ} = \frac{e^2}{4mc^2} \epsilon^{iab}\epsilon^{lcd} \sum_{nm} f_{n} \int_{BZ} \frac{d\textbf{k}}{(2\pi)^3} \text{Re}\Bigg[
    \delta_{bc} \xi^a_{nm} \xi^d_{mn}
    - \frac{m}{\hbar^2} \xi^a_{nm}\xi^d_{mn} \partial_b\partial_c E_{n\textbf{k}}
    \Bigg].
\end{split}
\end{equation}

The occupied term is identical to the contribution Ogata identifies as $\chi_\text{occ}$. We can manipulate our expressions (equation (\ref{ChiVanVleck}) and equation (\ref{ChiGeometric})) using equation (\ref{SpontMagRelation}):

\begin{equation}
\begin{split}
    \chi^{il}_\text{VV} =  -2\sum_{n\neq m} f_{n} \int_{BZ} \frac{d\textbf{k}}{(2\pi)^3} \frac{1}{E_{n\textbf{k}}-E_{m\textbf{k}}} \text{Re}\Bigg[ 
    \Big( \mathcal{M}^i_{nm} - \frac{e}{4\hbar c}(E_{n\textbf{k}}-E_{m\textbf{k}})\Omega^i_{nm} \Big)
    \\
    \times \Big( \mathcal{M}^l_{nm} - \frac{e}{4\hbar c} (E_{n\textbf{k}}-E_{m\textbf{k}}) \Omega^l_{nm} \Big)^*
    \Bigg]
    \\
    = -2\sum_{n\neq m} f_{n} \int_{BZ} \frac{d\textbf{k}}{(2\pi)^3} \text{Re}\Bigg[ \frac{ \mathcal{M}^i_{nm} \Big( \mathcal{M}^l_{nm} \Big)^*  }{E_{n\textbf{k}}-E_{m\textbf{k}}} -\frac{e}{4\hbar c} \Omega^i_{nm} \Big(\mathcal{M}^l_{nm}\Big)^* - \frac{e}{4\hbar c} \mathcal{M}^i_{nm} \Omega^l_{mn}
    \\
    + \frac{e^2}{16\hbar^2 c^2} (E_{n\textbf{k}}-E_{m\textbf{k}}) \Omega^i_{nm} \Omega^l_{mn} \Bigg],
\end{split}
\end{equation}
and
\begin{equation}
\begin{split}
    \chi^{il}_\text{geo} = -\frac{e}{2\hbar c} \sum_{nm} f_{n} \int_{BZ} \frac{d\textbf{k}}{(2\pi)^3} \text{Re}\Bigg[ 
    \Omega^i_{nm}\Bigg( \Big(\mathcal{M}^l_{nm} - \frac{e}{4\hbar c}(E_{n\textbf{k}}-E_{m\textbf{k}}) \Omega^l_{nm} \Big)^* + \frac{e}{8\hbar c} (E_{n\textbf{k}}-E_{m\textbf{k}})\Omega^l_{mn}  \Bigg)
    \\
    +\Big( \mathcal{M}^i_{nm} -\frac{e}{4\hbar c}(E_{n\textbf{k}}-E_{m\textbf{k}})\Omega^l_{nm} + \frac{e}{8\hbar c} (E_{n\textbf{k}}-E_{m\textbf{k}})\Omega^i_{nm} \Big) \Omega^l_{mn} \Bigg]
    \\
    = -\frac{e}{2\hbar c} \sum_{nm} f_{n} \int_{BZ} \frac{d\textbf{k}}{(2\pi)^3} \text{Re}\Bigg[ \Omega^i_{nm} (\mathcal{M}^l_{nm})^* + \mathcal{M}^i_{nm} \Omega^l_{mn}
    - \frac{e}{4\hbar c} (E_{n\textbf{k}}-E_{m\textbf{k}}) \Omega^i_{nm} \Omega^l_{mn} \Bigg].
\end{split}
\end{equation}
If we add these together we find:

\begin{equation}
\begin{split}
    \chi^{il}_\text{VV} + \chi^{il}_\text{geo} = -2 \sum_{n\neq m} f_{n} \int_{BZ} \frac{d\textbf{k}}{(2\pi)^3} \text{Re}\Bigg[ \frac{ \mathcal{M}^i_{nm} \Big(\mathcal{M}^l_{nm}\Big)^* }{E_{n\textbf{k}}-E_{m\textbf{k}}} \Bigg] 
    \\
    -\frac{e}{2\hbar c} \sum_{nn'} f_{n} \int_{BZ} \frac{d\textbf{k}}{(2\pi)^3} \text{Re}\Bigg[ \Omega^i_{nn'}\mathcal{M}^l_{n'n} + \mathcal{M}^i_{nn'} \Omega^l_{n'n} \Bigg] 
    \\
    = \chi^{il}_\text{inter} + \chi^{il}_\text{occ2},
\end{split}
\end{equation}
where $\chi^{il}_\text{inter}$ and $\chi^{il}_\text{occ2}$ are the two additional contributions beyond the occupied term that are present in insulators identified by Ogata \cite{OgataMagSus2017}. Note that the cancellation between the terms in $\chi^{il}_\text{VV}$ and $\chi^{il}_\text{geo}$ occurs because the sum is restricted in $\chi^{il}_\text{VV}$ to not include $m$ in the sum at \textbf{k} points where $n$ and $m$ are equal energy bands. We use the $n$ and $n'$ notation as a shortform for when the sums are restricted to only be when $E_{n\textbf{k}} = E_{n'\textbf{k}}$. If these are only points or lines of degeneracy the expression involves integration over a quantity with measure zero, and can be omitted, but in the case of fully degenerate bands the $n$ and $n'$ summation occur over the entire BZ. 

Thus we have shown that our 
result agrees with that of Ogata \cite{OgataMagSus2017} in the appropriate limit. 

\section{Terms with Equal Energy Elements of the
Berry Connection}\label{SectionDiagonalTerms}

Here we write out the occupied, interband, and the ``occ2"  terms in a way that separates out contributions involving the equal energy elements of the Berry connection. The ``purified" terms, which do not contain any equal energy elements, are denoted by an overset ring. When the same letter is used with `primes' it means the sums are constrained such that those band indices are equal energy. Below we consider bands degenerate over the entire BZ, which reduces to the case of non-degenerate bands if one removes the primes from indices. This is done since points or lines of degeneracy can be removed since we consider topologically trivial insulators and their contribution is an integration over a set of measure zero. We have 

\begin{equation}
\begin{split}
    \chi^{il}_\text{occ} = \frac{e^2}{4\hbar^2 c^2} \epsilon^{icd} \epsilon^{lab} \sum_{nm} f_{n} \int_{BZ} \frac{d\textbf{k}}{(2\pi)^3} \text{Re}\Bigg[ 
    \Bigg(  \partial_b \partial_d E_{n\textbf{k}} - \frac{\hbar^2}{m}\delta_{bd}  \Bigg) \xi^a_{nm}\xi^c_{mn}
    \Bigg]
    \\
    = \mathring{\chi}^{il}_\text{occ} + \frac{e^2}{4\hbar^2 c^2} \epsilon^{icd} \epsilon^{lab} \sum_{nn'n''} f_{n} \int_{BZ} \frac{d\textbf{k}}{(2\pi)^3} \text{Re}\Bigg[ 
    \Bigg(  \partial_b \partial_d E_{n\textbf{k}} - \frac{\hbar^2}{m}\delta_{bd}  \Bigg) \delta_{nn'} \xi^a_{n'n''}\xi^c_{n''n}
    \Bigg]
    \\
    = \mathring{\chi}^{il}_\text{occ} + \frac{e^2}{4\hbar c^2} \epsilon^{icd} \epsilon^{lab} \sum_{nn'n''} f_{n} \int_{BZ} \frac{d\textbf{k}}{(2\pi)^3} \text{Re}\Bigg[ \sum_{l\neq n}\Big( i\xi^b_{nl}v^d_{ln'} - iv^d_{nl}\xi^b_{ln'} \Big) \xi^a_{n'n''}\xi^c_{n''n} \Bigg] 
    \\
    = \mathring{\chi}^{il}_\text{occ} + \frac{e^2}{4\hbar^2 c^2} \epsilon^{icd} \epsilon^{lab} \sum_{nn'n''}\sum_{l\neq n} f_{n} \int_{BZ} \frac{d\textbf{k}}{(2\pi)^3} \text{Re}\Bigg[ (E_{n\textbf{k}}-E_{l\textbf{k}}) \Big(\xi^d_{nl}\xi^b_{ln'} + \xi^b_{nl}\xi^d_{ln'}\Big) \xi^a_{n'n''}\xi^c_{n''n} \Bigg],
\end{split}
\end{equation}
where we have used the diagonal inverse mass tensor sum rule to process the terms in brackets, and that $E_{n\textbf{k}} = E_{n'\textbf{k}}$. 

We write the off-diagonal spontaneous magnetization as

\begin{equation}
\begin{split}
    \mathcal{M}^l_{nm} = \frac{e}{2c} \epsilon^{lab} \Bigg( \sum_{l\neq n} \xi^a_{nl}v^b_{lm} + \frac{1}{\hbar} \partial_b E_{n\textbf{k}} \xi^a_{nm}\Bigg) + \frac{e}{mc} S^l_{nm} + \frac{ie}{2\hbar c} (E_{n\textbf{k}}-E_{m\textbf{k}}) \epsilon^{lab} \sum_{n'}\xi^a_{nn'} \xi^b_{n'm}
    \\
    = \mathring{\mathcal{M}}^l_{nm} + \frac{ie}{2\hbar c}(E_{n\textbf{k}}-E_{m\textbf{k}}) \epsilon^{lab} \sum_{n'}\xi^a_{nn'} \xi^b_{n'm}, 
\end{split}
\end{equation}
where the `purefied magnetization matrix element' is defined as
\begin{equation}
\label{PureMnm}
\begin{split}
    \mathring{\mathcal{M}}^l_{nm} = \frac{e}{2c} \epsilon^{lab} \Big(\sum_{l\neq n} \xi^a_{nl} v^b_{lm} + \frac{1}{\hbar} \partial_b E_{n\textbf{k}} \xi^a_{nm} \Big) + \frac{e}{mc} S^l_{nm}
\end{split}
\end{equation}
for $n$ and $m$ labelling distinct energy bands. When we use the notation $l\neq n$ we mean, at a given \textbf{k}, the sum l is only over distinct energy bands. Thus the interband term can be written
\begin{equation}
\begin{split}
    \chi^{il}_\text{inter} = -2 \sum_{n}\sum_{ m\neq n} f_{n} \int_{BZ} \frac{d\textbf{k}}{(2\pi)^3} \text{Re}\Bigg[ 
    \frac{ \mathring{\mathcal{M}}^i_{nm} \Big(\mathring{\mathcal{M}}^l_{nm}\Big)^* }{E_{n\textbf{k}}-E_{m\textbf{k}}} 
    \Bigg] 
    \\
    -2 \sum_{nn'} \sum_{m\neq n} f_{n} \int_{BZ} \frac{d\textbf{k}}{(2\pi)^3} \text{Re}\Bigg[ \frac{ie}{2\hbar c}(E_{n\textbf{k}}-E_{m\textbf{k}}) \frac{\Big(\epsilon^{icd} \xi^c_{nn'}\xi^d_{n'm}\Big( \mathring{\mathcal{M}}^l_{nm} \Big)^*-\epsilon^{lab} 
    \mathring{\mathcal{M}}^i_{nm} \xi^b_{mn'}\xi^a_{n'n} \Big)}{E_{n\textbf{k}}-E_{m\textbf{k}}}
    \Bigg] 
    \\
    -2\sum_{nn'n''} \sum_{m\neq n} f_{n} \int_{BZ} \frac{d\textbf{k}}{(2\pi)^3} \text{Re}\Bigg[ 
    \frac{e^2}{4\hbar^2 c^2} \epsilon^{icd}\epsilon^{lab} \xi^c_{nn'} \xi^d_{n'm} \xi^b_{mn''}\xi^a_{n''n} \frac{(E_{n\textbf{k}}-E_{m\textbf{k}})^2}{E_{n\textbf{k}}-E_{m\textbf{k}}}
    \Bigg].
\end{split}
\end{equation}
The middle line can be processed further to give

\begin{equation}
\begin{split}
     -2\sum_{nn'} \sum_{m\neq n} f_{n} \int_{BZ} \frac{d\textbf{k}}{(2\pi)^3} \text{Re}\Bigg[ \frac{ie}{2\hbar c}(E_{n\textbf{k}}-E_{m\textbf{k}}) \frac{\Big(\epsilon^{icd} \xi^c_{nn'}\xi^d_{n'm}\Big( \mathring{\mathcal{M}}^l_{nm} \Big)^*-\epsilon^{lab} 
    \mathring{\mathcal{M}}^i_{nm} \xi^b_{mn'}\xi^a_{n'n}   \Big)}{E_{n\textbf{k}}-E_{m\textbf{k}}}
    \Bigg] 
    \\
    = -\frac{e^2}{2\hbar c^2} \epsilon^{icd}\epsilon^{lab} \sum_{nn'} \sum_{m\neq n} f_{n} \int_{BZ} \frac{d\textbf{k}}{(2\pi)^3} \text{Re}\Bigg[ 
    i\xi^c_{nn'}\xi^d_{n'm} \Bigg(\sum_{l\neq n}  v^b_{ml}\xi^a_{ln} + \frac{1}{\hbar} \partial_b E_{n\textbf{k}}\xi^a_{mn}\Bigg)
    \Bigg] 
    \\
    +\frac{e^2}{2\hbar c^2} \epsilon^{icd}\epsilon^{lab} \sum_{nn'} \sum_{m\neq n} f_{n} \int_{BZ} \frac{d\textbf{k}}{(2\pi)^3} \text{Re}\Bigg[ 
    i\Big(\sum_{l\neq n}  \xi^c_{nl}v^d_{lm} + \frac{1}{\hbar} \partial_d E_{n\textbf{k}} \xi^c_{nm}\Big) \xi^b_{mn'} \xi^a_{n'n}
    \Bigg]
    \\
    -\frac{e^2}{m \hbar c^2} \sum_{nn'} \sum_{m\neq n} f_{n} \int_{BZ} \frac{d\textbf{k}}{(2\pi)^3} \text{Re}\Bigg[ 
    i\epsilon^{icd} \xi^c_{nn'}\xi^d_{n'm} S^l_{mn} - i\epsilon^{lab} S^i_{nm} \xi^b_{mn'} \xi^a_{n'n}
    \Bigg].
\end{split}
\end{equation}
Separating out the contribution of the equal energy Berry connection matrix elements to the equal energy magnetization matrix elements,

\begin{equation}
    \mathcal{M}^l_{nn'} = \mathring{\mathcal{M}}^l_{nn'} + \frac{e}{\hbar c} \epsilon^{lab} \partial_b E_{n\textbf{k}} \xi^a_{nn'},
\end{equation}
where
\begin{equation}\label{PureMnn}
\begin{split}
    \mathring{\mathcal{M}}^l_{nn'} = \frac{e}{2c} \epsilon^{lab} \sum_{l\neq n} \xi^a_{nl}v^b_{ln'} + \frac{e}{mc} S^l_{nn'}, 
\end{split}
\end{equation}
when $n$ and $n'$ are equal energy bands. 
Thus the ``occ2" term can be written as

\begin{equation}
\begin{split}
    \chi^{il}_\text{occ2} = -\frac{e}{2\hbar c} \sum_{n} f_{n} \int_{BZ} \frac{d\textbf{k}}{(2\pi)^3} \text{Re}\Bigg[ \Omega^i_{nn'} \mathcal{M}^l_{n'n} + \mathcal{M}^i_{nn'} \Omega^l_{n'n} \Bigg] 
    \\
    = \mathring{\chi}^{il}_\text{occ2}
    -\frac{e}{2\hbar c} \sum_{nn'n''} f_{n} \int_{BZ} \frac{d\textbf{k}}{(2\pi)^3} \text{Re}\Bigg[ 
    i\epsilon^{iab} \xi^a_{nn''}\xi^b_{n''n'} \mathring{\mathcal{M}}^l_{n'n} + i\epsilon^{lcd} \mathring{\mathcal{M}}^i_{nn'} \xi^c_{n'n''} \xi^d_{n''n}
    \Bigg]
    \\
    -\frac{e^2}{2\hbar^2 c^2} \epsilon^{iab} \epsilon^{lcd} \sum_{nn'} \sum_{m\neq n} f_{n} \int_{BZ} \frac{d\textbf{k}}{(2\pi)^3} \text{Re}\Bigg[ 
    i\xi^a_{nm}\xi^b_{mn'}\xi^c_{n'n}\partial_d E_{n\textbf{k}}  +    i\partial_b E_{n\textbf{k}} \xi^a_{nn'} \xi^c_{n'm}\xi^d_{mn} 
    \Bigg]
    \\
    -\frac{e^2}{2\hbar^2 c^2} \epsilon^{iab} \epsilon^{lcd} \sum_{nn'n''} f_{n} \int_{BZ} \frac{d\textbf{k}}{(2\pi)^3} \text{Re}\Bigg[ 
    i\xi^a_{nn''}\xi^b_{n''n'}\xi^c_{n'n}\partial_d E_{n\textbf{k}}  +    i\partial_b E_{n\textbf{k}} \xi^a_{nn'} \xi^c_{n'n''}\xi^d_{n''n} 
    \Bigg].
\end{split}
\end{equation}
The combination of the three contributions gives 
\begin{equation}
\label{PurefiedSusceptibility}
\begin{split}
    \chi^{il} = \mathring{\chi}^{il}_\text{inter} +  \mathring{\chi}^{il}_\text{occ}  +  \mathring{\chi}^{il}_\text{occ2} + \underline{\chi}^{il} .
\end{split}
\end{equation}
where 
\begin{equation}
\label{brevesusceptibility}
\begin{split}
    \underline{\chi}^{il}  = \frac{e^2}{4\hbar^2 c^2} \epsilon^{iab} \epsilon^{lcd} \sum_{n n' n''} \sum_{m\neq n} f_{n} \int_{BZ} \frac{d\textbf{k}}{(2\pi)^3} \text{Re}\Bigg[ 
    (E_{n\textbf{k}}-E_{m\textbf{k}}) \Big(\xi^b_{nm}\xi^d_{mn'}+\xi^d_{nm}\xi^b_{mn'}\Big)  \xi^c_{n'n''}\xi^a_{n''n}
    \Bigg]
    \\
    -\frac{e^2}{2\hbar^2 c^2} \epsilon^{iab} \epsilon^{lcd} \sum_{n n' n''}\sum_{m\neq n}  f_{n} \int_{BZ} \frac{d\textbf{k}}{(2\pi)^3} \int_{BZ} \frac{d\textbf{k}}{(2\pi)^3} \text{Re}\Bigg[ 
     (E_{n\textbf{k}}-E_{m\textbf{k}}) \xi^b_{n'm} \xi^d_{mn''} \xi^c_{n''n} \xi^a_{nn'}
    \Bigg]
    \\
    -\frac{e}{\hbar c} \sum_{n n'} \sum_{m\neq n} f_{n} \int_{BZ} \frac{d\textbf{k}}{(2\pi)^3} \text{Re}\Bigg[ 
    i\epsilon^{iab} \xi^a_{nn'}\xi^b_{n'm} \Big( 
    \frac{e}{2c}\epsilon^{lcd} \Big(\sum_{l\neq n} v^d_{ml}\xi^c_{ln} + \frac{1}{\hbar} \partial_d E_{n\textbf{k}} \xi^c_{mn}\Big)
    +\frac{e}{mc} S^l_{mn}
    \Big) 
    \\
    - i\epsilon^{lcd} \Big( \frac{e}{2c} \epsilon^{iab} \Big( \sum_{l\neq n} \xi^a_{nl}v^b_{lm} + \frac{1}{\hbar} \partial_b E_{n\textbf{k}} \xi^a_{nm} \Big) + \frac{e}{mc} S^i_{nm}  \Big)\xi^d_{mn'}\xi^c_{n'n}
    \Bigg]
    \\
    -\frac{e^2}{2\hbar^2 c^2} \epsilon^{iab} \epsilon^{lcd} \sum_{nn'} \sum_{m\neq n} f_{n} \int_{BZ} \frac{d\textbf{k}}{(2\pi)^3} \text{Re}\Bigg[ 
    i\xi^a_{nm}\xi^b_{mn'}\xi^c_{n'n}\partial_d E_{n\textbf{k}}  +    i\partial_b E_{n\textbf{k}} \xi^a_{nn'} \xi^c_{n'm}\xi^d_{mn} 
    \Bigg]
    \\
    -\frac{e^2}{2\hbar^2 c^2} \epsilon^{iab} \epsilon^{lcd} \sum_{nn'n''} f_{n} \int_{BZ} \frac{d\textbf{k}}{(2\pi)^3} \text{Re}\Bigg[ 
    i\xi^a_{nn''}\xi^b_{n''n'}\xi^c_{n'n}\partial_d E_{n\textbf{k}}  +    i\partial_b E_{n\textbf{k}} \xi^a_{nn'} \xi^c_{n'n''}\xi^d_{n''n} 
    \Bigg]
    \\
    -\frac{e^2}{4\hbar c^2} \epsilon^{iab} \epsilon^{lcd} \sum_{n n'n''} \sum_{l\neq n} f_{n} \int_{BZ} \frac{d\textbf{k}}{(2\pi)^3} \text{Re}\Bigg[ 
    i\xi^a_{nn''}\xi^b_{n''n'}   v^d_{n'l}\xi^c_{ln} + i\xi^a_{nl}v^b_{ln'} \xi^c_{n'n''}\xi^d_{n''n} 
    \Bigg]
    \\
    -\frac{e^2}{2m \hbar c^2} \sum_{n n'n''} f_{n} \int_{BZ} \frac{d\textbf{k}}{(2\pi)^3} \text{Re}\Bigg[ 
    i\epsilon^{iab} \xi^a_{nn''}\xi^b_{n''n'} S^l_{n'n}+ i\epsilon^{lcd}S^i_{nn'} \xi^c_{n'n''}\xi^d_{n''n} 
    \Bigg].
\end{split}
\end{equation}
We then make a few manipulations to process equation (\ref{brevesusceptibility}), like identifying the sum rule 
\begin{equation}
\label{newsumrule1}
\begin{split}
    \epsilon^{lab}\sum_{ml}\Big( \xi^d_{nm}v^b_{ml}\xi^a_{ln'} - \xi^a_{nl}v^b_{lm}\xi^d_{mn'}\Big) = \epsilon^{lab}\Big( \frac{i\hbar}{m} \delta_{bd} \xi^a_{nn'} + i\partial_a \Big(v^b_{nl}\xi^d_{ln'}\Big) - i\partial_d \Big(\xi^a_{nl}v^b_{ln'}\Big) \Big), 
\end{split}
\end{equation}
which can be derived by using the commutation relations of the position and velocity operators to write
\begin{equation}
\begin{split}
    \epsilon^{lab}\int d\textbf{x}  W^\dag_{\alpha\textbf{R}}
    \Bigg[(x^d-R'^d) \hat{v}^b(\textbf{x}) (x^a-R'^a) - (x^a-R'^a) \hat{v}^b(\textbf{x}) (x^d-R'^d) \Bigg]W_{\beta\textbf{R}'}(\textbf{x}) 
    \\
    = \epsilon^{lab} \int d\textbf{x} W^\dag_{\alpha\textbf{R}}  \Big[ (x^d-R'^d),\hat{v}(\textbf{x})^b \Big](x^a-R'^a) W_{\beta\textbf{R}'}(\textbf{x}).
\end{split}
\end{equation}
The left-hand-side of (\ref{newsumrule1}) can be identified with the real parts of the first expressions within the brackets in the third and fourth lines of (\ref{brevesusceptibility}), except that in that equation the sums over $m$ and $l$ are restricted. Therefore a set of terms that completes the sums had to added and subtracted. Making these substitutions we find
\begin{equation}
\label{chibar_intermediate}
\begin{split}
    \underline{\chi}^{il}
    = 
    \frac{e^2}{4\hbar^2 c^2} \epsilon^{iab} \epsilon^{lcd} \sum_{nn'n''} \sum_{m\neq n} f_{n} \int_{BZ} \frac{d\textbf{k}}{(2\pi)^3} \text{Re}\Bigg[ 
    (E_{n\textbf{k}}-E_{m\textbf{k}}) (\xi^d_{nm}\xi^b_{mn'}-\xi^b_{nm}\xi^d_{mn'}) \xi^c_{n'n''}\xi^a_{n''n}
    \Bigg]
    \\
    -\frac{e^2}{4\hbar c^2} \epsilon^{iab}\epsilon^{lcd} \sum_{nn'} \sum_{m\neq n} f_{n} \int_{BZ} \frac{d\textbf{k}}{(2\pi)^3} \text{Re}\Bigg[ 
    i \xi^a_{nn'} \Bigg( \frac{i\hbar}{m}\delta_{bd} \xi^c_{n'n} + i\partial_c \Big( v^d_{n'm}\xi^b_{mn} \Big) - i \partial_b \Big(\xi^c_{n'm}v^d_{mn} \Big) \Bigg)
    \\
    +i \xi^c_{nn'} \Bigg( 
    \frac{i\hbar}{m}\delta_{bd} \xi^a_{n'n} + i\partial_a \Big( v^b_{n'm}\xi^d_{mn}\Big) - i\partial_d\Big(\xi^a_{n'm}v^b_{mn}\Big)
    \Bigg)
    \Bigg]
    \\
    -\frac{e^2}{4\hbar^2 c^2} \epsilon^{iab}\epsilon^{lcd} \sum_{nn'} f_{n} \int_{BZ} \frac{d\textbf{k}}{(2\pi)^3} \text{Re}\Bigg[ 
    i \xi^a_{nn'} \Bigg( i\partial_c \Big( \partial_d E_{n\textbf{k}} \xi^b_{n'n} \Big) - i \partial_b \Big(\partial_d E_{n\textbf{k}} \xi^c_{n'n} \Big) \Bigg)
    \\
    +i \xi^c_{nn'} \Bigg( 
    i\partial_a \Big( \partial_b E_{n\textbf{k}}\xi^d_{n'n} \Big) - i\partial_d\Big( \partial_b E_{n\textbf{k}} \xi^a_{n'n}\Big)
    \Bigg)
    \Bigg]
    \\
    +\frac{e^2}{2\hbar c^2} \epsilon^{iab} \epsilon^{lcd} \sum_{nn'n''} \sum_{m\neq n} f_{n} \int_{BZ} \frac{d\textbf{k}}{(2\pi)^3}  \text{Re}\Bigg[ 
    i\xi^a_{nn'}\xi^b_{n'n''} v^d_{n''m} \xi^c_{mn} 
    +i\xi^c_{nn'}\xi^d_{n'n''} v^b_{n''m} \xi^a_{mn}
    \\
    + i \xi^c_{nn''}\xi^a_{n''n'}\Big(
    \xi^b_{n'm}v^d_{mn}
    -v^b_{n'm}\xi^d_{mn}\Big)
    \Bigg]
    \\
    +\frac{e^2}{2\hbar c^2} \epsilon^{iab} \epsilon^{lcd} \sum_{nn' n''} f_{n} \int_{BZ} \frac{d\textbf{k}}{(2\pi)^3} \text{Re}\Bigg[ 
    i\partial_d E_{n\textbf{k}}\xi^a_{nn'} \xi^b_{n'n''} \xi^c_{n''n} 
    +i\partial_b E_{n\textbf{k}} \xi^c_{nn'}\xi^d_{n'n''}\xi^a_{n''n}
    \Bigg] 
    \\
    -\frac{e^2}{2\hbar^2 c^2} \epsilon^{iab} \epsilon^{lcd} \sum_{nn'} \sum_{m\neq n} f_{n} \int_{BZ} \frac{d\textbf{k}}{(2\pi)^3} \text{Re}\Bigg[ 
    i\partial_d E_{n\textbf{k}} \Big(\xi^a_{nn'} \xi^b_{n'm} \xi^c_{mn} + \xi^c_{nn'}\xi^a_{n'm} \xi^b_{mn} \Big)
    \Bigg]
    \\
    +\frac{e^2}{2\hbar^2 c^2} \epsilon^{iab}\epsilon^{lcd} \sum_{nn'} \sum_{m\neq n}  f_{n} \int_{BZ} \frac{d\textbf{k}}{(2\pi)^3} \text{Re}\Bigg[ 
    i\partial_b E_{n\textbf{k}}\Big( \xi^a_{nm}\xi^d_{mn'}\xi^c_{n'n} - \xi^c_{nm}\xi^d_{mn'} \xi^a_{n'n} \Big)
    \Bigg]
    \\
    -\frac{e^2}{4\hbar c^2} \epsilon^{iab}\epsilon^{lcd} \sum_{nn'n''} \sum_{m \neq n} f_{n} \int_{BZ} \frac{d\textbf{k}}{(2\pi)^3}  \text{Re}\Bigg[ 
    i\xi^a_{nn''}\xi^b_{n''n'} v^d_{n'm} \xi^c_{mn} + i\xi^a_{nm}v^b_{mn'} \xi^c_{n'n''} \xi^d_{n''n}
    \Bigg]
    \\
    -\frac{e^2}{2\hbar^2 c^2} \epsilon^{iab} \epsilon^{lcd} \sum_{nn'n''} f_{n} \int_{BZ} \frac{d\textbf{k}}{(2\pi)^3} \text{Re}\Bigg[ 
    i\xi^a_{nn''}\xi^b_{n''n'}\xi^c_{n'n}\partial_d E_{n\textbf{k}}  +    i\partial_b E_{n\textbf{k}} \xi^a_{nn'} \xi^c_{n'n''}\xi^d_{n''n} 
    \Bigg]
    \\
    -\frac{e^2}{\hbar m c^2} \sum_{nn'} \sum_{m\neq n } f_{n} \int_{BZ} \frac{d\textbf{k}}{(2\pi)^3} \text{Re}\Bigg[ 
    i\epsilon^{iab} \xi^a_{nn'} \xi^b_{n'm}S^l_{mn} 
    -i\epsilon^{lcd} S^i_{nm}\xi^d_{mn'} \xi^c_{n'n}
    \Bigg]
    \\
    -\frac{e^2}{2\hbar m c^2} \sum_{nn'n''} f_{n} \int_{BZ} \frac{d\textbf{k}}{(2\pi)^3} \Bigg[ 
    i\epsilon^{iab} \xi^a_{nn''}\xi^b_{n''n'}S^l_{n'n} +i\epsilon^{lcd}  S^l_{nn'}\xi^c_{n'n''} \xi^d_{n''n} 
    \Bigg].
\end{split}
\end{equation}

Next we make use of an identity involving the derivatives of the spin matrix elements
\begin{equation}\label{Spinderivative}
    \sum_{m} \Big(i\xi^d_{nm}S^l_{mn} - iS^l_{nm}\xi^d_{mn}\Big) = \partial_d S^l_{nn},
\end{equation}
and the derivatives of the Berry connection
\begin{equation}\label{ConnectionDerivative}
\begin{split}
    \partial_a \xi^d_{nn} - \partial_d \xi^a_{nn} = \sum_{l} \Bigg(i\xi^a_{nl}\xi^d_{ln} - i\xi^d_{nl}\xi^a_{ln} \Bigg).
\end{split}
\end{equation}
Equations (\ref{Spinderivative}) and (\ref{ConnectionDerivative}) require the equality of the mixed partials of the cell periodic Bloch functions to hold. 

Note that in using the sum and commutation rule identities we have reintroduced equal energy Berry connection matrix elements. Rearranging the terms, using equation (\ref{Spinderivative}) and (\ref{ConnectionDerivative}), the inverse effective mass tensor sum rule, equation (\ref{chibar_intermediate}) can be rewritten as

\begin{equation}
\begin{split}
    \underline{\chi}^{il} =
    \frac{e^2}{4\hbar c^2} \epsilon^{iab} \epsilon^{lcd} \sum_{nn'} \sum_{m\neq\{n\}} f_{n} \int_{BZ} \frac{d\textbf{k}}{(2\pi)^3} \text{Re}\Bigg[ 
    \xi^a_{nn'}\Bigg( \partial_c \Big( v^d_{n'm} \xi^b_{mn}\Big) - \partial_b \Big( \xi^c_{n'm}v^d_{mn}\Big) \Bigg) 
    +\xi^c_{nn}\Bigg( \partial_a \Big( v^b_{n'm} \xi^d_{mn} \Big) - \partial_d \Big(\xi^a_{n'm}v^b_{mn}\Big) \Bigg)
    \Bigg]
    \\
    -\frac{e^2}{2\hbar^2 c^2} \epsilon^{iab} \epsilon^{lcd} \sum_{nn'} \sum_{m\neq\{n\}} f_{n} \int_{BZ} \frac{d\textbf{k}}{(2\pi)^3} \text{Re}\Bigg[ 
    i\partial_b E_{n\textbf{k}} \Big( \xi^c_{nm}\xi^d_{mn'}\xi^a_{n'n} - 2\xi^a_{nm}\xi^d_{mn'}\xi^c_{n'n}\Big)
    +i\partial_d E_{n\textbf{k}} \Big( \xi^c_{nn'}\xi^a_{n'm} \xi^b_{mn} + 2\xi^a_{nn'}\xi^b_{n'm}\xi^c_{mn} \Big)
    \Bigg]
    \\
    +\frac{e^2}{4\hbar c^2} \epsilon^{iab} \epsilon^{lcd} \sum_{nn'n''} \sum_{m\neq \{n\}} \int_{BZ} \frac{d\textbf{k}}{(2\pi)^3} \text{Re}\Bigg[ 
    i\xi^a_{nn'}\xi^b_{n'n''} v^d_{n''m} \xi^c_{mn}
    +i\xi^c_{nn'}\xi^d_{n'n''} v^b_{n''m}\xi^a_{mn}
    \\
    +i\xi^c_{nn''}\xi^a_{n''n'}\Big(\xi^b_{n'm} v^d_{mn}
    +v^d_{n'm}\xi^b_{mn} \Big)
    \Bigg]
    \\
    -\frac{e^2}{2\hbar^2 c^2} \epsilon^{iab} \epsilon^{lcd} \sum_{nn'n''} \sum_{m} f_{n} \int_{BZ} \frac{d\textbf{k}}{(2\pi)^3} \text{Re}\Bigg[ 
    i\partial_d E_{n\textbf{k}} \xi^a_{nn'} \xi^b_{n'n''}\xi^c_{n''n}
    +i\partial_b E_{n\textbf{k}} \xi^a_{n''n}\xi^c_{nn'}\xi^d_{n'n''} 
    \Bigg]
    \\
    -\frac{e}{2\hbar c} \sum_{nn'} f_{n} \int_{BZ} \frac{d\textbf{k}}{(2\pi)^3} \text{Re}\Bigg[ 
    \mathring{\Omega}^{i}_{nn'} S^l_{n'n} + S^i_{nn'} \mathring{\Omega}^l_{n'n}
    \Bigg],
\end{split}
\end{equation}
where we integrated by parts the spin term that used equation (\ref{Spinderivative}), and the last line of equation (\ref{chibar_intermediate}) allows for the replacement of the curl of the Berry connection $\Omega_{nn'}$ with the gauge-covariant curl $\mathring{\Omega}_{nn'}$. Equal energy Berry connection elements still remain, but to process this further it is useful to look at the specific Cartesian components, such as $\underline{\chi}^{xz}$ or $\underline{\chi}^{zz}$. We demonstrate here for $\underline{\chi}^{zz}$, for which the Levi-Civita pre-factor is $\delta_{ac}\delta_{bd} - \delta_{ad}\delta_{bc}$ ($a,b,c,d\neq z$), and we have

\begin{equation}
\begin{split}
    \underline{\chi}^{zz} = \frac{e^2}{2\hbar c^2} \sum_{nn'}\sum_{m\neq n} f_{n} \int_{BZ} \frac{d\textbf{k}}{(2\pi)^3} \text{Re}\Bigg[ 
    \xi^a_{nn'} \Big( \partial_a (v^b_{n'm}\xi^b_{mn}) - \partial_b (\xi^a_{n'm}v^b_{mn}) \Big) 
    -\xi^a_{nn'} \Big( \partial_b (v^a_{n'm}\xi^b_{mn}) - \partial_b (\xi^b_{n'm}v^a_{mn} ) \Big)
    \Bigg]
    \\
    -\frac{e^2}{2\hbar^2 c^2} \sum_{nn'} \sum_{m\neq n} f_{n} \int_{BZ} \frac{d\textbf{k}}{(2\pi)^3} \text{Re}\Bigg[ 
    -i\partial_b E_{n\textbf{k}} \Big( \xi^a_{n'n}\xi^a_{nm}\xi^b_{mn'} \Big) 
    +i \partial_b E_{n\textbf{k}} \Big( \xi^a_{nn'}\xi^b_{n'm}\xi^a_{mn} \Big)
    \\
    -i\partial_b E_{n\textbf{k}}\Big(\xi^a_{nn'}\xi^b_{n'm}\xi^a_{mn} - 2\xi^a_{nm}\xi^a_{mn'}\xi^b_{n'n} \Big)
    -i \partial_b E_{n\textbf{k}}\Big(\xi^a_{nn'}\xi^b_{n'm}\xi^a_{mn}
    +2\xi^b_{nn'}\xi^a_{n'm}\xi^a_{mn}\Big)
    \Bigg]
    \\
    -\frac{e^2}{2\hbar^2 c^2} \sum_{nn'n''} f_{n} \int_{BZ} \frac{d\textbf{k}}{(2\pi)^3} \text{Re}\Bigg[ 
    i\partial_b E_{n\textbf{k}} \xi^a_{nn'}\xi^b_{n'n''}\xi^a_{n''n} -i\partial_b E_{n\textbf{k}} \xi^b_{nn'}\xi^a_{n'n''}\xi^a_{n''n}
    +i\partial_b E_{n\textbf{k}} \xi^a_{nn'}\xi^a_{n'n''}\xi^b_{n''n}
    \\
    -i\partial_b E_{n\textbf{k}} \xi^a_{nn'}\xi^b_{n'n''}\xi^a_{n''n}
    \Bigg]
    \\
    + \frac{e^2}{4\hbar c^2} \sum_{nn'n''} \sum_{m\neq n} \int_{BZ} \frac{d\textbf{k}}{(2\pi)^3} \text{Re}\Bigg[ 
    i\xi^a_{nn'}\xi^b_{n'n''} \Big(v^b_{n''m}\xi^a_{mn} - v^a_{n''m}\xi^b_{mn}\Big)
    +i\xi^a_{nn'}\xi^b_{n'n''} v^b_{n''m}\xi^a_{mn} -i \xi^b_{nn'}\xi^a_{n'n''} v^b_{n''m}\xi^a_{mn}
    \\
    +i\xi^a_{nn'}\xi^a_{n'n''}\Big(\xi^b_{n'm}v^b_{mn} + v^b_{n'm}\xi^b_{mn}\Big)
    -i\xi^b_{nn'}\xi^a_{n'n''}\Big(\xi^b_{n''m}v^a_{mn} + v^a_{n''m}\xi^b_{mn} \Big) 
    \Bigg]
    \\
    -\frac{e}{2\hbar c} \sum_{nn'} f_{n} \int_{BZ} \frac{d\textbf{k}}{(2\pi)^3} \text{Re}\Bigg[ 
    \mathring{\Omega}^{z}_{nn'} S^z_{n'n} + S^z_{nn'} \mathring{\Omega}^z_{n'n}
    \Bigg]
\end{split}
\end{equation}

Using the fact that the expressions are under the real part operation the terms multiplying $\partial_b E_{n\textbf{k}}$ cancel each other. Using a change of indices $a \leftrightarrow b$, where $a,b\neq z$ are summed over, we can write 
\begin{equation}
\begin{split}
     \underline{\chi}^{zz} = \frac{e^2}{2\hbar c^2} \sum_{nn'} \sum_{m\neq n } f_{n} \int_{BZ} \frac{d\textbf{k}}{(2\pi)^3} \text{Re}\Bigg[ \xi^a_{nn'} \partial_b (\xi^b_{n'm}v^a_{mn}) \Bigg] 
    \\
    +\frac{e^2}{4\hbar c^2} \sum_{nn'n''} \sum_{m\neq n}\int_{BZ} \frac{d\textbf{k}}{(2\pi)^3} \text{Re}\Bigg[ 
    i\xi^a_{nn'}\xi^b_{n'n''} v^b_{n''m} \xi^a_{mn}
    -i\xi^b_{nn'}\xi^a_{n'n''} v^b_{n''m}\xi^a_{mn}
    \Bigg], 
    \\
    -\frac{e}{\hbar c} \sum_{nn'} f_{n} \int_{BZ} \frac{d\textbf{k}}{(2\pi)^3} \text{Re}\Bigg[ 
    \mathring{\Omega}^{z}_{nn'} S^z_{n'n}
    \Bigg]
\end{split}
\end{equation}
where we have used the identity 

\begin{equation}
\label{vzeta}
    \sum_{l\neq n} (v^b_{nl}\xi^a_{ln'}) = -\sum_{l\neq n}(\xi^b_{nl}v^a_{ln'})
\end{equation}
which follows from (\ref{velocity}). 

Finally, we 
integrate by parts to get
\begin{equation}
\begin{split}
    \underline{\chi}^{zz} = -\frac{e^2}{4\hbar c^2} \sum_{nn'} \sum_{m,l \neq n }  f_{n} \int_{BZ} \frac{d\textbf{k}}{(2\pi)^3} \text{Re}\Bigg[ 
    i(\xi^a_{nm}\xi^b_{mn'}-\xi^b_{nm}\xi^a_{mn'})v^b_{n'l}\xi^a_{ln} 
    \Bigg]
    \\
    -\frac{e}{\hbar c} \sum_{nn'} f_{n} \int_{BZ} \frac{d\textbf{k}}{(2\pi)^3} \text{Re}\Bigg[ 
    \mathring{\Omega}^{z}_{nn'} S^z_{n'n}
    \Bigg]
\end{split}
\end{equation}

This is in a form that is clearly gauge-invariant. Instead of depending on the equal energy Berry connection elements, it is written in terms of the gauge-invariant Berry curvature; this is explicit in the second line, and in the first line it follows from using (\ref{vzeta}). Within numerical factors, the two contributions to $\underline{\chi}^{zz}$ are the ``purified" versions of the orbital and spin contributions to the ``occ2" term in the susceptibility. But in fact they appear with different prefactors, 
\begin{equation}
    \underline{\chi}^{zz}  = \frac{1}{2} \mathring{\chi}^{zz}_\text{occ2:Orb} + \mathring{\chi}^{zz}_\text{occ2:Spin}.
\end{equation}

This clearly identifies a distinction in how the orbital and spin components of $\mathring{\chi}^{il}_\text{occ2}$ contribute to the total susceptibility.

\section{Comparison to Gao et al. \cite{GaoGeometricalSus}  and Blount \cite{BlountMagSus}}\label{SectionGaoBlountComparison}

In the work of Gao et al. \cite{GaoGeometricalSus}  the magnetic susceptibility is obtained from the second order energy corrections to a wave-packet free energy; the expression is their equation (6). Considering only the terms relevant for an insulator, that expression for the susceptibility reduces to
\begin{equation}
\begin{split}\label{GaoSusceptibility}
    &\chi^{il}_\text{Gao} = 
    \\
    &-\sum_{m} \int_{BZ} \frac{d\textbf{k}}{(2\pi)^3} 
    \frac{  ( \mathring{\mathcal{M}}^{l}_{0m} )^*_{\text{Orb}}(\mathring{\mathcal{M}}^{i}_{0m})_{\text{Orb}} +  (\mathring{\mathcal{M}}_{0m}^{l})_{\text{Orb}} ( \mathring{\mathcal{M}}^{i}_{0m})^*_{\text{Orb}} }{E_{0\textbf{k}}-E_{m\textbf{k}}}
    \\
    &-\frac{3e}{4 \hbar c}  \int_{BZ} \frac{d\textbf{k}}{(2\pi)^3} 
    \Big( \Omega^i_{00} (\mathring{\mathcal{M}}^{l}_{00})_{\text{Orb}} + (\mathring{\mathcal{M}}^{i}_{00})_{\text{Orb}}\Omega^l_{00} \Big)
    \\
    &- \frac{e}{4mc^2} \int_{BZ} \frac{d\textbf{k}}{(2\pi)^3} 
    \Big( ( g_{ss} \delta_{il} - g_{il} \Big) - \frac{m}{\hbar^2} \epsilon^{iab}\epsilon^{lcd} g_{ac} \alpha_{bd} \Big) .
\end{split}
\end{equation}
We have made some notational changes in transcribing their expression to equation (\ref{GaoSusceptibility}) and reinstated factors of $e$, $\hbar$ and $c$; the index $0$ denotes the band that is filled in the ground state. The magnetization matrix element Gao et al. \cite{GaoGeometricalSus} employ is exactly the orbital contribution of the `purified' matrix elements we defined earlier in equation (\ref{PureMnm}) and (\ref{PureMnn}). The final line of equation (\ref{GaoSusceptibility}) is written in terms of the quantum metric of \textbf{k}-space, $g_{il} = \sum_{m}\text{Re}\Big[\xi^i_{0m}\xi^l_{m0}\Big] - \xi^i_{00}\xi^l_{00}$, and the inverse effective mass tensor $\alpha_{bd} = \partial_b \partial_d E_{0\textbf{k}}$.

Extending this expression to a ground state with a set of n filled bands each with a \textbf{k}-independent filling factor ($f_n$) of unity, we can see that the first two lines are of the same form as the `purified' contributions $\mathring{\chi}^{il}_\text{inter}$ and $\mathring{\chi}^{il}_\text{occ2}$. The final line can be rewritten using the definitions of the quantum metric and inverse effective mass tensor as
\begin{equation}
\begin{split}
    = -\frac{e^2}{4mc^2}  \epsilon^{iab}\epsilon^{lcd} \sum_{n,m\neq n} f_{n} \int_{BZ} \frac{d\textbf{k}}{(2\pi)^3} \text{Re}\Bigg[ \xi^a_{nm}\xi^c_{mn}
    \Big(\delta_{bd} -\frac{m}{\hbar^2} \partial_b \partial_d E_{n\textbf{k}}\Big) \Bigg]. 
\end{split}
\end{equation}

Therefore, in the limit of an insulating crystal we see that the expression derived by Gao et al. \cite{GaoGeometricalSus} can be written as
\begin{equation}
    \chi^{il}_\text{Gao} = \mathring{\chi}^{il}_\text{inter:Orb} + \mathring{\chi}^{il}_\text{occ} 
    + \frac{3}{2} \mathring{\chi}^{il}_\text{occ2:Orb},
\end{equation}
where $\mathring{\chi}^{il}_\text{inter:Orb}$ indicates the contribution to $\mathring{\chi}^{il}_\text{inter}$ that is independent of spin.  Thus, neglecting spin contributions, in the limit of an insulating crystal the result of Gao et al. \cite{GaoGeometricalSus} is in agreement with the result of Ogata \cite{OgataMagSus2017} and with our result.  
Blount \cite{BlountMagSus} also does a free energy expansion and, reinstating factors of $e$, $\hbar$, and $c$, we can identify the terms that remain in his expansion in the insulating crystal limit as
\begin{equation}
\begin{split}
    &F_{VD} =\sum_n f_{n} \int_{BZ} \frac{d\textbf{k}}{(2\pi)^3} E'_{VD} 
    \\
    &F_{at} = \sum_n f_{n} \int_{BZ} 
    \frac{d\textbf{k}}{(2\pi)^3}  
    \sum_{l,b} 
    \frac{\mathcal{A}^b_{nl}\mathcal{A}^b_{ln}}{2m}  
    \\
    &F_{pa} = - \sum_n f_{n} \int_{BZ} \frac{d\textbf{k}}{(2\pi)^3} \sum_{l,b,d} \alpha_{bd} \frac{ \mathcal{A}^b_{nl} \mathcal{A}^d_{ln} }{2}  
    \\
    &F_\Omega = \frac{3e}{8\hbar c} \sum_{b,d} B^{b}B^{d} \sum_n f_{n} \int_{BZ} \frac{d\textbf{k}}{(2\pi)^3}  \Big[ \Omega^{b}_{nn'}(\mathring{\mathcal{M}}^{d}_{n'n})_{\text{Orb}}
    \\
    &\hspace{20pt}+(\mathring{\mathcal{M}}^{d}_{nn'})_{\text{Orb}}\Omega^{b}_{n'n}\Bigg].
\end{split}
\end{equation}
The matrix elements $\mathcal{A}$ are purely interband and are given by  
\begin{equation}
    \mathcal{A}^{b}_{nm} = \frac{e}{2c} B^i \epsilon^{iab} \xi^a_{nm}, 
\end{equation}
for $n\neq m$, $B^{i}$ indicates the components of the magnetic field, and  
\begin{equation}
    E_{VD}' = \sum_{n} \sum_{n\neq m} \frac{ |\Big(\mathcal{A}\cdot\textbf{V} + \{\textbf{v}\cdot\mathcal{A}\}_{+} \Big)_{nm}|^2}{E_{n}-E_{m}}, 
\end{equation}
where $V^a_{mn}$ are the interband matrix elements $v^a_{mn}$ in our notation, and $v^a_{nn}$ are the intraband matrix elements. So we can write these free-energy terms in our notation as

\begin{equation}
\label{FVD}
\begin{split}
    F_{VD} = & B^s B^t \epsilon^{sab}\epsilon^{tcd} \sum_{n \neq m} f_{n} \int_{BZ} \frac{d\textbf{k}}{(2\pi)^3} \frac{1}{E_{n\textbf{k}}-E_{m\textbf{k}}}
    \\
    &\frac{e}{2c}\Big( \sum_{s\neq n,m} \xi^a_{ns} v^b_{sm} + \frac{1}{\hbar} \partial_b(E_{n\textbf{k}}+E_{m\textbf{k}}) \xi^a_{nm} 
    \Big)
    \\
    &\times \frac{e}{2c}\Big( \sum_{s \neq n,m } \xi^c_{ns} v^d_{sm} + \frac{1}{\hbar} \partial_b (E_{n\textbf{k}}+E_{m\textbf{k}}) \xi^a_{nm} \Big)^*  ,
\end{split}
\end{equation}
and

\begin{equation}
\begin{split}
    F_{at} = \frac{e^2}{8mc^2} \epsilon^{sab}\epsilon^{tcd}  B^s B^t \sum_{n,m\neq n} f_{n} \int_{BZ} \frac{d\textbf{k}}{(2\pi)^3}  \delta_{bd} \xi^a_{nm} \xi^c_{mn} ,
\end{split}
\end{equation}

\begin{equation}
\begin{split}
    F_{pa} = - \frac{e^2}{8\hbar^2 c^2} B^s B^t \epsilon^{sab}\epsilon^{tcd} \sum_{n,m\neq n} f_{n} \int_{BZ} \frac{d\textbf{k}}{(2\pi)^3} \partial_b \partial_d E_{n\textbf{k}} 
    \\
    \times \Big(\xi^a_{nm}\xi^c_{mn} \Big)  ,
\end{split}
\end{equation}
and
\begin{equation}
\begin{split}
    F_\Omega = \frac{3e}{8\hbar c} B^s B^t \sum_{nn'} f_{n} \int_{BZ} \frac{d\textbf{k}}{(2\pi)^3} \Bigg[ \Omega^s_{nn'} (\mathring{\mathcal{M}}^{t}_{n'n})_{\text{Orb}}
    \\
    + (\mathring{\mathcal{M}}^{t}_{nn'})_{\text{Orb}} \Omega^s_{n'n} \Bigg].   
\end{split} 
\end{equation}
Here we used the fact that Blount's intraband magnetization matrix element is what we call the orbital part of the purified equal energy spontaneous magnetization matrix element $\mathring{\mathcal{M}}_{nn'}$; the purified spontaneous magnetization matrix elements can also be identified in equation (\ref{FVD}). The susceptibility is then found by taking the negative double derivative of the free energy with respect to the magnetic field, 
\begin{equation}
\begin{split}
    \chi^{il}_\text{Blount} = &-2\sum_{n\neq m} f_{n} \int_{BZ} \frac{d\textbf{k}}{(2\pi)^3} \text{Re}\Bigg[ \frac{ (\mathring{\mathcal{M}}^{i}_{nm})_{\text{Orb}}(\mathring{\mathcal{M}}^{l}_{nm})_{\text{Orb}}^*} {E_{n\textbf{k}}-E_{m\textbf{k}}}
    \Bigg] 
    \\
    &-\frac{e^2}{4mc^2} \epsilon^{iab}\epsilon^{lcd} \sum_{n \neq m} f_{n} \int_{BZ} \frac{d\textbf{k}}{(2\pi)^3} \text{Re}\Bigg[ 
    \xi^a_{nm}\xi^c_{mn}
    \\
    &\hspace{100pt}\times\Big( \delta_{bd} - \frac{m}{\hbar^2} \partial_b \partial_d E_{n\textbf{k}}\Big)
    \Bigg] 
    \\
    &-\frac{3e}{4\hbar c} \sum_{nn'} f_{n} \int_{BZ} \frac{d\textbf{k}}{(2\pi)^3} \text{Re}\Bigg[ 
    \Omega^i_{nn'} (\mathring{\mathcal{M}}^{l}_{n'n})_{\text{Orb}} \\
    &\hspace{100pt}+ (\mathring{\mathcal{M}}^{i}_{nn'})_{\text{Orb}}\Omega^l_{n'n} 
    \Bigg] 
    \\
    = &\mathring{\chi}^{il}_\text{inter:Orb} + \mathring{\chi}^{il}_\text{occ} + \frac{3}{2} \mathring{\chi}^{il}_\text{occ2:Orb} 
    \\
    = &\chi^{il}_\text{Gao} 
\end{split}
\end{equation}

So despite the different strategies, our result (which agree with those of Ogata \cite{OgataMagSus2017}, see section \ref{SectionOgataComparison}), agrees with those of Gao et al. \cite{GaoGeometricalSus}  and Blount \cite{BlountMagSus} when the spin effects that they neglected are omitted. 

\end{widetext}

\bibliographystyle{apsrev4-2}
\bibliography{references.bib}

\end{document}